\documentclass[a4paper,11pt]{amsart}
\newcommand{\emailadd}{2} 
\usepackage{t1enc}

\def\Bbb{\mathbb}
\def\Cal{\mathcal}

\newcommand{\newc}{\newcommand}

\let\ccdot\cdot
\def\cdot{\hbox to 2.5pt{\hss$\ccdot$\hss}}

\newcommand{\de}{\delta}

\newcommand{\om}{\omega}
\renewcommand{\phi}{\varphi}

\newcommand{\si}{\sigma}

\newcommand{\De}{\Delta}

\newcommand{\Om}{\Omega}

\newc{\aR}{\mbox{\boldmath{$ R$}}}
\newc{\aS}{\mbox{\boldmath{$ S$}}}
\newc{\aDeR}{\mbox{\boldmath{$ U$}}_B{}^P{}_C{}^Q}
\newc{\aDe}{\mbox{\boldmath$ \Delta$}}
\newc{\aNd}{\mbox{\boldmath$ \nabla$}}

\newc{\aK}{\mbox{\boldmath{$ K$}}}
\newc{\aL}{\mbox{\boldmath{$ L$}}}

\newtheorem{theorem}{Theorem}[section]
\newtheorem{lemma}[theorem]{Lemma}
\newtheorem{proposition}[theorem]{Proposition}
\newtheorem{corollary}[theorem]{Corollary}

\newcommand{\cg}{{\Cal G}}
\newcommand{\cv}{{\Cal V}}
\newcommand{\cw}{{\Cal W}}
\newcommand{\ce}{{\Cal E}}
\newcommand{\cq}{{\Cal Q}}

\newcommand{\ct}{{\Cal T}}

\usepackage{amssymb}
\usepackage{amscd}

\newcommand{\nd}{\nabla}

\newcommand{\Rho}{{\mbox{\sf P}}}
\newcommand{\Up}{\Upsilon}

\newcommand{\nn}[1]{(\ref{#1})}

\newcommand{\bD}{\mbox{\boldmath{$ D$}}}
\newcommand{\D}{\mbox{\boldmath{$ D$}}}
\newcommand{\X}{\mbox{\boldmath{$ X$}}}
\newcommand{\sX}{\mbox{\scriptsize\boldmath{$X$}}}        
\newcommand{\h}{\mbox{\boldmath{$ h$}}}
\newcommand{\bg}{\mbox{\boldmath{$ g$}}}

\newcommand{\cce}{\tilde{\ce}}                          
\newcommand{\tM}{\tilde{M}}
\newcommand{\tS}{\tilde{S}}

\newcommand{\tf}{\tilde{f}}
\newcommand{\tW}{R}
\newcommand{\tV}{\tilde{V}}
\newcommand{\tU}{\tilde{U}}

\newcommand{\vol}{\mbox{\boldmath $ \epsilon$}}

\let\s=\sigma
\let\t=\tau
\let\m=\mu

\newcommand{\V}{{\mbox{\sf P}}}                   
\newcommand{\J}{{\mbox{\sf J}}}

\newc{\strutdd}{\rule{0mm}{5mm}}

\newcommand{\rpl}                         
{\mbox{$
\begin{picture}(12.7,8)(-.5,-1)
\put(0,0.2){$+$}
\put(4.2,2.8){\oval(8,8)[r]}
\end{picture}$}}

\newcommand{\lpl}                         
{\mbox{$
\begin{picture}(12.7,8)(-.5,-1)
\put(2,0.2){$+$}
\put(6.2,2.8){\oval(8,8)[l]}
\end{picture}$}}

\usepackage{ifthen}

\newc{\tensor}[1]{#1}
\newc{\Mvariable}[1]{\mbox{#1}}
\newc{\down}[1]{{}_{
\ifthenelse{\equal{#1}{;}}{|}{#1}}}
\newc{\up}[1]{{}^{#1}}
\newc{\C}{C}

\newc{\JulyStrut}{\rule{0mm}{6mm}}
\newc{\midtenPan}{\mbox{\sf S}}
\newc{\midten}{\mbox{\sf T}}
\newc{\midtenEi}{\mbox{\sf U}}
\newc{\ATen}{\mbox{\sf E}}
\newc{\BTen}{\mbox{\sf F}}
\newc{\CTen}{\mbox{\sf G}}

\def\sideremark#1{\ifvmode\leavevmode\fi\vadjust{\vbox to0pt{\vss
 \hbox to 0pt{\hskip\hsize\hskip1em
 \vbox{\hsize3cm\tiny\raggedright\pretolerance10000
 \noindent #1\hfill}\hss}\vbox to8pt{\vfil}\vss}}}%
                        
\begin{document}

\begin{abstract}
  We describe an elementary algorithm for expressing, as explicit
  formulae in tractor calculus, the conformally invariant GJMS
  operators due to C.R. Graham et alia. These differential operators
  have leading part a power of the Laplacian. Conformal tractor
  calculus is the natural induced bundle calculus associated to the
  conformal Cartan connection.  Applications discussed include
  standard formulae for these operators in terms of the Levi-Civita
  connection and its curvature and a direct definition and formula for
  T. Branson's so-called $Q$-curvature (which integrates to a global
  conformal invariant) as well as generalisations of the operators and
  the $ Q$-curvature.  Among examples, the operators of order 4, 6 and
  8 and the related $Q$-curvatures are treated explicitly.  The
  algorithm exploits the ambient metric construction of Fefferman and
  Graham and includes a procedure for converting the ambient curvature
  and its covariant derivatives into tractor calculus expressions.
  This is partly based on \cite{CapGoFG}, where the relationship of the
  normal standard tractor bundle to the ambient construction is described.
%
%
%
\ifthenelse{\emailadd=1}
{ 
\vspace{3mm}
\\
\newlength{\eaddwid}
\settowidth{\eaddwid}{Authors' e-mail addresses:\ \ }
\parbox[t]{\eaddwid}{Authors' e-mail addresses:}
\parbox[t]{55mm}{
\texttt{gover@math.auckland.ac.nz}\\
\texttt{lawrence.peterson@und.nodak.edu}
}
}  
{} 
%
%
\end{abstract}

\title{Conformally
  invariant powers of the Laplacian, $Q$-curvature, and tractor
  calculus}
\author{A. Rod Gover and Lawrence J. Peterson}
\date{}
\address{Department of Mathematics\\
  The University of Auckland\\
  Private Bag 92019\\
  Auckland 1\\
  New Zealand} \email{gover@math.auckland.ac.nz}
\address{Department of Mathematics\\
  The University of North Dakota\\
  Grand Forks, ND 58202-8376\\
  USA} \email{lawrence.peterson@und.nodak.edu}

\maketitle

\pagestyle{myheadings}
\markboth{A.R. GOVER AND L.J. PETERSON}{CONFORMALLY INVARIANT POWERS}

\section{Introduction}

Conformally invariant differential operators have long been known to
play an important role in physics and the geometry of many structures
related to and including Riemannian and conformal geometries. For
example, the classical field equations describing massless particles,
including the Maxwell and Dirac (neutrino) equations, depend only on
conformal structure \cite{bateman,dirac}. More recently string theory
and quantum gravity have motivated several developments in mathematics
where conformally invariant operators play a key role.  Many of these
could be said to fall under the umbrella of geometric spectral theory
where, broadly, one attempts to relate global geometry to the spectrum
of some natural operators on the manifold. For example, on compact
manifolds there are programmes to find extremal metrics for functional
determinants of natural operators. Conformally invariant operators
yield determinants with a workable formula (a so called Polyakov
formula) for the conformal variation of the determinant thus leading to
significant progress \cite{BO3,BCY,Brsharp}. In another direction there
is new progress \cite{GrZ2} in relating scattering matrices on
conformally compact Einstein manifolds with conformal objects on their
boundaries at infinity. This falls within the framework of the AdS/CFT
correpondence of quantum gravity \cite{Witten98,HS1,HS2,GrW}.

In these areas it seems an especially important role is played by
natural conformally invariant operators with principal part a power of
the Laplacian $ \Delta$. The earliest known of these is the
conformally invariant wave operator which was first constructed for
the study of massless fields on curved spacetime. More recently its
Riemannian signature variation, usually called the Yamabe operator,
has played a large role in the Yamabe problem on compact Riemannian
manifolds. As an operator on functions it is given by the formula
$\Delta -(n-2)R/(4(n-1))$,
and it governs the transformation of the
scalar curvature $ R$ under conformal rescaling.  An operator with
principal part $ \Delta^2$ is due to Paneitz \cite{Pan} (see also
\cite{Rie,ES}), and then sixth-order analogues were constructed in
\cite{Br85,Wuensch86}. Graham, Jenne, Mason and Sparling (GJMS) solved
a major existence problem in \cite{GJMS} where they used a formal
geometric construction to show the existence of conformally invariant
differential operators $ P_{2k}$ (to be referred to as the GJMS
operators) with principal part $ \Delta^k$.  In odd dimensions, $
k$ is any positive integer, while in dimension $ n$ even, $k $ is a
positive integer no more than $ n/2$.  The $ k=1$ and $ k=2$ cases
recover, respectively, the Yamabe and Paneitz operators.

In dimension 2 the transformation of the scalar curvature can also be
deduced from the Yamabe operator by a dimensional continuation
argument, and the curvature fixing problem corresponding to the Yamabe
problem is usually known as Gauss curvature prescription.  In the late
1980's Branson \cite{Brpriv,BO3} observed that the Paneitz operator
$P$ is formally self-adjoint and can be expressed in the form
$P^1+((n-4)/2)Q_4$, where $ P^1$ annihilates constant functions and $
Q_4$ is a scalar curvature invariant which could play a role parallel
to the scalar curvature in higher order analogues of the Gauss
curvature prescription programme.  In dimension 4 the conformal
transformation of $ Q_4$ is given by the Paneitz operator, and it
follows that the integral of $ Q_4$ over compact 4-manifolds is a
global conformal invariant. On conformally flat structures this is a
multiple of the Euler characteristic. It has recently been established
by Graham and Zworski and Fefferman and Graham
\cite{GrZ1,GrZ2,FeffGr01} that the GJMS operators $ P_{2k}$ are
formally self-adjoint, and so \cite{Brsharp} shows that these
operators yield an analogous local Riemannian invariant $ Q_{n}$ for
each even-dimensional manifold. There has been considerable recent
interest and progress in understanding Branson's $ Q$-curvatures,
especially in low dimensions and on conformally flat structures
\cite{CQY,CY}.

In \cite{GJMS} the GJMS operators are derived from the Laplacian of
the ambient metric of Fefferman and Graham \cite{FG,FG2}.  This
construction is very valuable not only in itself but also because of
the close links with the Poincar\'{e} metrics of the conformally
compact Einstein theory.  On the other hand there is another way to
generate a conformally invariant operator with principal part $
\Delta^k$.  The result is usually presented as a simple formulae,
first due to M.G. Eastwood, as given in \nn{defboxk}. (See \cite{Gosrni}
for a derivation and some further related developments.)  
  Underlying this formula are two related key tools.
The first is a geometric construction developed by Eastwood and others
\cite{EastRice,Esrni} known as the curved translation principle.  This
construction is a generalised and geometric variant of the 
translation functor due to Zuckerman and others \cite{Z}. The second
is a machinery known as tractor calculus
\cite{BEGo,Goadv,Cap-Gover,Cap-Gover2}. This calculus brings the
conformally invariant Cartan connection to induced bundles and also
involves other fundamental conformally invariant operators (such as
the ones used in this formula).  The combination is potent since on
the one hand it is very easy to expand these tractor formulae in terms
of the Levi-Civita connection and its curvature (which is useful for
the investigation of issues such as positivity of the operators), and
on the other hand the link with representation theory means one easily
obtains rules for generalising the operators and how they may be
composed with certain other conformally invariant operators. See for
example \nn{strongboxes}.  It should be pointed out that the tractor
formulae are themselves complete and explicit formulae and can be
readily worked with directly without using any knowledge of the
representation theory aspects. That is essentially the approach below.
See also \cite{BrGo}, for example, where these tractor formulae for
conformally invariant powers of the Laplacian are used to construct
formally self-adjoint conformally invariant boundary problems, higher
order conformally invariant Dirichlet-to-Neumann operators, and
related constructions.

One problem with the tractor approach up until now has been that, on
even dimension $ n$ manifolds, this had failed to yield the operators
of order $ n$ except for a quotient construction in dimension 4
\cite{Gosrni}.  Here we give a similar quotient tractor construction
for a sixth-order operator and show that we have in fact recovered
$ P_4$ and $ P_6$.  This brings us to one of the main purposes of this
paper, which is to explicitly relate the tractor calculus approach to
the GJMS construction.  This is achieved in Section~\ref{GJMS}, where
an algorithm is described for finding a tractor formula for any of the
GJMS operators $ P_{2k}$. Remarkably this algorithm does not require
solving the Fefferman-Graham ambient construction.  For low order
operators it is essentially trivial and quickly recovers the simple
tractor formulae for $ P_4$ and $ P_6$ and yields a corresponding
tractor formula for $ P_8$. See Section~\ref{examples} and
Proposition~\ref{p468}.
In Proposition \ref{psfsa} we use these formulae to prove directly
that these operators are formally self-adjoint (verifying directly for
these cases the general results of \cite{GrZ2,FeffGr01}).  Expanding these
formulae into formulae in terms of the Levi-Civita connection and its
curvature simply requires repeated use of the Leibniz rule and the
definitions of the tractor objects.  This is easily automated and is
done in Section~\ref{convent}. The nature of the formulae we use mean
the calculations have a large number of built-in self-checks which
ensure that the formulae used are entered and used correctly by the
software.  Thus overall this demonstrates an effective means to obtain
explicit formulae for the GJMS operators.  It should be pointed out
that the formulae in Section~\ref{convent} are not in fact the raw
output from the expansion of the tractor formulae, but rather this
output manipulated into the canonical form described in \cite{EGoTN}.
The authors performed these expansions and manipulations mainly by
using \textit{Mathematica} and J. Lee's Ricci programme
\cite{JackLee}; this work was performed under the 
assumption of a Riemannian signature metric, but the resulting
formulae are independent of the signature.

The most important outcomes of Section~\ref{GJMS} are
Proposition~\ref{tract4gjms} and Theorem~\ref{L24Maybb}. The first of
these establishes important features about the form of the tractor
formulae for the GJMS operators, and the latter exploits this to
provide some new invariant operators closely linked to the GJMS
operators. There are several applications of these. One is a direct
tractor based construction of Branson's $ Q$-curvatures.  See
Proposition~\ref{newQ}. In fact, this also gives a new
definition for these invariants.  This gives an effective
way to calculate these ($ Q_4$ and $ Q_6$ are treated as examples), and
it sheds light on their remarkable transformation properties.  Another
application of Theorem~\ref{L24Maybb} is Corollary~\ref{strongp}. In
words this states that except for the $ k=n/2$ case, the theorem
yields generalisations of the GJMS operators $ P_{2k}$ that are
``strongly invariant'' in the sense of \cite{Esrni}. That is, operators
that can be composed with tractor bundle valued operators to yield
further conformally invariant operators. This is one of the key ideas
of the curved translation principle. Finally, Theorem~\ref{L24Maybb}
is a crucial ingredient in the general construction in \cite{BrGogau}
of an elliptic conformally invariant operator on 1-forms with close
connections to the first de Rham cohomology.

There are other results presented. For example, in
Section~\ref{Qcurvsect} we describe how to proliferate Riemannian
invariants which are not conformally invariant but have a
transformation formula similar to the Branson $Q $-curvatures. These
can be viewed as representing terms that could be added to the $
Q$-curvature without affecting its key properties and so play a role
in generating new curvature prescription problems. 

There are also many other potential applications for this work not
touched upon in this article. For example, the tractor formulae for the
GJMS operators could immediately be used in a construction parallel
to that in \cite{BrGo} to produce alternative conformally invariant
boundary problems and non-local operators based around the GJMS
operators.

It should also be pointed out that the results and ideas in this paper
should have analogues for CR structures, where one would instead be
involved with CR-invariant powers of the sub-Laplacian \cite{GoGr} and
the ambient construction of C.\ Fefferman \cite{Feff}.  The
construction presented in this article is in part an application of
ideas developed in the joint work of one of the authors with A.
\u{C}ap. See \cite{CapGoFG} where it is described explicitly how to
relate the Cartan/tractor approach to the ambient construction of
Fefferman and Graham and its applications to invariant theory.  The
relevant aspects of this theory are summarised in
Section~\ref{recovtract}.  There is a corresponding theory for the CR
case \cite{CapGoCRamb}.

The authors are indebted to Tom Branson, Andi \u{C}ap, Mike Eastwood,
and Robin Graham for several illuminating conversations.  The
authors would also like to thank the Mathematical Sciences Research
Institute and the organisers of Spring session in 2001 for helping to
make this research possible.

\section{Conformal geometry and tractor calculus}\label{tractorsect}

We summarise here an approach to local conformal geometry that is
rather useful for our applications. This is broadly based on the
development presented in \cite{Cap-Gover2}, but many of the ideas and
tools had their origins in \cite{T}, \cite{BEGo}, and \cite{Goadv}.
The notation and conventions in general follow the last two sources.

We shall work on a real conformal $n$-manifold $M$, where $n\geq 3$.
 That is, we have a pair
$(M,[g])$, where $M$ is a smooth $n$-manifold and $[g]$ is a conformal
equivalence class of metrics of signature $ (p,q)$.  Two metrics $g$
and $\widehat{g}$ are said to be {\em conformally equivalent} if
$\widehat{g}$ is a positive scalar function multiple of $g$. In this
case it is convenient to write $\widehat{g}=\Omega^2 g$ for some
positive smooth function $\Omega$.  Although we assume that the
metrics have some fixed signature, all considerations below
will be signature independent.  For a given conformal manifold
$(M,[g])$, we shall denote by $\Cal Q$ the bundle of metrics. That is,
$\Cal Q$ is a subbundle of $S^2T^*M$ with fibre $\Bbb R^+$.  The points
correspond to values of metrics in the conformal class.
  
Let $\ce^a$ denote the space of smooth sections of the tangent bundle
$ TM$, and similarly let $\ce_a$ be the smooth sections of the
cotangent bundle $ T^*M$. In fact, we will generally abuse notation and
also use these symbols to indicate the sheaves of germs of smooth
sections and even the bundles themselves.  These conventions will be
carried through to all bundles that we discuss. We write $\ce$ to
denote the trivial bundle over $M$. Penrose's abstract index notation
is embraced throughout, so tensor products of these bundles will be
indicated by adorning the symbol $\ce$ with appropriate abstract
indices.  For example, in this notation $\otimes^2T^*M$ is written
$\ce_{ab}$.  An index which appears twice, once raised and once
lowered, indicates a contraction.  These conventions will be extended
in an obvious way to the tractor bundles described below. In all
settings indices may also be ``suppressed'' (omitted) if superfluous
by context. 
   
  The bundle $\Cal Q$ is a principal bundle with group ${\Bbb R}_+$,
  so there are natural line bundles on $ (M,[g])$ induced from the
  irreducible representations of ${\Bbb R}_+$. We write $\ce[w]$ for
  the line bundle induced from the representation of weight $-w/2$ on
  ${\Bbb R}$ (that is ${\Bbb R}_+ \ni x\mapsto x^{-w/2}\in {\rm
    End}(\Bbb R)$).  Thus a section of $\ce[w]$ corresponds to a real-valued
  function $ f$ on $\Cal Q$ with the homogeneity property $f(x, \Omega^2
  g)=\Omega^w f(x,g)$, where $\Omega$ is a positive function on $M$,
  $x\in M$, and $g$ is a metric from the conformal class $[g]$. 
  We use the notation $\ce_a[w]$ for $\ce_a\otimes \ce[w]$ and so on.
  Note that for consistency with \cite{BEGo}, this convention differs
  in sign from the one of \cite[Section~4.15]{Cap-Gover}.
  
  Let $\ce_+[w]$ be the fibre subbundle of $\ce[w]$ corresponding to
  ${\Bbb R}_+\subset {\Bbb R}$. Choosing a metric $g$ from the
  conformal class defines a function $f:\Cal Q\to\Bbb R$ by
  $f(\hat{g},x)=\Omega^{-2}$, where $\hat g=\Omega^2g$, and this
  clearly defines a smooth section of $\ce_+[-2]$. Conversely, if $f$
  is such a section, then $f(g,x)g$ is constant up the fibres of $\Cal
  Q$ and so defines a metric in the conformal class. Thus $\ce_+[-2]$
  is canonically isomorphic to $\Cal Q$, and the {\em conformal
    metric\/} ${\mbox{\boldmath $g$}}_{ab}$ is the tautological
  section of $\ce_{ab}[2]$ that represents the map
  $\ce_+[-2]\cong{\Cal Q}\to\ce_{(ab)}$. From this there is a
  canonical section $\bg^{ab}$ of $\ce^{ab}[-2]$ such that
  $\bg_{ab}\bg^{bc}=\de_a{}^c$ (where $\de_a{}^c $ is the section of $
  \ce_a{}^c$
  corresponding to the identity endomorphism of the
  tangent bundle).  The conformal metric (and its inverse $\bg^{ab}$)
  will be used to raise and lower indices without further mention.
  Given a choice of metric $ g$ from the conformal class, we write $
  \nabla_a$ for the corresponding Levi-Civita connection. With these
  conventions the Laplacian $ \Delta$ is given by
  $\Delta=\bg^{ab}\nd_a\nd_b= \nd^b\nd_b\,$. In view of the
  isomorphism $\ce_+[-2]\cong{\Cal Q}$, a choice of metric also
  trivialises the bundles $ \ce[w]$. In particular we will write
  $\xi^g$ for the canonical section of $ \ce[1]$ satisfying
  $g=(\xi^g)^{-2}\bg $. Conversely a  section
  of $ \ce_+[1]$ clearly determines a metric by this relation, so such a
  $ \xi^g$ is termed a choice of conformal scale.  This determines a
  connection on $ \ce[w]$ via the corresponding trivialisation of $
  \ce[w]$ and the exterior derivative on functions. We shall also
  denote such a connection by $ \nabla_a$ and refer to it as the
  Levi-Civita connection. Note in particular then that, by definition,
  $ \nd_a \xi^g=0$, so $ \nd_a$ also preserves the conformal metric.
  The curvature $R_{ab}{}^c{}_d$ of the Levi-Civita connection is
  known as the Riemannian curvature, and is defined by
$$
(\nd_a\nd_b-\nd_b\nd_a)v^c=R_{ab}{}^c{}_dv^d .
$$
\pagebreak
This can be decomposed into the totally trace-free Weyl curvature
$C_{abcd}$ and a remaining part described by the symmetric {\em
  Rho-tensor} $\Rho_{ab}$, according to
$$
R_{abcd}=C_{abcd}+2\bg_{c[a}\Rho_{b]d}+2\bg_{d[b}\Rho_{a]c},
$$
where $[\cdots]$ indicates the antisymmetrization over the enclosed indices.
The Rho-tensor is a trace modification of the Ricci tensor $R_{ab}$. We write 
$ \J$ for the trace $ \V^a{}_{a}$ of $ \V$.

 Under a {\em conformal transformation} we replace our
  choice of metric $ g$ by the metric $ \hat{g}=\Omega^2 g$, where
  $\Omega$ is a positive smooth function.  The Levi-Civita connection
  then transforms as follows:
\begin{equation}\label{transform}
\widehat{\nd_a u_b}=\nd_a u_b -\Up_a u_b-\Up_b u_a +\bg_{ab} \Up^c u_c \quad
 \widehat{\nd_a \si} = \nd_a \si +w\Up_a \si.
\end{equation}
Here $ u_b\in \ce_b$, $ \si\in \ce[w]$, and 
$\Upsilon_a:=\Omega^{-1}\nabla_a\Omega$.
The Weyl curvature is
conformally invariant, that is $\widehat{C}_{abcd}=C_{abcd}$, and the
Rho-tensor transforms by
\begin{equation}\label{Rhotrans}
\textstyle \widehat{\V}_{ab}=\V_{ab}-\nd_a \Up_b +\Up_a\Up_b
-\frac{1}{2} \Up^c\Up_c \bg_{ab} .
\end{equation}

For the density bundle $\ce[1]$, we have the jet exact sequence at
2-jets,
$$
0\to \ce_{(ab)}[1]\to J^2(\ce[1])\to J^1(\ce[1])\to 0,
$$
where $ (\cdots)$ indicates symmetrization over the enclosed
indices.  Note we have a bundle homomorphism $ \ce_{(ab)}[1] \to \ce[-1]$
given by complete contraction with $\bg^{ab}$. This is split via 
$ \rho\mapsto  \frac{1}{n}\rho\bg_{ab}$ and so 
the conformal structure decomposes $\ce_{(ab)}[1]$ into the
direct sum $\ce_{(ab)_0}[1]\oplus\ce[-1]$.  
Clearly then  $\ce_{(ab)_0}[1]$ is a smooth subbundle of $J^2(\ce[1])$, and we
define $\ce^A$ to be the quotient bundle.  That is, the {\em standard
  tractor bundle} $\ce^A$ is defined by the exact sequence
\begin{equation}\label{trdef}
0\to \ce_{(ab)_0}[1]\to J^2(\ce[1])\to \ce^A\to 0.
\end{equation}
{}The jet exact sequence at 2-jets and the corresponding sequence at
1-jets, viz $ 0\to \ce_{a}[1]\to J^1(\ce[1])\to \ce[1]\to 0 , $
determine a composition series for $\ce^A$ which we can summarise 
via a self-explanatory semi-direct sum notation $ \ce^A= \ce[1]\lpl
\ce_a[1]\lpl\ce[-1]$. We denote by $ X^A$ the canonical section of $
\ce^A[1]:=\ce^A\otimes \ce[1]$ corresponding to the mapping $ \ce[-1]
\to \ce^A$.

Composing the canonical projection $J^2(\ce[1])\to\ce^A$ with the
2-jet operator $ j^2$ yields an invariant differential operator
$\tfrac1{n}D^A:\ce[1]\to \ce^A$.
On the other hand, if we choose a
metric $ g$ from the conformal class, then the map
$$
j^2_x\si \mapsto [\tfrac1{n}{D^A\si}(x)]_g:=(\si(x),\nabla_a\si(x),
-\tfrac{1}{n}(\Delta+\J)\si(x)), 
$$
induces an isomorphism $\ce^A\to
\ce[1]\oplus\ce_a[1]\oplus\ce[-1]=:[\ce^A]_g$ of vector bundles.
Tautologically the displayed formula for $\tfrac1{n}[{D^A\si}(x)]_g $
gives the operator $D^A$ in terms of this decomposition. If the image
of $ V^A\in\ce^A$ is $[V^A]_g=(\si,\mu_a,\tau) $, then from the change
in the  Levi-Civita connection \nn{transform} we get
$$
[V^A]_{\hat{g}}=\widehat{(\si,\mu_a,\tau)}=(\si,\mu_a+\si\Up_a,\tau-
\Up_b\mu^b- \tfrac{1}{2}\si\Up_b\Up^b).
$$
This transformation formula characterises sections of $\ce^A$ in terms
of triples in $ \ce[1]\oplus\ce_a[1]\oplus\ce[-1]$.
With a fixed rescaling of the map $\ce[-1]\to\ce^A$, we have
$ [X^A]_g=(0,0,1)$. 
It is convenient to introduce
scale-dependent sections $Z^A{}^b\in\ce^{Ab}[-1]$ and
$Y^A\in\ce^A[-1]$ mapping into the other slots of these triples so
that $[V^A]_g=(\si,\mu_a,\tau)$ is equivalent to
$$
V^A=Y^A\s+Z^{Ab}\m_b+X^A\t.
$$
If $\hat{Y}^A$ and $ \hat{Z}^A{}_b$ are the corresponding quantities in terms of the metric $ \hat{g}=\Omega^2g$ then we have   
\begin{equation}\label{XYZtrans}
\textstyle
\begin{array}{rl}
\hat Z^{Ab}=Z^{Ab}+\Up^bX^A, &
\hat Y^A=Y^A-\Up_bZ^{Ab}-\frac12\Up_b\Up^bX^A.
\end{array}
\end{equation}

The standard tractor bundle has an invariant metric $ h_{AB}$ of
signature $(p+1,q+1)$ and an invariant connection, which we shall also
denote by $ \nabla_a$, preserving $h_{AB}\,$. If $ V^A$ is as above
and $\underline{V}^B \in \ce^B$ is given by
$[\underline{V}^B]_g=(\underline{\s},\underline{\m}_b,
\underline{\t})$, then
$$
h_{AB}V^A\underline{V}^B=\m^b\underline{\m}_b+\s\underline{\t}+
\t\underline{\s}.
$$
Using $h_{AB}$ and its inverse to raise and lower indices, we
immediately see that 
$$
Y_AX^A=1,\ \ Z_{Ab}Z^A{}_c=\bg_{bc}
$$
and that all other quadratic combinations that contract the tractor
index vanish. This is summarised in Figure~\ref{Larry22April01a}.  
\begin{figure}
$$
\begin{array}{l|ccc}
& Y^A & Z^{Ac} & X^{A}
\\
\hline
Y_{A} & 0 & 0 & 1
\\
Z_{Ab} & 0 & \de_{b}{}^{c} & 0
\\
X_{A} & 1 & 0 & 0
\end{array}
$$
\caption{Tractor inner product}
\label{Larry22April01a}
\end{figure}
Thus we also have
$
Y_AV^A=\t,\ \ X_AV^A=\s,\ \ Z_{Ab}V^A=\m_b\, $
and the metric may be decomposed into a sum of projections, $
h_{AB}=Z_A{}^cZ_{Bc}+X_AY_B+Y_AX_B\,$.

If for a metric $ g$ from the conformal class $V^A \in\ce^A$ is given by 
$[V^A]_g=(\si,\mu_a,\tau)$, then the invariant  connection is given by
\renewcommand{\arraystretch}{1}
\begin{equation}\label{stdtracconn}
[\nabla_a V^B]_g =
\left(\begin{array}{c} \nabla_a \si-\mu_a \\
                       \nabla_a \mu_b+ \bg_{ab} \tau +\V_{ab}\si \\
                       \nabla_a \tau - \V_{ab}\mu^b  \end{array}\right) . 
\end{equation}
\renewcommand{\arraystretch}{1.5}

The tractor metric will be used to raise and lower indices without
further comment.  We shall use either ``horizontal'' (as in
$[V^B]_g=(\si,\mu_b,\tau)$) or ``vertical'' (as in \nn{stdtracconn})
notation, depending on which is clearer in each given situation.

Tensor products of the standard tractor bundle, skew or symmetric
parts of these and so forth are all termed tractor bundles.  The
bundle tensor product of such a bundle with $ \ce[w]$, for some real
number weight $ w$, is termed a weighted tractor bundle. For example
$\ce_{A_1A_2\cdots A_\ell}[w]= \ce_{A_1}\otimes \cdots \otimes
\ce_{A_\ell}\otimes \ce[w]$ is a weighted tractor bundle.  Given a  choice
of conformal  scale we have the corresponding Levi-Civita connection on tensor and density bundles.  In this
setting we can use the coupled Levi-Civita tractor connection to act
on sections of the tensor product of a tensor bundle with a tractor
bundle. This is defined by the Leibniz rule in the usual way.  For
example if
$ u^b V^C \sigma\in \ce^b\otimes \ce^C\otimes \ce[w]=:
\ce^{bC}[w]$,
then
$ \nd_a u^b V^C \sigma = (\nd_a u^b) V^C \sigma +
u^b(\nd_a V^C) \sigma + u^b V^C \nd_a \sigma$.
Here $\nd$ means the Levi-Civita
connection on $ u^b\in \ce^b$ and $ \si\in \ce[w]$,
while it denotes the tractor
connection on $ V^C\in \ce^C$. In particular with this convention we have 
\begin{equation}\label{connids}
\begin{array}{rcl}
\nd_aX_A=Z_{Aa}\,, &
\nd_aZ_{Ab}=-\V_{ab}X_A-Y_A\bg_{ab}\,, & \nd_aY_A=\V_{ab}Z_A{}^b ,
\end{array}
\end{equation}
which for the purposes of automating calculations is a very useful
description of the tractor connection.

Note that if $V$ is a section of $ \ce^\Phi[w]$, which means simply some
tractor bundle of weight $ w$, then the coupled Levi-Civita tractor
connection is not confomally invariant but transforms just as the
Levi-Civita connection transforms on densities of the same weight.
That is
$$
\widehat{\nd}_a V = \nd_a V + w\Up_a V
$$
under the conformal rescaling $ g\mapsto \hat{g}=\Omega^2 g$ (cf.
\nn{transform}). It is an elementary  exercise using the last
transformation formulae and  \nn{XYZtrans} to show that,
for $V\in \ce^\Phi[w]$, the formula
\begin{equation}\label{doubleD}
D^{AP}V:= 2w X^{[P}Y^{A]}V + 2X^{[P}Z^{A]b}\nd_b V  
\end{equation}
determines an invariant operator $D^{AP} : \ce^\Phi[w]\to
\ce^{[AP]}\otimes\ce^\Phi[w]$. (This was first developed in early
versions of \cite{Goadv} and is closely related to the ``fundamental $
D$'' operator developed in \cite{Cap-Gover}.) Since we can vary the
weight and the
tractor bundle $ \ce^\Phi$, $D^{AP}$ is really an entire family of
operators. The point is that with the way we have defined $ \nd$, the
same formula works for the entire family, and so it is reasonable to
let the single symbol $ D^{AP}$ denote all of these operators.  We
abuse terminology and describe it as an operator.  (The Levi-Civita
connection  is usually  used this way.)  If we have a single formula Op
that gives a family of conformally invariant operators
$$
{\rm Op}: \ce^\Phi \otimes \cv \to \ce^\Phi \otimes \cw
$$
as we range over all tractor bundles $ \ce^\Phi$ then, following
\cite{Esrni}, we describe Op as a {\em strongly invariant} operator.
For example $ D^{AP}$ is strongly invariant. As already pointed out $
D^{AP}$ is rather more universal since we can vary the weight $ w$ as
well. Thus we can form compositions of this operator with itself, and
in particular we consider $ h^{AB}D_{A(Q}D_{|B|P)_0} V $ for $V\in
\ce^\Phi[w]$ some weighted tractor.  Expanding this out using
\nn{connids}, \nn{doubleD}, and the Leibniz rule for $\nabla$,
it is easily verified that it may be re-expressed in the form
$$
h^{AB}D_{A(Q}D_{|B|P)_0}V= -X_{(Q}D_{P)_0}V,
$$
where $D$ is some operator determined explicitly in the calculation.  
Since the map $\ce_P[w-1] \to \ce_{(PQ)_0}[w]$ given by $ S_P\mapsto
X_{(Q}S_{P)_0} $ is injective, this establishes
$ D_A :\ce^\Phi[w] \to \ce_A \otimes \ce^\Phi[w-1] $
as a conformally invariant differential operator on weighted tractor
bundles.  For $ V\in\ce^\Phi[w]$, this is given by
\begin{equation}\label{Dform}
D^A V:=(n+2w-2)w Y^A V+ (n+2w-2)Z^{Aa}\nabla_a V -X^A\Box V,
\end{equation} 
 where
\begin{equation} \label{box} 
\Box V :=\nd_p\nd^p V+w\J V.
\end{equation}
 So $D_A$ is in fact precisely the tractor $D$-operator in \cite{BEGo}.
Note the identity
\begin{equation} \label{Dident}
D_AX^A V = (n+2w+2)(n+w)V,
\end{equation}
which we will use later.

The curvature $ \Omega$ of the tractor connection 
is defined by 
$$
[\nd_a,\nd_b] V^C= \Omega_{ab}{}^C{}_EV^E 
$$
for $ V^C\in\ce^C$ and is precisely the local obstruction to
conformal flatness. (That is locally there is a flat metric in the
conformal class if and only if this curvature vanishes.)  Using
\nn{connids} and the usual formulae for the curvature of the
Levi-Civita connection we calculate (cf. \cite{BEGo})
\begin{equation}\label{tractcurv}
\Omega_{abCE}= Z_C{}^cZ_E{}^e C_{abce}-4X_{[C}Z_{E]}{}^e\nd_{[a}\Rho_{b]e}. 
\end{equation}
It is straightforward to use this and \nn{connids} to show that if $
V\in \ce_{CE\cdots F}[w]$, then
\begin{equation}\label{Dcomm}
\begin{array}{lll}
\lefteqn{[D_{A},D_{B}] V_{CE\cdots F}=}&&
\\
&&
(n+2w-2)[W_{ABC}{}^Q V_{QE\cdots F} +2w \Omega_{ABC}{}^Q V_{QE\cdots F}
\\
&&
+4X_{[A}\Omega_{B]}{}^{s}{}_C{}^Q \nd_sV_{QE\cdots F}+\cdots
+ W_{ABF}{}^Q V_{CE\cdots Q}
\\&&
+2w \Omega_{ABF}{}^Q V_{CE\cdots Q}
+4X_{[A}\Omega_{B]}{}^{s}{}_F{}^Q \nd_sV_{CE\cdots Q}] 
.
\end{array}
\end{equation} 
Here $\Omega_{ABCE}=Z_A{}^aZ_B{}^b\Omega_{abCE} $,
$\Omega_{BsCE}=Z_B{}^b\Omega_{bsCE} $, and
\begin{equation}
\label{WDef}
W_{ABCE}=(n-4)\Omega_{ABCE}-2X_{[A}Z_{B}{}_{]}^b\nd^p \Omega_{pbCE}.
\end{equation}
It follows that on conformally flat structures $[D_{A},D_{B}]
V_{CE\cdots F}=0$.  Similarly
it is easily verified that $[D_{B},D_{C}]$ annihilates densities.

We should point out some features of $ W_{ABCE}$. Firstly, it
is conformally invariant. One can already see this from \nn{Dcomm} by
setting $ w=0$ and then considering sections $ V_C$ of $ \ce_C$ such
that $ \nd_a V_C$ vanishes at a given point.  This is also immediately
clear from the formula $
W_{AB}{}^{K}{}_{L}:=\frac{3}{n-2}D^PX_{[P}\Omega_{AB]}{}^{K}{}_{L}$
(see \cite{Goadv}), which is readily verified. From this several
things are immediately clear.
Firstly, $W_{ABCE} $ vanishes on conformally flat structures. Next, we have
 that $
W_{ABCE}=W_{[AB][CE]}$ and that it is trace-free (since $\Omega_{ABCE}
$ is annihilated by contraction with $ X^P$ on any index). Furthermore
expanding \nn{WDef} reveals that $W_{[ABC]E}=0$. Thus $ W_{ABCE}$ has
``Weyl tensor symmetries''. Whence it is immediately clear that $
W_{ABCE}$ is also annihilated upon contraction with $ X^P$.

Finally we should comment on the uniqueness of this tractor calculus.
In sections 2.6 and 2.7 of \cite{Cap-Gover2} it is shown that the
transformation properties \nn{XYZtrans} and the form of the connection
\nn{connids} identify $ \ce^A$ and its tractor connection $ \nd_a$ as
above as a normal tractor bundle and connection corresponding to the
defining representation of SO$(p+1,q+1)$. Let $ \Bbb V$ be $ {\Bbb
  R}^{n+2}$ as the representation space for the standard (or defining)
representation of SO$(p+1,q+1)$. We can construct \cite{Cap-Gover}
from the pair ($\ce^A,\nd_a)$ a principal bundle $ \cg$ which is the
frame bundle for $ \ce^A$ corresponding to the metric and filtration.
This has fibre $P$, a certain parabolic subgroup of SO$(p+1,q+1)$.  A
Cartan connection $\omega $ on $ \cg$ is determined by $ \nd$. This is
the normal Cartan connection on $ \cg$ such that $ \nd_a$ is the
vector bundle connection induced from $ \om$. That is the normality condition
on the pair $(\ce^A,\nd_a)$ is equivalent to the pair $( \cg,\omega)$
being a normal Cartan bundle and connection in the sense of
\cite{Cartan}.

\subsection{Conformally invariant powers of the Laplacian} \label{powersLap}

Since the tractor-$D$ operator constructed above is well-defined on any
weighted tractor bundle, we can compose the tractor-$D$ operators. It is
clear from the formula for the tractor-$D$ operator that any such
composition will yield a {\em natural} operator, that is an operator
which can be written as a polynomial formula in terms of a
representative metric, its inverse, the metric connection and its curvature. 
On densities of the
appropriate weight and with some minor adjustment a composition of this form 
will lead to
conformally invariant operators with principal part a power of the Laplacian.
 
First let us observe how the conformal Laplacian arises from the
tractor machinery. Let $f\in \ce[1-n/2]$. Then observe that
immediately from \nn{Dform} we have $ D_A f= - X_A \Box f$.  Since
$D_A$ is conformally invariant we have immediately that for $f\in
\ce[1-n/2]$, $\Box f$ is conformally invariant. From \nn{box} this is
$(\nd^a\nd_a+\frac{2-n}{2}\J)f$ -- the usual conformal Laplacian. 
 
Now
suppose that instead we have $f\in \ce^\Phi[1-n/2]$, where here and
below $ \ce^\Phi[w]$ will be used to indicate any tractor bundle of
weight $ w$. We still have
\begin{equation}\label{key}
 D_A f=
-X_A \Box f
,
\end{equation}
 but now in $\Box f= (\nd^a\nd_a+\frac{2-n}{2}\J) f $,
$\nd $ means the Levi-Civita tractor coupled connection.
In particular this establishes a strongly invariant
generalisation of the Yamabe operator on
tractor sections of the said weight. 

It is clear from our observations that  that there is a
conformally invariant operator
$$
\Box D_{A_1}D_{A_2}\cdots D_{A_{k-1}}:\ce[k-n/2]\to
\ce_{A_1A_2\cdots A_{k-1}}[-1-n/2] .
$$
In the conformally flat case this already yields an operator
between densities (cf.\ \cite{Gosrni}).
\begin{proposition} \label{tractflatcase} 
On conformally flat structures, if $f\in \ce[k-n/2]$,  
then 
  $$
  \Box D_{A_{k-1}}\cdots D_{A_1} f =(-1)^{k-1}
  X_{A_1}\cdots X_{A_{k-1}} \Box_{2k} f
  ,
  $$
  where $ \Box_{2k}:\ce[k-n/2]\to \ce[-k-n/2]$ is a conformally
  invariant operator. Locally we can choose a flat metric from the
  conformal class. This determines a connection in terms of which  we have
  $\Box_{2k} f= \Delta^k f $.
\end{proposition}
\begin{proof}
  In any choice of conformal scale, 
  expand out $ \Box D_{A_{k-1}}\cdots D_{A_1} f$ via the
  formula \nn{Dform} and move the $X,Y,Z$'s to the left of all
  $\nd$'s via the identities \nn{connids}. It is immediately clear
  that the highest order term is precisely $(-1)^{k-1}X_{A_1}\cdots
  X_{A_{k-1}} \Delta^k f $ and that any other  coefficient of
  $X_{A_1}\cdots X_{A_{k-1}}$ involves the curvature $\Rho_{ab}$ or its trace. 
    
    On the other hand, on conformally flat
    structures, $ [D_A,D_B] V=0$ for $ V$ any weighted tractor field.
    Thus $ D_{A_\ell} \cdots D_{A_1} f$ is completely symmetric for
    any $ \ell\in {\Bbb Z}_+$.  In particular $\Box D_{A_{k-1}} \cdots
    D_{A_1} f\in \ce_{(A_1 \cdots A_{k-1})}[-1-n/2]$ and $D_{A_k}
    D_{A_{k-1}} \cdots D_{A_1} f\in \ce_{(A_1 \cdots A_{k})}[-n/2]$.
    Consequently it must be that $0= D_{[A_k}D_{A_{k-1]}}\cdots
    D_{A_1} f$. But $ D_{A_{k-1}}\cdots D_{A_1} f$ has weight $ 1-n/2$,
    so from \nn{key} this implies $X_{[A_k}\Box
    D_{A_{k-1}]}D_{A_{k-2}}\cdots D_{A_1} f=0 $. It follows immediately
      that $\Box D_{A_{k-1}}D_{A_{k-2}}\cdots D_{A_1} f = (-1)^{k-1}
        X_{A_1}\cdots X_{A_{k-1}} \Box_{2k} f$ for some operator
    $ \Box_{2k}$. With the above we are done.  
\end{proof}
If we are happy to work in the scale of a flat metric then
there is an even simpler proof along the lines of the proof of
Proposition~\ref{flatcase}. We leave this for the reader. 

By the same ideas as in the proof above, it is easy to use \nn{Dcomm}
and \nn{Dform} to show that, if $k\geq 3$, then $X_{[A_k}\Box
D_{A_{k-1}]}D_{A_{k-2}}\cdots D_{A_1} f \neq 0$ for $ f\in \ce[k-n/2]$
  on a general conformally curved manifold. Thus the proposition fails
  if we remove the requirement of conformal flatness. One way to
  generalise the $ \Box_{2k}$ is as follows. 

Consider 
$$
D^{A_1}\cdots D^{A_{k-1}} \Box D_{A_{k-1}}\cdots D_{A_1} f
$$
for $ f\in \ce^\Phi[k-n/2]$. This is manifestly strongly
conformally invariant in all dimensions and for all positive integers
$ k$. Furthermore by the
identity \nn{Dident} we have that, on conformally flat structures,
\begin{equation}\label{defboxk}
D^{A_1}\cdots D^{A_{k-1}} \Box D_{A_{k-1}}\cdots D_{A_1} f
= \left(\prod_{i=2}^k (n-2i)(i-1)\right) \Box_{2k} f.
\end{equation}
On the other hand for general conformally curved structures,
suppose that the dimension $n$ is odd or satisfies $2k<n$.
Then we can {\em define} $ \Box_{2k} f$ by \nn{defboxk}, and this gives
a conformally invariant operator
\begin{equation}\label{strongboxes}
\Box_{2k}:\ce^\Phi[k-n/2]\to \ce^\Phi[-k-n/2]
\end{equation}
with principal part $ \Delta^k$. Here, as usual, $ \ce^\Phi[k-n/2]$
indicates any tractor bundle of weight $ k-n/2$. In these dimensions
this generalises the operator $ \Box_{2k}$ of the proposition.
Although we do not wish to describe the curved translation principle
\cite{EastRice,Esrni}, it is worth pointing out that it is partly
illustrated here.  The tractor formula \nn{defboxk} manifests $
\Box_{2k}$ as a ``translate'' of the Yamabe operator $ \Box$. In fact
proceeding in smaller steps it demonstrates $ \Box_{2k}$ as a
translate of $ \Box_{2k-2}$.

Before we move on, let us demonstrate that the operators $ \Box_{2k}$
are formally self-adjoint. We summarise some results we need from
Section~7 of \cite{BrGo} in the following proposition. These results
can be verified easily using the definitions above, and it is important
for our needs to note that this works in a rather formal manner.
That is, we can leave the dimension and weight as unknown in the
calculations.
\begin{proposition}\label{intparts} On a conformal manifold $ M$ we have\\
(i) If $\psi^{B}\in\Gamma \ce^{B}[w]$ and
$\phi\in \Gamma\ce [1-n-w]$ is compactly supported on $M$, then
$$
\int_M \phi D_A \psi^A =\int_M (D_A \phi) \psi^A.
$$
(ii) If $\ce^\Phi$ is any tractor bundle, then $
\ce^\Phi$ is canonically isomorphic to its dual $ \ce_\Phi$ via 
the tractor metric and for any pair
$\psi^\Phi,\phi^\Phi \in \Gamma \ce^\Phi[1-n/2]$,  
($\ce^\Phi[1-n/2]:=\ce^\Phi\otimes\ce[1-n/2]$) we have
$$
\int_M \phi^\Phi \Box \psi_\Phi = \int \psi^\Phi \Box
\phi_\Phi.
$$
\end{proposition}
Since $\ce[-n]$ is naturally identified with the space of volume
densities the integrals are well-defined.
Now part (ii)
of the proposition asserts $ \Box_{2}=\Box$ is formally self-adjoint,
while the same result for $D^{A_1}\cdots D^{A_{k-1}} \Box
D_{A_{k-1}}\cdots D_{A_1} $ follows immediately from this and repeated
use of (i). So the formal self-adjoint property of $ \Box_{2k}$ is
proved.  It should be pointed out that as well as observing that
$D^{A_1}\cdots D^{A_{k-1}} \Box D_{A_{k-1}}\cdots D_{A_1} f $ recovers
a conformally invariant power of the Laplacian, M.G.\ Eastwood also
observed the formal self-adjoint property.  It is clear from
\nn{defboxk} that, unfortunately, this formula does not yield a
conformally invariant operator of order $n$ on even dimensional
structures, yet the existence of such an operator is guaranteed by the
construction of \cite{GJMS}.

Recall from Section~\ref{tractorsect} that if $ f\in \ce[w]$, then $
[D_A,D_B]f=0$, and so the $ k=2$ case of
the proposition does hold on general conformal structures.  In
particular, as observed in \cite{Gosrni}, for $f\in \ce[2-n/2] $ we
can define $ P'_4 f$ by the quotient formula
$$
\Box D_A f= -X_A P'_4 f.
$$
Then $ P'_4$ has principle part $ \Delta^2$. This construction works even when
$n=4$, and in other dimensions $ P'_4=\Box_4$ as defined above. It is
not hard to do the next even order in a similar way.  If now $f\in
\ce[3-n/2]$, then $ [D_B,D_C]f=0$ and hence $ D_BD_C f=D_{(B}D_{C)}f$. Now it
is a short exercise, using \nn{Dcomm} and the definition of $
W_{ABCD}$ once more, to show that
$$
(n-4)X_{[A} \Box D_{B]}D_C f= -2 X_{[A}W_{B]}{}^S{}_C{}^T D_S D_T f .
$$
Now from the fact that $D_SD_T f$ is symmetric and that $ W_{BSCT}$ has 
Weyl tensor type symmetries, we can deduce that in dimensions $n\neq 4$,
\begin{equation}\label{dbc}
P_{BC} f:= \Box D_B D_C f +\frac{2}{n-4} W_{B}{}^S{}_C{}^T D_S D_T f 
\end{equation}
is symmetric (i.e.\ $P_{BC} f\in \ce_{(BC)}[-1-n/2]$). On the other hand,
from the previous display $X_{[A}P_{B]C} f=0 $.
Thus, for $ n\neq 4$,
$$
P_{BC} f= X_B X_C P'_{6} f ,
$$
where $ P'_6$ is a conformally invariant operator $\ce[3-n/2]\to
\ce[-3-n/2]$ generalising (for the allowed dimensions) the sixth-order 
operator of Proposition~\ref{tractflatcase}. In particular this works in
dimension 6. 

We should point out that although $ \Box_{2k}$ as defined by \nn{defboxk} is
manifestly strongly invariant, we cannot conclude this for $ P'_6$ as
defined here. The operator $ P_{BC}$ defined above is clearly
invariant when acting on weighted tractors, but the argument here to
deduce that $ P_{BC} f$ has the form $ X_B X_C P'_{6} f $ relies on
the vanishing of $ [D_A,D_B]f$.

We will establish in Section~\ref{GJMS} (see in particular Subsection
\ref{examples}) the following result.
\begin{proposition}
\label{p468}
The operators $ P'_4$ and $ P'_6$ defined by the tractor expressions
above are precisely the fourth-order and sixth-order GJMS
operators. That is $ P'_4=P_4$, $ P'_6=P_6$.  A tractor expression for
the eighth-order GJMS operator $ P_8$ is as follows:
$$
\begin{array}{rll}
        \lefteqn{X_A X_B X_C P_8 f=}&&
\\
        &&
        -\Box D_A D_B D_C f
        -\frac{2}{n-4}W_A{}^P{}_B{}^Q D_P D_Q D_C f
     -\frac{2}{n-4}W_A{}^P{}_C{}^Q D_P D_B D_Q f
        \strutdd
\\
 &&
       -
        \frac{4}{n-6}X_A
     U_B{}^P{}_C{}^Q D_P D_Q f
        \strutdd
        +
        \frac{2}{(n-4)(n-6)}X_AD^E W_B{}^P{}_C{}^Q D_E D_P D_Q f
        \strutdd
\\
&&        +
        4\frac{n-2}{(n-4)^2(n-6)} X_A
        W_B{}^P{}_C{}^Q W_P{}^S{}_Q{}^T D_S D_T f
        \strutdd
        ,
\end{array}
$$
where all operators act on everything to their right, in a given term, and
$U_B{}^P{}_C{}^Q$ is the tractor field
$$
\begin{array}{l}
        \hspace{2.98mm}
        \frac{2}{(n-4)^2}
        \left(
        W^{AP}{}_B{}^F W_{FAC}{}^Q
        +
        W^{AP}{}_C{}^F W_{BAF}{}^Q
        +
        W^{APQF}
        W_{BACF}
        \right)
        \rule{0mm}{4.5mm}    
        .
\end{array}
$$
\end{proposition}
\noindent Here $D^E W_B{}^P{}_C{}^Q D_E D_P D_Q f$ is taken to mean
$ D^E (W_B{}^P{}_C{}^Q D_E D_P D_Q f)$ (and {\em not} $ (D^E
W_B{}^P{}_C{}^Q) D_E D_P D_Q f $).  This and similar conventions for
other operators and situations will apply throughout the paper.

We should emphasise at this point that the tractor formulae for $
P_4$, $ P_6$ and $ P_8$ above, and similar ones for the higher order
$P_{2k}$ that we could easily construct via the algorithm of
Section~\ref{GJMS}, are genuine formulae for the GJMS operators.  No
further algorithm is required.  They are valid on any conformal
manifold where the given GJMS operators exist. In this tractor form
they are already suitable for many applications, such as establishing
strong invariance or constructing related operators.  The remainder
of the section will demonstrate this.

We begin by using the tractor formulae directly to show that the operators $
P_4,P_6$ and $P_8 $ are formally self-adjoint (FSA).  We treat these
in order. For $ f\in\ce[2-n/2]$, we have 
\begin{equation}\label{p4}
\Box D_A f =-X_A P_4 f \quad \mbox{which
implies}\quad  D^A \Box D_A f=(n-4)P_4 f.
\end{equation}
We have already observed that $D^{A_1}\cdots D^{A_{k-1}} \Box
D_{A_{k-1}}\cdots D_{A_1}$ is FSA on $ \ce[k-n/2]$.  
So from the second of these it is
clear that $ P_4$ is FSA in dimensions other than 4. From the
expressions \nn{Dform} and \nn{box} it follows that $ D^A \Box D_A f$ and
$\Box D_A f$, as expressions in terms of Levi-Civita covariant
derivatives of $f$, $ \V_{ab}$ and $ \J$, are polynomial in $ n$. So
from \nn{p4} it is clear that $(4-n)$ divides this expression for
$D^A \Box D_A f$ and so $ P_4 f$ is also given as a formula polynomial
in $ n$ and the Levi-Civita covariant derivatives of $f$, $ \V_{ab}$
and $ \J$. Working among tensors of this form a calculation to verify
the FSA property of $ P_4$ (in dimensions greater than 4) can be
carried out formally in dimension $ n$, since
Proposition~\ref{intparts} is established that way. It follows
immediately that
the same calculation must work when we set $ n=4$. Thus $ P_4$ is also
FSA in dimension 4.  Now for $ P_6$, let $ f\in \ce[3-n/2]$ and note
that $\Box D_B D_C f$ and $W_{B}{}^S{}_C{}^T D_S D_T f $ are
polynomial in $ n$. Thus $ P_{BC} f$ is rational in $ n$ with a
singularity only at $ n=4$. From $P_{BC} f= X_B X_C P_{6} f$ we have
$$
\begin{array}{lll}
        \lefteqn{(n-4)D^CD^BP_{BC} f}
        &&
\\
        &&
        =
        (n-4)D^CD^B\Box D_BD_C f+
        2D^CD^BW_{B}{}^S{}_C{}^T D_S D_T f
\\
        &&
        =
        2(n-4)^2(n-6)P_{6} f.
\end{array}
$$
Now since
$ W_{BSCT}$ has the Weyl tensor symmetries (in fact here we just need
$W_{BSCT}=W_{CTBS}=W_{TCSB}$), it follows from
Proposition~\ref{intparts} that $D^CD^BW_{B}{}^S{}_C{}^T D_S D_T $ is FSA on
$\ce[3-n/2] $. We know $D^CD^B\Box D_BD_C f $ is also FSA and as
expressions in terms of Levi-Civita covariant derivatives of $f$, $
C_{abcd}$, $ \V_{ab}$, and $ \J$, both of these and $(n-4)P_{BC}$ are
polynomial in $ n$. Thus the expression like this for $D^CD^BP_{BC} f
$ is divisible by $( n-6) $, so reasoning as for $ P_4$, we
quickly conclude that $ P_6$ is FSA in all dimensions for which it is
defined.  Finally, since $ U_{ABCD}$ also has Weyl tensor symmetry (as
readily verified directly or since it corresponds to $ \aDe \aR_{ABCD}$
as in Section~\ref{ambsect}), it follows that $D^CD^B
U_B{}^P{}_C{}^QD_P D_Q f $ is FSA for $ f\in \ce[4-n/2]$. A similar comment applies to the  other terms on the right-hand side of the formula,
$$
\begin{array}{rll}
        \lefteqn{6(n-4)^2(n-6)(n-8) P_8 f=}&&
\\
&&        (n-4)D^CD^BD^A \Box D_AD_BD_C f
        +2D^CD^BD^AW_A{}^P{}_B{}^QD_PD_QD_C f
        \strutdd
\\
&&        +2D^CD^BD^AW_A{}^P{}_C{}^QD_PD_BD_Q f
        -
        4\frac{(n-4)^2}{n-6}D^CD^B
        U_B{}^P{}_C{}^QD_P D_Q f
        \strutdd
\\
 &&       +
        2\frac{n-4}{n-6} D^CD^BD^E W_B{}^P{}_C{}^QD_ED_P D_Q f
        \strutdd
\\
&&
        +
        4\frac{n-2}{n-6}D^CD^B
        W_B{}^P{}_C{}^Q W_P{}^S{}_Q{}^TD_SD_T f,
\end{array}
$$
which follows from the earlier display for $P_8$. This shows
immediately that $ P_8 f$ is FSA in dimensions other than 8, and then,
arguing as in the previous cases, we can deduce that it is also FSA in
dimension 8. We have directly proved the following.
\begin{proposition}\label{psfsa}
The GJMS operators $ P_{4}, P_6$ and $ P_8$ are formally self-adjoint.
\end{proposition}
In fact the result is also immediate from the formulae for these
operators in Section~\ref{convent}. The formulae there are given in
terms of the Levi-Civita connection and its curvature and are in a
canonical form that manifests the formal self-adjoint symmetry. (It
should be pointed out that in deriving those formulae formal
self-adjointness was not assumed.)

Recently the entire family of operators $ P_{2k}$ have been shown to
be formally self-adjoint by other means \cite{GrZ2,FeffGr01}.

In Section~\ref{GJMS} we will show that there are similar tractor
formulae for all of the GJMS operators and that these tractor formulae
share some of the qualitative features of the examples above. In particular
we will prove the following theorem. 
\begin{theorem}\label{L24Maybb} 
(i) In each dimension $n$ and
  for each integer $2 \leq k$, with $ k\leq n/2$ if $ n$ is even, there
  is a  conformally invariant differential operator
$$
E_{CD}{}^{AB} : \ce_{AB}[k-2-n/2]\to \ce_{CD}[2-k-n/2] 
$$
such that 
$$
E_{CD}{}^{AB} D_AD_B f= X_C X_D P_{2k} f
$$
for $ f\in \ce[k-n/2]$, where $P_{2k} :\ce[k-n/2]\to \ce[-k-n/2]$ is
the order $ 2k$ GJMS operator. The operator is given by a formula
which is a partial contraction polynomial in $\Box$, $D_A$, $ W_{ABCD}$, $
X_A$, $h_{AB}$ and its inverse $ h^{AB}$.  
\\
(ii)~With $ n$ and $ k$ as in (i) and for $ \ce^\Phi$ any tractor
bundle, there is a conformally invariant differential operator
$$
E_{CD}{}^{AB} : \ce^\Phi\otimes\ce_{AB}[k-2-n/2]\to \ce^\Phi\otimes \ce_{CD}[2-k-n/2] 
$$
which generalises the operator of part (i).\\ 
(iii) On conformally flat
structures the operator $ E$ is, up to a non-zero scale, $
\Box_{2k-4}$. In this case, given a choice of flat metric from the
conformal class, $ E$ is, up to a non-zero scale, $\Delta^{k-2}$. 
\end{theorem} 
\begin{proof} We have already observed that for $ f\in\ce[2-n/2]$, we 
  have (see \nn{p4}) $\Box D_A f =-X_A P_4 f $.  Since then $ D_A f\in
\ce_A[1-n/2]$ we have $X_B \Box D_A f=-D_BD_A f$, and so we have (i)
for $ k=2$.  Otherwise establishing part (i) is the primary purpose of
Section~\ref{GJMS}. More precisely, from the discussion there we
obtain Proposition~\ref{tract4gjms}, which asserts that
$$
X_{A_{k-1}}\cdots X_{A_1}P_{2k} f =(-1)^{k-1}\Box D_{A_{k-1}}
\cdots D_{A_1} f + \Psi_{A_{k-1} \cdots A_{1}}{}^{PQ}D_PD_Q f
.
$$
Applying $D^{A_3} \cdots D^{A_{k-1}}$ to both sides of this and
using \nn{Dident}, we obtain
$$
X_{A_1}X_{A_2} P_{2k} f = E_{A_1A_2}{}^{PQ}D_PD_Q f,
$$
where 
\begin{equation}\label{Eform}
\begin{array}{lll}
        E_{A_1A_2}{}^{PQ} &=&
        \Box_{2(k-2)}\delta_{A_1}{}^{\!P}\delta_{A_2}{}^{\!Q} +
\\&&
        (\prod_{i=2}^{k-2} 
        (2i-n)(i-1))^{-1} D^{A_3} \cdots D^{A_{k-1}}\Psi_{A_{k-1} \cdots
        A_3A_2 A_{1}}{}^{\!PQ} .
        \strutdd
\end{array}
\end{equation}
As explained in Proposition~\ref{tract4gjms}, $ \Psi$ is given
explicitly by a sum of terms each of which is a monomial in $D_A$, $
W_{ABCD}$, $ X_A$, $h_{AB}$, and its inverse $ h^{AB}$. Each such
monomial is thus a composition of strongly conformally invariant
operators. 
So $E_{A_1 A_2}{}^{PQ}$ is a sum of compositions of strongly
conformally invariant operators.  Just knowing that $E_{A_1
  A_2}{}^{PQ}$ is a sum of compositions of conformally invariant
operators of this form gives part (i). Then part (ii) is immediate
from the fact that these are strongly invariant
operators.
 
Next we show part (iii). From Proposition~\ref{tract4gjms} each
term in the expression for $ \Psi$ is of degree at least 1 in $
W_{ABCD}$. The latter vanishes on conformally flat structures.  Thus,
from \nn{Eform}, on such structures $ E$ is just $ \Box_{2k-4}$.  From
Proposition~\ref{tractflatcase}, given a choice of flat metric from
the conformal class, we have $ \Box_{2k-4}=\Delta^{k-2}$.

\vspace{-.4cm}

\end{proof}

\noindent{\bf Remarks:} In regard to part (i) of the theorem we should point 
out that Section~\ref{GJMS} not only establishes the existence of a
formula for $ E$ which is a partial contraction polynomial in $D_A$, $
W_{ABCD}$, $ X_A$, $h_{AB}$, and its inverse $ h^{AB}$, but describes an
algorithm for finding such a formula. 

It seems likely that the operator $ E$ in the theorem is
formally self-adjoint.  Note for example, if we write $ E^*$ for the
formal adjoint of $ E$, then from Proposition~\ref{intparts}, the
identity \nn{Dident}, and the result that $ P_{2k}$ is
formally self-adjoint, we have $(\prod_{i=k-1}^{k} (n-2i)(i-1))P_{2k}=
D^AD^BE^*_{AB}{}^{PQ}D_PD_Q $ on $ \ce[k-n/2]$.

Finally we should add that the terms which distinguish the $ P_{2k}$
from the $ \Box_{2k}$ do not vanish in general. At least we have
verified by direct calculation that for $f\in \ce[3-n/2] $ (and $ n\neq 4$) 
the leading term of
$D^CD^BW_{B}{}^S{}_C{}^T D_S D_T f$ is a non-zero scalar multiple of
$C^a{}_{cde}C^{bcde}\nd_a\nd_b f$. Thus $P_6$ is not simply a scalar multiple 
of $ \Box_6$.

\vspace{.5cm}

It is a non-trivial matter to know when conformally invariant
operators have strongly invariant generalisations. Some do not. For
example in dimension 4 we know there is a conformally invariant
operator $ P_4: \ce \to \ce[-4]$ with principal part $ \Delta^2$.
Suppose there were a strongly invariant generalisation of this. Then,
in particular, it would give a conformally invariant operator
$H_A{}^B: \ce_B \to \ce_A[-4]$ with principal part $ \Delta^2$. (Here
we mean the principal part as an operator between the reducible
bundles indicated.)  Then, in the case of Riemannian signature
conformal 4-manifolds, using the ellipticity of this, \nn{Dident},
Proposition~\ref{intparts} and the differential operator existence
results in the conformally flat setting (cf.\ \cite{ES}) we can
conclude that we would have a conformally invariant operator
$D^AH_A{}^BD_B :\ce[1]\to \ce[-5]$ with principal part $ \Delta^3$ (on
arbitrary conformal 4-manifolds).  This contradicts C.R. Graham's
non-existence result \cite{Grno}, and so we can conclude the operator
$ H_A{}^B$ does not exist. (See also \cite{EastRice}.) However $ P_4$
does have a strongly invariant generalisation in all other dimensions.
This is just $ \Box_4$ as a special case of \nn{strongboxes}.  More
generally, a consequence of part (ii) of the theorem is that the
GJMS operators $ P_{2k}$ admit strongly invariant generalisations
except in the critical dimension $ n=2k$.  That is, we have the
following proposition on $ n$-dimensional conformal manifolds:
\begin{corollary} \label{strongp}
  For each integer $ k\geq 1$, with $ 2k< n$ if $ n$ is even, there is
  a (tractor) formula that gives, for each tractor bundle $
  \ce^\Phi$, a formally self-adjoint differential operator
$$
{P}^\Phi_{2k} : \ce^\Phi[k-n/2]\to \ce^\Phi[-k-n/2],
$$
where $\ce^\Phi[w]:=\ce^\Phi\otimes\ce[w] $. The operator has
principal part $ \Delta^{k}$ and can be expressed as a sum of $
\Box_{2k}$ and a contraction polynomial in $D_A$, $ W_{ABCD}$, $ X_A$,
$h_{AB}$, and its inverse $ h^{AB}$. In the conformally flat case the operator
is $ \Box_{2k}$. In the case that $\ce^\Phi=\ce$ then ${P}^\Phi_{2k}=P_{2k}$.
\end{corollary}
\begin{proof}
  Since $ D_A$ is strongly invariant and since also, from (ii) of the theorem, 
$E_{AB}{}^{PQ}$ is strongly invariant, it follows that there is a
  conformally invariant operator $(F:=(\prod_{i=k-1}^{k} 
(n-2i)(i-1))^{-1}D^AD^B
  E_{AB}{}^{PQ}D_PD_Q) :\ce^\Phi[k-n/2]\to \ce^\Phi[-k-n/2] $ for
  any tractor bundle. By \nn{Dident} this precisely recovers the GJMS
  operator $ P_{2k}$ if $ \ce^\Phi[k-n/2]$ is simply the density
  bundle $ \ce[k-n/2]$.  Now consider the formal adjoint $ F^*$ of $F
  $. This is another conformally invariant operator
  $\ce^\Phi[k-n/2]\to \ce^\Phi[-k-n/2] $ (where as usual we
  identify $ \ce^\Phi$ with its dual via the tractor metric). Since
  $ P_{2k}$ is formally self-adjoint, it is clear that,
  when applied to $\ce[k-n/2] $, $ F^*$ also recovers the GJMS
  operator. Thus $(F+F^*)/2$ is the required formally self-adjoint
  operator.
  
  It is clear from Proposition~\ref{tract4gjms} that we can
  express $F $ by a formula which is a sum of $ \Box_{2k}$ and a
  contraction polynomial in $D_A$, $ W_{ABCD}$, $ X_A$, $h_{AB}$, and
  its inverse $ h^{AB}$. From that proposition we also have that each
  term in the latter polynomial expression is of degree at least 1 in
  $W_{ABCD} $.  Using Proposition~\ref{intparts} and the formal
  self-adjoint property of $ \Box_{2k}$, we see that there is an
  expression for $ F^*$ as a sum of $ \Box_{2k}$ and a contraction
  polynomial in $D_A$, $ W_{ABCD}$, $ X_A$, $h_{AB}$, and its inverse
  $ h^{AB}$. Again each term in the latter polynomial is of degree at
  least 1 in $W_{ABCD} $. So the final part of the corollary follows
  from these observations and Proposition~\ref{tractflatcase}.
 \end{proof}

\subsection{Conventional formulae} \label{convent}

There are circumstances where it is useful to have explicit formulae
for the GJMS-related operators and invariants in terms of the
Levi-Civita connection and its curvature. These formulae are
generally cumbersome.
But the various curvature terms are
closely related to the spectrum of the operator, so
it is important to be able to
extract these explicitly. In particular, for example, issues of
positivity can be investigated directly in this setting.  Moreover such
formulae are ready to be mechanically rewritten in local coordinates
should this be required. 

Here we will describe how to re-express tractor formulae for $ P_{2k}
f$ into formulae which are polynomial in $\bg$, its inverse,
$\nd$ (meaning the Levi-Civita connection), $C$, $ \Rho$ and $\J $,
and of course linear in $f$. 

For the most part, the process is simply an expansion of the tractor
formulae using the definitions above. Consider the Paneitz operator $
P_4$ first.  We observed in Proposition~\ref{p468} that for $ f\in
\ce[2-n/2]$, $ -X_A P_4 f =\Box D_A f$. So $ P_4 f=-Y^A\Box D_A f$, and
we could simply calculate this scalar quantity $Y^A\Box D_A f $. In
fact we prefer to expand the entire tractor valued quantity $\Box D_A
f $ using \nn{Dform} and \nn{box}. According to its definition, $ D_A$
lowers weight by 1.  Thus $\Box D_A f$ is given by
$$
        (\nd_b\nd^b +(1-\frac{n}{2})\J) ((4-n) Y_A f+ 2Z_A{}^{a}\nabla_a f
        -X_A(\nd_c\nd^c+(2-\frac{n}{2})\J) f).
$$
Now we simply move the $ X_A, Y_A$ and $ Z_{A}{}^a$ to the left of
all other operators by repeated use of \nn{connids}.  This is easily
done by hand and simplified via the Bianchi identity to yield
$$
  \begin{array}{lll}
-\Box D_A f &=&
        X_A(\De^2 f
        -(n-2)\J\De f
        +4\,\V^{ij}\nd_i\nd_j f
        -(n-6)(\nd^i\J)\nd_if
\\
&&
        \mbox{}-\frac{n-4}{2}(\De\J)f
        +
        \frac{n(n-4)}{4}\J^2 f
        -
        (n-4)\V_{ij}\V^{ij}f)
        .
\end{array} 
$$
The coefficient of $X_A$ on the right-hand side is a formula
for the Paneitz operator. Note that the coefficient of $ Y_A$ and the
coefficient of $ Z_A{}^a$ both turned out to be zero. Of course this is
exactly as predicted by our formula $ -X_A P_4 f =\Box D_A f$, but it
provides a very useful check of the formulae to verify this. So this
is all there is to producing the required formula for $ P_4$ from the
tractor formula. Before we continue with the general case let us just
reorganise the result.

For any linear differential operator on densities of the appropriate
weight there is a canonical form for the formula which, among other
features, manifests the symmetry in the formally self-adjoint and
formally anti-self-adjoint parts \cite{EGoTN}. As already observed, the
Paneitz operator is formally self-adjoint.  Applying this idea to
the formula above yields
\begin{equation}
\label{PanOne}
\begin{array}{c}
       P_4f= \nd_i\nd_j\midtenPan^{ijkl}_4\nd_k\nd_l f
        +
        \nd_i\midtenPan^{ij}_2\nd_j f
        +
        \frac{n-4}{2}Q_{4,n}^g f
        .
\end{array}
\end{equation}
Here $\midtenPan^{ijkl}_4$ and $\midtenPan^{ij}_2$ are the tensors
$(1/3)(\bg^{il}\bg^{jk}+\bg^{ik}\bg^{jl}+\bg^{ij}\bg^{kl})$ and
$$
\begin{array}{c}
\frac{4-3n}{3}\bg^{ij}\J-\frac{2(n-8)}{3}\V^{ij},
\end{array}
$$
respectively, and $Q^{g}_{4,n}$ denotes the scalar
\begin{equation}\label{Q4sc}
\begin{array}{c}
        \frac{n}{2}\J^2
        -2\V_{ij}\V^{ij}
        -\De\J
        .
\end{array}
\end{equation}
In (\ref{PanOne}), the $\nd$'s  act on all tensors to
their right within the given term.

We now discuss the general case. Explicit tractor formulae are readily
produced by the algorithm described in Section~\ref{GJMS}, and so we
shall suppose we are beginning with a formula for $ P_{2k}$ as
described in Proposition~\ref{tract4gjms}. The formulae for $ P_6$ and
$ P_8$ above (see Proposition \ref{p468}) give explicit examples that
can be kept in mind.  These formulae are polynomial in $ \Box$, $D_A$,
$ W_{ABCD}$, $ X_A$, $h_{AB}$, and its inverse $ h^{AB}$. We replace
each of these with its formula in terms of the coupled
tractor-Levi-Civita connection $ \nd$, $ X_A$ and so forth according
to the formulae \nn{Dform}, \nn{box}, \nn{tractcurv}, and \nn{WDef}.
In doing so we note that $W$ has weight $-2$ and that $D$ lowers the
weight of a tractor by $1$. Next we move all occurrences of $X_A,
Z_A{}^a$, and $ Y_A$ to the left of the $ \nd$ via repeated use of
\nn{connids}. At the end of this process all tractor valued objects
are to the left of the remaining $ \nd$'s, and so at this point these
$ \nd$'s are simply Levi-Civita covariant derivative operators.  Next
we use the inner product rules of Figure~\ref{Larry22April01a} to
simplify the resulting expression.  The formula for $P_{2k}f$ is then
simply the overall coefficient of $X_{A_1}X_{A_2}\cdots X_{A_{k-1}}$.
From Proposition~\ref{tract4gjms} all other slots of the tractor
expression vanish. That is, the sum of the terms that do \textit{not}
contain $X_{A_1}X_{A_2}\cdots X_{A_{k-1}}$ is zero.  Verifying this or
even partly verifying this provides a very serious check of all
formulae and any software that are used in the calculation.  For
example, one can verify that the sum of the terms containing
$Z_{A_1}{}^a X_{A_2}\cdots X_{A_{k-1}}$ vanishes.

This procedure is very simple.  But there are many terms involved, as
the next examples will illustrate. Thus it becomes very useful to be
able calculate via a suitable computer algebra system.  For the
examples below the authors used \textit{Mathematica} and J. Lee's {\em Ricci}
program \cite{JackLee}, which proved to be very effective.
Certainly the $ P_8$ case is beyond a reasonable hand calculation. The
use of software and the self-checking nature of the formulae as
discussed above mean that one can be confident of the final result.

As a technical point for these calculations, we describe a simple
technique which can considerably reduce the computing time they
require.  One can implement this technique by developing a short
computer programme.  We begin by noting that certain steps in the
computation may produce tractor inner products of the form $\Psi_{B_1
B_2 \ldots B_\ell}\Phi^{B_1 B_2 \ldots B_\ell}$, where the indices
$B_1$, $B_2$,\ldots, and $B_\ell$ appear as subscripts or superscripts
attached to the tractors $Y$, $Z$, and $X$.  Suppose that $\ell$ is
large and that the tractors $\Psi_{B_1 B_2 \ldots B_\ell}$ and
$\Phi^{B_1 B_2 \ldots B_\ell}$ are the sums of many terms.  Suppose
also that no derivatives of $Y$, $Z$, or $X$ occur.  The tractor
$\Psi_{B_1 B_2 \ldots B_\ell}$ is a linear combination of the
following $3^{\ell}$ terms:
\begin{quote}
$Y_{B_1}Y_{B_2}\cdots Y_{B_{\ell-1}}Y_{B_\ell}$\\
$Y_{B_1}Y_{B_2}\cdots Y_{B_{\ell-1}}Z_{B_\ell}{}^{b_{\ell}}$\\
$Y_{B_1}Y_{B_2}\cdots Y_{B_{\ell-1}}X_{B_\ell}$\\
$Y_{B_1}Y_{B_2}\cdots Z_{B_{\ell-1}}{}^{b_{\ell-1}}Y_{B_\ell}$\\
$Y_{B_1}Y_{B_2}\cdots Z_{B_{\ell-1}}{}^{b_{\ell-1}}Z_{B_\ell}{}^{b_{\ell}}$\\
$Y_{B_1}Y_{B_2}\cdots Z_{B_{\ell-1}}{}^{b_{\ell-1}}X_{B_\ell}$\\
\hspace*{17mm}\vdots\\
$X_{B_1}X_{B_2}\cdots X_{B_{\ell-1}}X_{B_\ell}$
\end{quote}
The coefficients of these terms may, of course, be very complicated.
By raising indices we may write $\Phi^{B_1 B_2 \ldots B_\ell}$ as a
similar linear combination.  Each term of each linear combination may
be paired off with at most one term in the other linear combination so
as to give a nonzero inner product.  We compute the $3^{\ell}$
possible inner products and add the results.

We conclude this section with the calculation of $ P_6$ and $ P_8$ via
these methods, beginning with the tractor formulae indicated in
Proposition~\ref{p468}.  As a check, the authors verified the
vanishing of the overall coefficient of the $Z_{B}{}^{i} X_{C}$ term
in the expansion of \nn{dbc}.  In a similar fashion, they also
verified the vanishing of the overall coefficients of the
$X_{B}Z_{C}{}^{i}$ and $Z_{A}{}^{i} X_{B} X_{C}$ terms in the
expansions for $P_6$ and $P_8$, respectively.  This involved the use
of the Bianchi identities, tensor symmetries, and changes in the order
in which covariant derivatives are taken.  The authors also
manipulated the resulting formulae for the GJMS operators into the
canonical form suggested in \cite{EGoTN}. Here are the results:
$$
\begin{array}{rcl}
P_6&=&\nd_i\nd_j\nd_k \midten^{ijklmp}_{6}\nd_l\nd_m\nd_p f+
\nd_i\nd_j \midten^{ijkl}_{4}\nd_k\nd_l f +
\nd_i \midten^{ij}_{2}\nd_j f
\\
&& +\frac{n-6}{2}Q^g_{6,n} f
.
\JulyStrut
\end{array}
$$
Here $\midten^{ijklmp}_6$ and $\midten^{ijkl}_4$ are the symmetrizations
of the tensors $\bg^{ij}\bg^{kl}\bg^{mp}$ and
$$
\frac{2-3n}{2}\J \bg^{ij}\bg^{kl}
+(20-2n)P^{ij}\bg^{kl},
$$
respectively, and $\midten^{ij}_2$ is the tensor
\begin{equation}
\label{MTwoPSix}
\begin{array}{l}
\mbox{}
-\frac{\left( 88 - 86\,n + n^2 \right) }{15\,\left( n-4 \right) } 
   \tensor{\V}{\up{i}\up{j}\down{;}\down{k}\up{k}} - 
  \frac{2\,\left( 2176 - 768\,n + 82\,n^2 + n^3 \right) }
    {15\,\left( n-4 \right) } 
   \tensor{\V}{\up{i}\down{k}} \tensor{\V}{\up{j}\up{k}}
\\
\JulyStrut
\mbox{}
- 
  \frac{2\,\left( 320 - 218\,n + 27\,n^2 \right) }
    {15\,\left( n-4  \right) } 
   \tensor{\bg}{\up{i}\up{j}} \tensor{\V}{\down{k}\down{l}} 
     \tensor{\V}{\up{k}\up{l}} + 
  \frac{\left( 3\,n-2 \right) \,\left(5\,n-54 \right) }{15} 
   \tensor{\V}{\up{i}\up{j}} \Mvariable{\J}
\\
\JulyStrut
\mbox{}
+ 
  \frac{-164 - 120\,n + 45\,n^2}{60} 
   \tensor{\bg}{\up{i}\up{j}} \Mvariable{\J}^{2} - 
  \frac{2\,\left( 5\,n-22 \right) }{15} 
   \tensor{\bg}{\up{i}\up{j}} \tensor{\Mvariable{\J}}{\down{;}\down{k}\up{k}} \
- \frac{744 - 250\,n + 31\,n^2}{15\,\left( n-4\right) } 
   \tensor{\Mvariable{\J}}{\down{;}\up{i}\up{j}}
\\
\JulyStrut
\mbox{}
+ 
  \frac{2\,\left( 296 - 26\,n + 3\,n^2 \right) }{15\,\left( n-4  \right) } 
   \tensor{\V}{\down{k}\down{l}} 
    \tensor{\C}{\up{i}\up{k}\up{j}\up{l}} - 
  \frac{4}{15} \tensor{\C}{\up{i}\down{k}\down{l}\down{m}} 
    \tensor{\C}{\up{j}\up{k}\up{l}\up{m}}
.
\end{array}
\end{equation}
Here and below, for typesetting convenience, we write
$\tensor{\V}{\up{i}\up{j}\down{;}\down{k}\up{k}} $ as an alternative
notation for $ \nd^k\nd_k \Rho^{ij}$ and so forth.  Finally,
$Q^g_{6,n}$ denotes the scalar
\begin{equation}
\label{QSix}
\begin{array}{l}
\mbox{}
-8 \tensor{\V}{\down{i}\down{j}\down{;}\down{k}} 
    \tensor{\V}{\up{i}\up{j}\down{;}\up{k}} - 
  \frac{8\,\left( n-2 \right) }{n-4} 
   \tensor{\V}{\down{i}\down{j}} 
    \tensor{\V}{\up{i}\up{j}\down{;}\down{k}\up{k}} + 
  \frac{64}{n-4} \tensor{\V}{\down{i}\down{j}} 
    \tensor{\V}{\up{i}\down{k}} \tensor{\V}{\up{j}\up{k}}
\\
\JulyStrut
\mbox{}
+
  \frac{4\,\left( -4 - 4\,n + n^2 \right) }{n-4} 
   \tensor{\V}{\down{i}\down{j}} \tensor{\V}{\up{i}\up{j}} \Mvariable{\J}
- 
  \frac{\left( n-2  \right) \,\left( n+2 \right) }{4} 
   \Mvariable{\J}^{3} + \left( n-6 \right)  
   \tensor{\Mvariable{\J}}{\down{;}\down{i}} 
    \tensor{\Mvariable{\J}}{\down{;}\up{i}}
\\
\JulyStrut
\mbox{}
+ 
  \frac{3\,n-2}{2} \Mvariable{\J}
    \tensor{\Mvariable{\J}}{\down{;}\down{i}\up{i}}
- 
  \frac{8\,\left( n-6 \right) }{n-4} 
   \tensor{\V}{\down{i}\down{j}} 
    \tensor{\Mvariable{\J}}{\down{;}\up{i}\up{j}}
-
  \tensor{\Mvariable{\J}}{\down{;}\down{i}\up{i}\down{j}\up{j}} - 
  \frac{32}{n-4} \tensor{\V}{\down{i}\down{j}} 
    \tensor{\V}{\down{k}\down{l}} 
     \tensor{\C}{\up{i}\up{k}\up{j}\up{l}}
.
\end{array}
\end{equation}
%
%

We find that $P_8f$ is given by
$$
\begin{array}{l}
        \nd_i\nd_j\nd_k\nd_l
        \midtenEi_8^{ijklmpqr}
        \nd_m\nd_p\nd_q\nd_r
        f
        +
        \nd_i\nd_j\nd_k
        \midtenEi_6^{ijklmp}
        \nd_l\nd_m\nd_p
        f
\\
        +
        \nd_i\nd_j
        \midtenEi_4^{ijkl}
        \nd_k\nd_l
        f
        +
        \nd_i
        \midtenEi_2^{ij}
        \nd_j
        f
        +
        \frac{n-8}{2}Q^g_{8,n}
        f.
        \JulyStrut
\end{array}
$$
In this formula, $\midtenEi_8^{ijklmpqr}$ and
$\midtenEi_6^{ijklmp}$ denote the symmetrizations of the tensors
$\bg^{ij}\bg^{kl}\bg^{mp}\bg^{qr}$ and
$$
-2n\J \bg^{ij}\bg^{kl}\bg^{mp}-4(n-12)\bg^{ij}\bg^{kl}\V^{mp},
$$
respectively, and $\midtenEi_4^{ijkl}$ denotes the symmetrization of
the tensor
$$
\begin{array}{l}
\frac{-8\,\left( -12 + n \right) }{5} 
   \tensor{\V}{\up{i}\up{j}\down{;}\up{k}\up{l}} + 
  \frac{4\,\left( -12 + n \right) \,\left( -64 + 5\,n \right) }{15} 
   \tensor{\V}{\up{i}\up{j}} \tensor{\V}{\up{k}\up{l}} + 
  \frac{24\,n}{-4 + n} \tensor{\bg}{\up{i}\up{j}} 
    \tensor{\V}{\up{k}\up{l}\down{;}\down{m}\up{m}}
\JulyStrut
\\
- 
  \frac{16\,\left( 1536 - 530\,n + 59\,n^2 \right) }
    {15\,\left( -4 + n \right) } 
   \tensor{\bg}{\up{i}\up{j}} \tensor{\V}{\up{k}\down{m}} 
     \tensor{\V}{\up{l}\up{m}} - 
  \frac{2\,\left( 480 - 568\,n + 67\,n^2 \right) }
    {15\,\left( -4 + n \right) } 
   \tensor{\bg}{\up{i}\up{j}} \tensor{\bg}{\up{k}\up{l}} 
     \tensor{\V}{\down{m}\down{p}} \tensor{\V}{\up{m}\up{p}}
\JulyStrut
\\
+ 
  \frac{4\,n\,\left( -64 + 5\,n \right) }{5} 
   \tensor{\bg}{\up{i}\up{j}} \tensor{\V}{\up{k}\up{l}} \Mvariable{\J} + 
  \frac{-96 - 20\,n + 15\,n^2}{10} 
   \tensor{\bg}{\up{i}\up{j}} \tensor{\bg}{\up{k}\up{l}} \Mvariable{\J}^{2} + 
  \frac{24 - 5\,n}{5} \tensor{\bg}{\up{i}\up{j}} 
\tensor{\bg}{\up{k}\up{l}} \tensor{\Mvariable{\J}}{\down{;}\down{m}\up{m}} \
\JulyStrut
\\
- \frac{8\,\left( 120 - 31\,n + 4\,n^2 \right) }{5\,\left( -4 + n \right) } 
   \tensor{\bg}{\up{i}\up{j}} \tensor{\Mvariable{\J}}{\down{;}\up{k}\up{l}} + 
  \frac{16\,\left( 192 - 7\,n + n^2 \right) }{15\,\left( -4 + n \right) } 
   \tensor{\bg}{\up{i}\up{j}} \tensor{\V}{\down{m}\down{p}} 
     \tensor{\C}{\up{k}\up{m}\up{l}\up{p}} - 
  \frac{16}{15} \tensor{\C}{\up{i}\down{m}\up{j}\down{p}} 
    \tensor{\C}{\up{k}\up{m}\up{l}\up{p}}
\JulyStrut
\\
- 
  \frac{32}{15} \tensor{\bg}{\up{i}\up{j}} 
    \tensor{\C}{\up{k}\down{m}\down{p}\down{q}} 
     \tensor{\C}{\up{l}\up{p}\up{m}\up{q}}
        .
\JulyStrut
\end{array}
$$
%
%
We let $\midtenEi_2^{ij}$ denote the symmetrization of the tensor
$\ATen^{ij}+\BTen^{ij}+\CTen^{ij}$, where $\ATen^{ij}$, $\BTen^{ij}$,
and $\CTen^{ij}$ are as given in Figures~\ref{TwoTenAA},
\ref{TwoTenBB}, and \ref{TwoTenCC}.
\begin{figure}
$$
\begin{array}{l}
\frac{-4\,\left( -19200 + 8468\,n - 980\,n^2 + 27\,n^3 \right) }
    {45\,\left( -6 + n \right) \,\left( -4 + n \right) } 
   \tensor{\bg}{\up{i}\up{j}} \tensor{\V}{\down{k}\down{l}\down{;}\down{m}} 
     \tensor{\V}{\up{k}\up{l}\down{;}\up{m}}
\JulyStrut
\\
- 
  \frac{4\,\left( 2592 - 5262\,n + 625\,n^2 + 14\,n^3 \right) }
    {315\,\left( -4 + n \right) } 
   \tensor{\bg}{\up{i}\up{j}} \tensor{\V}{\down{k}\down{l}} 
     \tensor{\V}{\up{k}\up{l}\down{;}\down{m}\up{m}}
\JulyStrut
\\
- 
  \frac{4\,\left( 74928 - 21908\,n - 968\,n^2 + 283\,n^3 \right) }
    {315\,\left( -6 + n \right) \,\left( -4 + n \right) } 
   \tensor{\bg}{\up{i}\up{j}} \tensor{\V}{\down{k}\down{l}\down{;}\down{m}} 
     \tensor{\V}{\up{k}\up{m}\down{;}\up{l}}
\JulyStrut
\\
+ 
  \frac{8\,\left( 311616 - 146460\,n + 27484\,n^2 - 2167\,n^3 + 5\,n^4 + 
        7\,n^5 \right) }{315\,\left( -6 + n \right) \,
      \left( -4 + n \right) } \tensor{\bg}{\up{i}\up{j}} 
    \tensor{\V}{\down{k}\down{l}} 
     \tensor{\V}{\up{k}\down{m}} \tensor{\V}{\up{l}\up{m}}
\JulyStrut
\\
+ 
  \frac{2\,\left( -44928 - 10224\,n + 14400\,n^2 - 3206\,n^3 + 
        203\,n^4 \right) }{45\,\left( -6 + n \right) \,
      \left( -4 + n \right) } \tensor{\bg}{\up{i}\up{j}} 
    \tensor{\V}{\down{k}\down{l}} \tensor{\V}{\up{k}\up{l}}
\Mvariable{\J}
\JulyStrut
\\
+ 
  \frac{-2560 + 1568\,n + 420\,n^2 - 105\,n^3}{210} 
   \tensor{\bg}{\up{i}\up{j}} \Mvariable{\J}^{3}
+ 
  \frac{14820 - 3650\,n + 231\,n^2}{315} 
   \tensor{\bg}{\up{i}\up{j}} \tensor{\Mvariable{\J}}{\down{;}\down{k}} 
     \tensor{\Mvariable{\J}}{\down{;}\up{k}}
\JulyStrut
\\
+ 
  \frac{4\,\left( -6 - 31\,n + 5\,n^2 \right) }{15} 
   \tensor{\bg}{\up{i}\up{j}} \Mvariable{\J} 
     \tensor{\Mvariable{\J}}{\down{;}\down{k}\up{k}}
\JulyStrut
\\
+ 
  \frac{2\,\left( 480384 - 238464\,n + 44040\,n^2 - 3346\,n^3 + 
        91\,n^4 \right) }{315\,\left( -6 + n \right) \,
      \left( -4 + n \right) } \tensor{\bg}{\up{i}\up{j}} 
    \tensor{\V}{\down{k}\down{l}} 
     \tensor{\Mvariable{\J}}{\down{;}\up{k}\up{l}}
- 
  \frac{3\,\left( -48 + 7\,n \right) }{35} 
   \tensor{\bg}{\up{i}\up{j}} \tensor{\Mvariable{\J}}{
     \down{;}\down{k}\up{k}\down{l}\up{l}}
\JulyStrut
\\
+ 
  \frac{8\,\left( 188 + 21\,n \right) }{315} 
   \tensor{\bg}{\up{i}\up{j}} \tensor{\V}{
      \down{k}\down{l}\down{;}\down{m}\down{p}} 
     \tensor{\C}{\up{k}\up{m}\up{l}\up{p}}
\JulyStrut
\\
- 
  \frac{8\,\left( 79488 - 20160\,n + 1020\,n^2 + 50\,n^3 + 7\,n^4 \right) }
    {315\,\left( -6 + n \right) \,\left( -4 + n \right) } 
   \tensor{\bg}{\up{i}\up{j}} \tensor{\V}{\down{k}\down{l}} 
     \tensor{\V}{\down{m}\down{p}} 
      \tensor{\C}{\up{k}\up{m}\up{l}\up{p}} + 
  \frac{16}{45} \tensor{\bg}{\up{i}\up{j}} 
    \Mvariable{\J} \tensor{\C}{
       \down{k}\down{l}\down{m}\down{p}} 
      \tensor{\C}{\up{k}\up{m}\up{l}\up{p}}
\JulyStrut
\\
- 
  \frac{16}{45} \tensor{\bg}{\up{i}\up{j}} 
    \tensor{\C}{\down{k}
       \down{l}\down{m}\down{p}\down{;}\down{q}} 
     \tensor{\C}{\up{k}\up{m}\up{p}\up{q}\down{;}\up{l}} + 
  \frac{16\,\left( -4 + n \right) }{45} 
   \tensor{\bg}{\up{i}\up{j}} \tensor{\V}{\down{k}\down{l}} 
     \tensor{\C}{\up{k}\down{m}\down{p}\down{q}} 
      \tensor{\C}{\up{l}\up{p}\up{m}\up{q}}
\JulyStrut
\\
- 
  \frac{16}{45} \tensor{\bg}{\up{i}\up{j}} 
    \tensor{\C}{\down{k}\down{l}\down{m}\down{p}} 
     \tensor{\C}{\up{k}\down{q}\up{m}\down{r}} 
      \tensor{\C}{\up{l}\up{q}\up{p}\up{r}} - 
  \frac{16}{45} \tensor{\bg}{\up{i}\up{j}} 
    \tensor{\C}{\down{k}\down{l}\down{m}\down{p}} 
     \tensor{\C}{\up{k}\up{m}\down{q}\down{r}} 
      \tensor{\C}{\up{l}\up{q}\up{p}\up{r}}
\JulyStrut
\end{array}
$$
\caption{\label{TwoTenAA}The tensor $\ATen^{ij}$}
\end{figure}
\begin{figure}
$$
\begin{array}{l}
\frac{8\,\left( 63648 - 43740\,n + 8300\,n^2 - 405\,n^3 + 37\,n^4 \right) }
    {315\,\left( -6 + n \right) \,\left( -4 + n \right) } 
\tensor{\V}{\down{k}\down{l}} \tensor{\V}{\up{i}\up{j}\down{;}\up{k}\up{l}} \
\\
- \frac{2\,\left( -1728 + 1164\,n - 292\,n^2 + n^3 \right) }
    {105\,\left( -6 + n \right) \,\left( -4 + n \right) } 
   \tensor{\V}{\up{i}\up{j}\down{;}\down{k}\up{k}\down{l}\up{l}} + 
  \frac{8\,\left( -11108 + 5601\,n - 447\,n^2 + 14\,n^3 \right) }
    {315\,\left( -4 + n \right) } 
   \tensor{\V}{\down{k}\down{l}\down{;}\up{j}} 
    \tensor{\V}{\up{i}\up{k}\down{;}\up{l}}
\JulyStrut
\\
- 
  \frac{16\,\left( -57312 + 18900\,n - 1240\,n^2 - 5\,n^3 + 27\,n^4 \
\right) }{315\,\left( -6 + n \right) \,\left( -4 + n \right) } 
\tensor{\V}{\down{k}\down{l}} \tensor{\V}{\up{i}\up{k}\down{;}\up{j}\up{l}} \
\JulyStrut
\\
- \frac{4\,\left( -23040 + 98136\,n - 18188\,n^2 + 134\,n^3 + 93\,n^4 \
\right) }{315\,\left( -6 + n \right) \,\left( -4 + n \right) } 
\tensor{\V}{\up{i}\down{k}\down{;}\down{l}\up{l}} \tensor{\V}{\up{j}\up{k}} \
\JulyStrut
\\
- \frac{4\,\left( -7754 - 1224\,n + 29\,n^2 \right) }{315} 
   \tensor{\V}{\up{i}\down{k}\down{;}\down{l}} 
    \tensor{\V}{\up{j}\up{k}\down{;}\up{l}}
\JulyStrut
\\
+ 
  \frac{8\,\left( 589824 - 282624\,n + 54804\,n^2 - 4472\,n^3 + 17\,n^4 + 
        6\,n^5 \right) }{105\,\left( -6 + n \right) \,
      \left( -4 + n \right) } \tensor{\V}{\down{k}\down{l}} 
    \tensor{\V}{\up{i}\up{k}} \tensor{\V}{\up{j}\up{l}}
\JulyStrut
\\
- 
  \frac{4\,\left( -51384 + 36938\,n - 9529\,n^2 + 662\,n^3 + 93\,n^4 \
\right) }{315\,\left( -6 + n \right) \,\left( -4 + n \right) } 
   \tensor{\V}{\up{i}\down{k}\down{;}\down{l}} 
    \tensor{\V}{\up{j}\up{l}\down{;}\up{k}}
\JulyStrut
\\
+ 
  \frac{4\,\left( 47232 - 112776\,n + 38488\,n^2 - 4444\,n^3 + 
        165\,n^4 \right) }{105\,\left( -6 + n \right) \,
      \left( -4 + n \right) } \tensor{\V}{\down{k}\down{l}} 
    \tensor{\V}{\up{i}\up{j}} \tensor{\V}{\up{k}\up{l}}
\JulyStrut
\\
+ 
  \frac{2\,\left( -53720 + 22678\,n - 3194\,n^2 + 63\,n^3 \right) }
    {315\,\left( -4 + n \right) } 
   \tensor{\V}{\down{k}\down{l}\down{;}\up{i}} 
    \tensor{\V}{\up{k}\up{l}\down{;}\up{j}}
\JulyStrut
\\
+ 
  \frac{4\,\left( 184896 - 132840\,n + 30900\,n^2 - 2390\,n^3 + 
        59\,n^4 \right) }{315\,\left( -6 + n \right) \,
      \left( -4 + n \right) } \tensor{\V}{\down{k}\down{l}} 
    \tensor{\V}{\up{k}\up{l}\down{;}\up{i}\up{j}}
\JulyStrut
\\
+ 
  \frac{2\,\left( -2880 - 37008\,n + 17564\,n^2 - 2432\,n^3 + 21\,n^4 \
\right) }{315\,\left( -6 + n \right) \,\left( -4 + n \right) } 
   \tensor{\V}{\up{i}\up{j}\down{;}\down{k}\up{k}} \Mvariable{\J}
\JulyStrut
\\
+ 
  \frac{4\,\left( -33408 - 425736\,n + 228064\,n^2 - 37810\,n^3 + 
        2094\,n^4 + 21\,n^5 \right) }{315\,\left( -6 + n \right) \,
      \left( -4 + n \right) } \tensor{\V}{\up{i}\down{k}} 
    \tensor{\V}{\up{j}\up{k}} \Mvariable{\J}
\JulyStrut
\\
+ 
  \frac{-9216 - 1264\,n + 1568\,n^2 - 105\,n^3}{105} 
   \tensor{\V}{\up{i}\up{j}} \Mvariable{\J}^{2} + 
  \frac{9336 - 8206\,n + 875\,n^2}{315} 
   \tensor{\Mvariable{\J}}{\down{;}\up{i}} 
    \tensor{\Mvariable{\J}}{\down{;}\up{j}}
\JulyStrut
\\
+ 
  \frac{8\,\left( -45432 + 19134\,n - 2965\,n^2 + 138\,n^3 \right) }
    {315\,\left( -6 + n \right) } 
   \tensor{\V}{\up{i}\down{k}\down{;}\up{j}} 
    \tensor{\Mvariable{\J}}{\down{;}\up{k}}
\JulyStrut
\\
- 
  \frac{2\,\left( 24768 - 1404\,n - 2584\,n^2 + 215\,n^3 \right) }
    {315\,\left( -6 + n \right) } 
   \tensor{\V}{\up{i}\up{j}\down{;}\down{k}} 
    \tensor{\Mvariable{\J}}{\down{;}\up{k}} + 
  \frac{4\,\left( 6108 - 1487\,n + 70\,n^2 \right) }{315} 
\tensor{\V}{\up{i}\up{j}}
\tensor{\Mvariable{\J}}{\down{;}\down{k}\up{k}} \
\JulyStrut
\\
+ \frac{2\,\left( -40320 - 77904\,n + 52144\,n^2 - 9826\,n^3 + 
        651\,n^4 \right) }{315\,\left( -6 + n \right) \,
      \left( -4 + n \right) } \Mvariable{\J} 
    \tensor{\Mvariable{\J}}{\down{;}\up{i}\up{j}}
\JulyStrut
\\
+ 
  \frac{4\,\left( 934848 - 481176\,n + 106800\,n^2 - 11644\,n^3 + 
        457\,n^4 \right) }{315\,\left( -6 + n \right) \,
      \left( -4 + n \right) } \tensor{\V}{\up{i}\down{k}} 
    \tensor{\Mvariable{\J}}{\down{;}\up{j}\up{k}}
\JulyStrut
\\
- 
  \frac{4\,\left( -15984 + 8490\,n - 1411\,n^2 + 85\,n^3 \right) }
    {105\,\left( -6 + n \right) \,\left( -4 + n \right) } 
   \tensor{\Mvariable{\J}}{\down{;}\up{i}\up{j}\down{k}\up{k}}
\JulyStrut
\end{array}
$$
\caption{\label{TwoTenBB}The tensor $\BTen^{ij}$}
\end{figure}
\begin{figure}
$$
\begin{array}{l}
  \frac{8\,\left( -3984 + 4492\,n - 296\,n^2 + 13\,n^3 \right) }
    {315\,\left( -6 + n \right) \,\left( -4 + n \right) } 
   \tensor{\V}{\down{k}\down{l}\down{;}\down{m}\up{m}} 
    \tensor{\C}{\up{i}\up{k}\up{j}\up{l}}
\JulyStrut
\\
- 
  \frac{4\,\left( 33984 - 21624\,n + 4540\,n^2 - 346\,n^3 + 21\,n^4 \
\right) }{105\,\left( -6 + n \right) \,\left( -4 + n \right) } 
   \tensor{\V}{\down{k}\down{l}} 
    \Mvariable{\J} \tensor{\C}{\up{i}\up{k}\up{j}\up{l}}
\JulyStrut
\\
- 
  \frac{8\,\left( -19392 + 14302\,n - 2403\,n^2 + 128\,n^3 \right) }
    {315\,\left( -6 + n \right) \,\left( -4 + n \right) } 
   \tensor{\Mvariable{\J}}{\down{;}\down{k}\down{l}} 
    \tensor{\C}{\up{i}\up{k}\up{j}\up{l}}
\JulyStrut
\\
+ 
  \frac{4\,\left( -33408 + 13680\,n - 1082\,n^2 + 95\,n^3 \right) }
    {315\,\left( -6 + n \right) \,\left( -4 + n \right) } 
   \tensor{\V}{\down{k}\down{l}\down{;}\down{m}} 
    \tensor{\C}{\up{i}\up{k}\up{j}\up{m}\down{;}\up{l}} - 
  \frac{80}{63} \tensor{\C}{\down{k}\down{l}\down{m}\down{p}} 
    \tensor{\C}{\up{i}\up{k}\up{j}\up{m}\down{;}\up{l}\up{p}}
\JulyStrut
\\
- 
  \frac{16\,\left( -52992 + 13248\,n - 472\,n^2 - 20\,n^3 + 3\,n^4 \right) }
    {63\,\left( -6 + n \right) \,\left( -4 + n \right) } 
   \tensor{\V}{\down{k}\down{l}} 
    \tensor{\V}{\up{j}\down{m}} 
     \tensor{\C}{\up{i}\up{k}\up{l}\up{m}}
\JulyStrut
\\
+ 
  \frac{4\,\left( 69168 - 25940\,n + 1472\,n^2 + 15\,n^3 \right) }
    {315\,\left( -6 + n \right) \,\left( -4 + n \right) } 
   \tensor{\V}{\down{k}\down{l}\down{;}\down{m}} 
    \tensor{\C}{\up{i}\up{k}\up{l}\up{m}\down{;}\up{j}}
\JulyStrut
\\
+ 
  \frac{8\,\left( 17088 - 16264\,n + 652\,n^2 + 94\,n^3 + 5\,n^4 \right) }
    {315\,\left( -6 + n \right) \,\left( -4 + n \right) } 
   \tensor{\V}{\down{k}\down{l}} 
    \tensor{\V}{\up{k}\down{m}} 
     \tensor{\C}{\up{i}\up{l}\up{j}\up{m}} + 
  \frac{128\,\left( 31 + 2\,n \right) }{315} 
   \tensor{\V}{\down{k}\down{l}\down{;}\up{i}\down{m}} 
    \tensor{\C}{\up{j}\up{k}\up{l}\up{m}}
\JulyStrut
\\
- 
  \frac{8}{105} \tensor{\C}{
     \up{i}\down{k}\down{l}\down{m}\down{;}\down{p}} 
    \tensor{\C}{\up{j}\up{k}\up{l}\up{p}\down{;}\up{m}} + 
  \frac{16\,\left( -7 + 3\,n \right) }{45} 
   \Mvariable{\J} \tensor{\C}{\up{i}\down{k}\down{l}\down{m}} 
     \tensor{\C}{\up{j}\up{l}\up{k}\up{m}} - 
  \frac{68}{63} \tensor{\C}{
     \up{i}\down{k}\down{l}\down{m}\down{;}\down{p}} 
    \tensor{\C}{\up{j}\up{l}\up{k}\up{p}\down{;}\up{m}}
\JulyStrut
\\
+ 
  \frac{32\,\left( 231 + 2\,n \right) }{315} 
   \tensor{\V}{\up{i}\down{k}\down{;}\down{l}\down{m}} 
    \tensor{\C}{\up{j}\up{m}\up{k}\up{l}}
+ 
  \frac{8\,\left( 8016 - 1576\,n - 506\,n^2 + 21\,n^3 \right) }
    {315\,\left( -6 + n \right) \,\left( -4 + n \right) } 
   \tensor{\V}{\down{k}\down{l}} 
    \tensor{\C}{\up{i}\down{m}\up{k}\down{p}} 
     \tensor{\C}{\up{j}\up{m}\up{l}\up{p}}
\JulyStrut
\\
+
  \frac{64}{35} \tensor{\C}{\down{k}\down{l}\down{m}\down{p}} 
    \tensor{\C}{\up{i}\up{k}\up{l}\down{q}} 
     \tensor{\C}{\up{j}\up{m}\up{p}\up{q}}
- 
  \frac{16\,\left( 22200 - 7798\,n + 215\,n^2 + 8\,n^3 \right) }
    {315\,\left( -6 + n \right) \,\left( -4 + n \right) } 
   \tensor{\V}{\down{k}\down{l}} 
    \tensor{\C}{\up{i}\down{m}\up{k}\down{p}} 
     \tensor{\C}{\up{j}\up{p}\up{l}\up{m}}
\JulyStrut
\\
+
  \frac{176}{105} \tensor{\C}{\down{k}\down{l}\down{m}\down{p}
     } \tensor{\C}{\up{i}\up{k}\up{m}\down{q}} 
     \tensor{\C}{\up{j}\up{p}\up{l}\up{q}} 
- 
  \frac{32}{45} \tensor{\V}{\up{i}\up{j}} 
    \tensor{\C}{\down{k}\down{l}\down{m}\down{p}} 
     \tensor{\C}{\up{k}\up{m}\up{l}\up{p}}
\JulyStrut
\\
- 
  \frac{16\,\left( 2692 - 143\,n + 25\,n^2 \right) }
    {315\,\left( -4 + n \right) } 
   \tensor{\V}{\down{k}\down{l}} 
    \tensor{\C}{\up{i}\down{m}\up{j}\down{p}} 
     \tensor{\C}{\up{k}\up{m}\up{l}\up{p}}
- 
  \frac{16\,\left( -33 + 20\,n \right) }{315} 
   \tensor{\V}{\up{i}\down{k}} \tensor{\C}{
      \up{j}\down{l}\down{m}\down{p}} 
     \tensor{\C}{\up{k}\up{m}\up{l}\up{p}}
\JulyStrut
\\
- 
  \frac{32}{45} \tensor{\C}{\down{k}\down{l}\down{m}\down{p}} 
    \tensor{\C}{\up{i}\down{q}\up{j}\up{k}} 
     \tensor{\C}{\up{l}\up{m}\up{p}\up{q}}
\JulyStrut
\end{array}
$$
\caption{\label{TwoTenCC}The tensor $\CTen^{ij}$}
\end{figure}
Finally, $Q_{8,n}^g$ denotes the scalar given in Figure~\ref{QEight}.
\begin{figure}
$$
\begin{array}{l}
%
%
\frac{-12\,\left( 8 - 4\,n + n^2 \right) }{{\left( -4 + n \right) }^2} 
   \tensor{\V}{\down{i}\down{j}\down{;}\down{k}\up{k}} 
    \tensor{\V}{\up{i}\up{j}\down{;}\down{l}\up{l}} - 
  24 \tensor{\V}{\down{i}\down{j}\down{;}\down{k}\down{l}} 
    \tensor{\V}{\up{i}\up{j}\down{;}\up{k}\up{l}} - 
  \frac{48\,\left( -18 + 7\,n \right) }{-6 + n} 
   \tensor{\V}{\down{i}\down{j}} 
    \tensor{\V}{\down{k}\down{l}}
   \tensor{\V}{\up{i}\up{j}\down{;}\up{k}\up{l}} \
\\
- \frac{48\,\left( -2 + n \right) }{-4 + n} 
   \tensor{\V}{\down{i}\down{j}\down{;}\down{k}} 
    \tensor{\V}{\up{i}\up{j}\down{;}\up{k}\down{l}\up{l}} - 
  \frac{12\,\left( -2 + n \right) }{-6 + n} 
   \tensor{\V}{\down{i}\down{j}} 
    \tensor{\V}{\up{i}\up{j}\down{;}\down{k}\up{k}\down{l}\up{l}} + 
  \frac{192\,n}{-6 + n} \tensor{\V}{\down{i}\down{j}} 
    \tensor{\V}{\down{k}\down{l}} \tensor{\V}{\up{i}\up{k}\down{;}\up{j}\up{l}} \
\JulyStrut
\\
- \frac{48\,\left( -32 + 44\,n - 12\,n^2 + n^3 \right) }
    {\left( -6 + n \right) \,\left( -4 + n \right) } 
   \tensor{\V}{\down{i}\down{j}} 
    \tensor{\V}{\down{k}\down{l}\down{;}\up{i}} 
     \tensor{\V}{\up{j}\up{k}\down{;}\up{l}} + 
  \frac{384\,\left( -16 + 3\,n \right) }
    {\left( -6 + n \right) \,\left( -4 + n \right) } 
   \tensor{\V}{\down{i}\down{j}} 
    \tensor{\V}{\up{i}\down{k}\down{;}\down{l}} 
     \tensor{\V}{\up{j}\up{k}\down{;}\up{l}}
\JulyStrut
\\
+ 
  \frac{384\,\left( 4 - 6\,n + n^2 \right) }
    {\left( -6 + n \right) \,{\left( -4 + n \right) }^2} 
   \tensor{\V}{\down{i}\down{j}} 
    \tensor{\V}{\up{i}\down{k}} \tensor{\V}{\up{j}\up{k}\down{;}\down{l}\up{l}} \
+ \frac{192\,\left( 144 - 32\,n - 4\,n^2 + n^3 \right) }
    {\left( -6 + n \right) \,{\left( -4 + n \right) }^2} 
   \tensor{\V}{\down{i}\down{j}} 
    \tensor{\V}{\down{k}\down{l}} 
     \tensor{\V}{\up{i}\up{k}} \tensor{\V}{\up{j}\up{l}}
\JulyStrut
\\
+ 
  \frac{48\,\left( -32 + 20\,n - 8\,n^2 + n^3 \right) }
    {\left( -6 + n \right) \,\left( -4 + n \right) } 
   \tensor{\V}{\down{i}\down{j}} 
    \tensor{\V}{\up{i}\down{k}\down{;}\down{l}} 
     \tensor{\V}{\up{j}\up{l}\down{;}\up{k}}
\JulyStrut
\\
+ 
  \frac{6\,\left( -8 + n \right) \,
      \left( 256 - 32\,n - 18\,n^2 + 3\,n^3 \right) }{\left( -6 + n \
\right) \,{\left( -4 + n \right) }^2} 
   \tensor{\V}{\down{i}\down{j}} 
    \tensor{\V}{\down{k}\down{l}} 
     \tensor{\V}{\up{i}\up{j}} \tensor{\V}{\up{k}\up{l}}
\JulyStrut
\\
+ 
  \frac{24\,\left( -176 + 80\,n - 14\,n^2 + n^3 \right) }
    {\left( -6 + n \right) \,\left( -4 + n \right) } 
   \tensor{\V}{\down{i}\down{j}} 
    \tensor{\V}{\down{k}\down{l}\down{;}\up{i}} 
     \tensor{\V}{\up{k}\up{l}\down{;}\up{j}} + 
  \frac{4\,\left( 48 + 80\,n - 46\,n^2 + 5\,n^3 \right) }
    {\left( -6 + n \right) \,\left( -4 + n \right) } 
   \tensor{\V}{\down{i}\down{j}\down{;}\down{k}} 
    \tensor{\V}{\up{i}\up{j}\down{;}\up{k}} \Mvariable{\J}
\JulyStrut
\\
+ 
  \frac{4\,\left( 12 - 20\,n + 5\,n^2 \right) }{-6 + n} 
   \tensor{\V}{\down{i}\down{j}} 
    \tensor{\V}{\up{i}\up{j}\down{;}\down{k}\up{k}} \Mvariable{\J} - 
  \frac{384\,n}{-6 + n} \tensor{\V}{\down{i}\down{j}} 
    \tensor{\V}{\up{i}\down{k}} \tensor{\V}{\up{j}\up{k}} \Mvariable{\J}
\JulyStrut
\\
- 
  \frac{96 - 80\,n - 30\,n^2 + 5\,n^3}{-6 + n} 
   \tensor{\V}{\down{i}\down{j}} \tensor{\V}{\up{i}\up{j}} \Mvariable{\J}^{2} \
+ \frac{\left( -4 + n \right) \,n\,\left( 4 + n \right) }{8} 
   \Mvariable{\J}^{4}
\JulyStrut
\\
+ \left( 2 + 13\,n - 2\,n^2 \right)  
   \Mvariable{\J} \tensor{\Mvariable{\J}}{\down{;}\down{i}} 
     \tensor{\Mvariable{\J}}{\down{;}\up{i}} - 
  \frac{\left( -2 + n \right) \,\left( 200 - 38\,n + 3\,n^2 \right) }
    {-4 + n} \tensor{\V}{\down{i}\down{j}} 
    \tensor{\Mvariable{\J}}{\down{;}\up{i}} 
     \tensor{\Mvariable{\J}}{\down{;}\up{j}}
\JulyStrut
\\
+ 
  \frac{24\,\left( -2 + n \right) \,\left( 88 - 20\,n + n^2 \right) }
    {\left( -6 + n \right) \,\left( -4 + n \right) } 
   \tensor{\V}{\down{i}\down{j}} 
    \tensor{\V}{\up{i}\down{k}\down{;}\up{j}} 
     \tensor{\Mvariable{\J}}{\down{;}\up{k}} + 
  \frac{16\,\left( -48 + 70\,n - 17\,n^2 + n^3 \right) }
    {\left( -6 + n \right) \,\left( -4 + n \right) } 
   \tensor{\V}{\down{i}\down{j}} 
    \tensor{\V}{\up{i}\up{j}\down{;}\down{k}} 
     \tensor{\Mvariable{\J}}{\down{;}\up{k}}
\JulyStrut
\\
+ 
  \frac{16 + 4\,n - 3\,n^2}{2} 
   \Mvariable{\J}^{2} \tensor{\Mvariable{\J}}{\down{;}\down{i}\up{i}} + 
  \frac{-8 + 3\,n}{2} \tensor{\Mvariable{\J}}{\down{;}\down{i}\up{i}} 
    \tensor{\Mvariable{\J}}{\down{;}\down{j}\up{j}} + 
  \frac{2\,\left( 32 - 40\,n + 5\,n^2 \right) }{-4 + n} 
   \tensor{\V}{\down{i}\down{j}} 
    \tensor{\V}{\up{i}\up{j}} \tensor{\Mvariable{\J}}{\down{;}\down{k}\up{k}} \
\JulyStrut
\\
- \frac{16\,\left( 4 - 8\,n + n^2 \right) }{{\left( -4 + n \right) }^2} 
   \tensor{\V}{\down{i}\down{j}\down{;}\down{k}\up{k}} 
    \tensor{\Mvariable{\J}}{\down{;}\up{i}\up{j}} + 
  \frac{4\,n\,\left( -46 + 5\,n \right) }{-6 + n} 
   \tensor{\V}{\down{i}\down{j}} 
    \Mvariable{\J} \tensor{\Mvariable{\J}}{\down{;}\up{i}\up{j}}
\JulyStrut
\\
+ 
  \frac{2\,\left( -336 + 136\,n - 20\,n^2 + n^3 \right) }
    {{\left( -4 + n \right) }^2} 
   \tensor{\Mvariable{\J}}{\down{;}\down{i}\down{j}} 
    \tensor{\Mvariable{\J}}{\down{;}\up{i}\up{j}} + 
  \frac{24\,\left( 1088 - 608\,n + 144\,n^2 - 18\,n^3 + n^4 \right) }
    {\left( -6 + n \right) \,{\left( -4 + n \right) }^2} 
   \tensor{\V}{\down{i}\down{j}} 
    \tensor{\V}{\up{i}\down{k}} \tensor{\Mvariable{\J}}{\down{;}\up{j}\up{k}} \
\JulyStrut
\\
+ \left( -38 + 5\,n \right)  \tensor{\Mvariable{\J}}{\down{;}\down{i}} 
    \tensor{\Mvariable{\J}}{\down{;}\up{i}\down{j}\up{j}} - 
  \frac{32\,\left( -7 + n \right) }{-4 + n} 
   \tensor{\V}{\down{i}\down{j}\down{;}\down{k}} 
    \tensor{\Mvariable{\J}}{\down{;}\up{i}\up{j}\up{k}} + 
  2\,\left( -1 + n \right)  \Mvariable{\J} 
    \tensor{\Mvariable{\J}}{\down{;}\down{i}\up{i}\down{j}\up{j}}
\JulyStrut
\\
- 
  \frac{4\,\left( -54 + 7\,n \right) }{-6 + n} 
   \tensor{\V}{\down{i}\down{j}} 
    \tensor{\Mvariable{\J}}{\down{;}\up{i}\up{j}\down{k}\up{k}} - 
  \tensor{\Mvariable{\J}}{\down{;}\down{i}\up{i}\down{j}\up{j}\down{k}\up{k}
   } - \frac{192\,\left( 4 - 6\,n + n^2 \right) }
    {\left( -6 + n \right) \,{\left( -4 + n \right) }^2} 
   \tensor{\V}{\down{i}\down{j}} 
    \tensor{\V}{\down{k}\down{l}\down{;}\down{m}\up{m}} 
     \tensor{\C}{\up{i}\up{k}\up{j}\up{l}}
\JulyStrut
\\
+ 
  \frac{96\,n}{-6 + n} \tensor{\V}{\down{i}\down{j}} 
    \tensor{\V}{\down{k}\down{l}} 
     \Mvariable{\J} \tensor{\C}{\up{i}\up{k}\up{j}\up{l}} - 
  \frac{24\,\left( -8 + n \right) \,\left( -16 - 2\,n + n^2 \right) }
    {\left( -6 + n \right) \,{\left( -4 + n \right) }^2} 
   \tensor{\V}{\down{i}\down{j}} 
    \tensor{\Mvariable{\J}}{\down{;}\down{k}\down{l}} 
     \tensor{\C}{\up{i}\up{k}\up{j}\up{l}}
\JulyStrut
\\
- 
  \frac{384\,\left( -5 + n \right) }
    {\left( -6 + n \right) \,\left( -4 + n \right) } 
   \tensor{\V}{\down{i}\down{j}} 
    \tensor{\V}{\down{k}\down{l}\down{;}\down{m}} 
     \tensor{\C}{\up{i}\up{k}\up{j}\up{l}\down{;}\up{m}} - 
  \frac{192}{-4 + n} \tensor{\V}{\down{i}\down{j}\down{;}\down{k}} 
    \tensor{\V}{\down{l}\down{m}\down{;}\up{k}} 
     \tensor{\C}{\up{i}\up{l}\up{j}\up{m}}
\JulyStrut
\\
+ 
  \frac{192\,\left( 112 - 16\,n - 6\,n^2 + n^3 \right) }
    {\left( -6 + n \right) \,{\left( -4 + n \right) }^2} 
   \tensor{\V}{\down{i}\down{j}} 
    \tensor{\V}{\down{k}\down{l}} 
     \tensor{\V}{\up{i}\down{m}} 
      \tensor{\C}{\up{j}\up{k}\up{l}\up{m}} - 
  \frac{96\,n}{-4 + n} \tensor{\V}{\down{i}\down{j}\down{;}\down{k}} 
    \tensor{\V}{\up{i}\down{l}\down{;}\down{m}} 
     \tensor{\C}{\up{j}\up{l}\up{k}\up{m}}
\JulyStrut
\\
+ 
  \frac{192}{-6 + n} \tensor{\V}{\down{i}\down{j}} 
    \tensor{\V}{\down{k}\down{l}} 
     \tensor{\C}{\up{i}\down{m}\up{k}\down{p}} 
      \tensor{\C}{\up{j}\up{m}\up{l}\up{p}} - 
  \frac{384}{-6 + n} \tensor{\V}{\down{i}\down{j}} 
    \tensor{\V}{\down{k}\down{l}} 
     \tensor{\C}{\up{i}\down{m}\up{k}\down{p}} 
      \tensor{\C}{\up{j}\up{p}\up{l}\up{m}}
\JulyStrut
\\
+ 
  \frac{192\,\left( -6 + n \right) }{{\left( -4 + n \right) }^2} 
   \tensor{\V}{\down{i}\down{j}} 
    \tensor{\V}{\down{k}\down{l}} 
     \tensor{\C}{\up{i}\down{m}\up{j}\down{p}} 
      \tensor{\C}{\up{k}\up{m}\up{l}\up{p}}
\JulyStrut
\end{array}
$$
\caption{\label{QEight}
The invariant $Q^g_{8,n}$}
\end{figure}
%
%

%
\vspace{5mm}
\subsection{Branson's Q-curvature} \label{Qcurvsect}

We have used $ P_{2k}$ to indicate a conformally invariant operator
between densities, $ P_{2k}:\ce[k-n/2]\to \ce[-k-n/2]$.  Suppose we
choose a metric from the conformal class.  Then we can trivialise
these density bundles, and so $ P_{2k}$ gives an operator $P_{2k}^g$
between functions on the Riemannian (or pseudo-Riemannian) structure
given by the choice of $ g$.  If we write $ (\xi^g)^w$ for the
operator given by multiplication by $ (\xi^g)^w\in\ce[w]$, then $
P_{2k}^g f= (\xi^g)^{k+n/2} P_{2k} (\xi^g)^{k-n/2k}f$. The GJMS
operators as discussed, for example, in \cite{Brsharp} are the $
P_{2k}^g$.  In this form the operators are not invariant but rather
covariant (see below), and conformally invariant operators are often
discussed entirely in this setting. For many purposes the difference
between $P_{2k}$ and $P_{2k}^g$ is rather small.  In particular, since
the Levi-Civita connection corresponding to $g$ annihilates $ \xi^g$,
the formulae above for the $ P_{2k}$ also serve as formulae for the
operators $ P_{2k}^g$.  From Theorem~\ref{L24Maybb} and the formulae
\nn{Dform}, \nn{box},\nn{tractcurv} and \nn{WDef}, there is a
universal expression for $ P_{2k}^g$ which is polynomial in $g$,
$g^{-1}$, $C$, $\V$, $\nd$, and the $ \nd$ covariant derivatives of $
C$ and $ \V$, and the coefficients in this universal expression are
real rational functions of the dimension $ n$ which are regular for
all odd $n$ and all $n\geq 2k$.

Let $\tilde{Q}_{2k,n}$ be the local invariant $ P^g_{2k}1$ on a
conformal $ n$-manifold.  Since $ P_{2k}^g$ is formally self-adjoint
(FSA) it is clear we can write it in the form $
P^g_{2k}=P^{g,1}_{2k}+\tilde{Q}_{2k,n}$, where $P^{g,1}_{2k}$ has the form $
\delta S^g_{2k}d$ with $ \delta$ the formal adjoint of $ d$ and $
S^g_{2k}$ an order $2k-2$ differential operator.

By setting $ w=0$ in the formulae \nn{Dform} and \nn{box} we see that,
as an operator on $ \ce$, $ D_A$ factors through the exterior
derivative $ d$. At least this is true given a choice of metric $ g$
from the conformal class.  Thus in dimension $n_0=2k$ it is clear from
Theorem~\ref{L24Maybb} that $ P^g_{n_0}$ is also a composition with $
d$.  Thus the term $ \tilde{Q}_{2k,n}$ vanishes in dimension $
2k=n_0$. Using this and a careful use of classical invariant theory  
one can conclude that in fact $\tilde{Q}_{2k,n}=\frac{n-2k}{2}Q^g_{2k,n}$.  
In the previous section we gave explicit formulae for 
$ Q^g_{4,n}, Q^g_{6,n}$ and $Q^g_{8,n}$.

Clearly $ Q^g_{2k,n}$ is also given by a formula rational in $ n$ and
regular at $ n=n_0:=2k$. In dimension $n_0 $, $
Q^g_{n_0}:=Q^g_{n_0,n_0}$ is by definition (modulo a sign $ (-1)^k$) Branson's Q-curvature, and for compact
conformal $ n_0$-manifolds, $\int_{M} Q^g_{n_0} $ is a global conformal
invariant. To see this, observe that the conformal invariance of $ P_{2k}$
is equivalent to the covariance law
$$
\Omega^{\frac{n+2k}{2}}P^{\hat{g}}_{2k}= P^g_{2k}\Omega^{\frac{n-2k}{2}},
$$
where $ \hat{g}=\Omega^2 g$ and we regard the powers of $ \Omega$
as multiplication operators. Applying both sides to the constant
function 1 we obtain $\frac{n-2k}{2}\Omega^{\frac{n+2k}{2}} Q^{\hat{g}}_{2k}=
\frac{n-2k}{2}Q^g_{2k}\Omega^{\frac{n-2k}{2}}+
P^{g,1}_{2k}\Omega^{\frac{n-2k}{2}} $. Expanding this out yields a
universal transformation formula for $ Q^g_{2k}$. Since $P^{g,1}_{2k} $
is a composition with $ d$ it is clear that we can divide this formula by
$\frac{n-2k}{2} $. Then in dimension $n_0=2k$ we obtain 
$$
\Omega^{n_0} Q^{\hat{g}}_{2k}= Q^g_{2k} + \delta S^g_{2k}d \Upsilon,
$$
where $ \Upsilon=\log \Omega$.
Then, if we denote by $\vol_g$ the volume density associated with a
metric $g$, we have $\vol_{\hat{g}}=\Omega^{n_0}\vol_{g}$, and the
conformal invariance of $\int_M Q^g_{2k}$ is clear.
Recall that a choice of metric $ g$ determines a
canonical section $ \xi^g$ of $ \ce[1]$ by 
$(\xi^g)^{-2}\bg=g$. It is convenient to redefine $ Q^g_{2k}$ to be $
(\xi^g)^{-n_0}$ times $ Q^g_{2k}$ as above. Then $ Q^g_{2k}$ is valued
in $ \ce[-n]$ and the transformation law simplifies to
\begin{equation}\label{Qtrans}
Q^{\hat{g}}_{2k}= Q^g_{2k} + \delta S^g_{2k}d \Upsilon, 
\end{equation}
where $ \delta$ and $ S^g_{2k}$ are now also density valued. Note that
we can also write this as $Q^{\hat{g}}_{2k}= Q^g_{2k} + P_{2k} \Up $
as $P^{g,1}_{2k}$ agrees with $P^{g}_{2k} $ in dimension $2k$.

The discussion above for $ Q^g_{n_0}$ and its properties are a minor
adaption of the arguments presented in Branson's \cite{Brsharp}. It is
clear that given explicit formulae for the $ P_{2k}$ as in the
previous section we can extract a formula for the $Q$-curvature as
follows: Take the order 0 part of the formula, divide by $(n-2k)/2 $
and then set $ n=2k$. For example $ Q^g_8$ is obtained by setting $
n=8$ in the formula given in Figure \ref{QEight}. In \cite{Brsharp} it
is also shown that the global invariant is not trivial. In fact, it is
established there that, on conformally flat structures, $ Q^g_{n_0}$
is given by a multiple of the Pfaffian plus a divergence, and so
$\int_M Q^g_{n_0} $ is a multiple of the Euler characteristic $
\chi(M)$.

One of the keys to the importance of $ Q_{2k}$ is the remarkable
transformation formula \nn{Qtrans}. We will describe a new
definition and construction for $ Q_{2k}$ and proof of this
transformation formula.  This leads to a direct formula for $Q_{2k}$.
Here to prove the transformation law we will use a dimensional
continuation argument. (This plays a minor role and can in fact can be 
replaced by a direct proof \cite{BrGogau}). The
construction is then adapted to proliferate other curvature quantities
with transformation formulae of the general form \nn{Qtrans}, and in
these cases dimensional continuation is not used at all.  See
Proposition~\ref{Qthings}.  (Since the original writing of this
Fefferman and Graham \cite{FeffGr01} have given another alternative
 construction and generalisation of the $ Q$-curvature which
involves the Poincar\'{e} metric.)

We work on a conformal manifold of dimension $n_0= 2k$. 
For a choice of metric $ g$ from the conformal structure let $I^g_A $
be the section of $ \ce_A[-1]$ defined by $ I^g_A:=(n-2)Y_A-\J X_A$,
where, recall, $Y_A\in\ce_A[-1] $ gives the splitting of the tractor
bundle corresponding to the metric $ g$ (as in section
\ref{tractorsect}). We can write this as a triple $
[I^g_A]_g=((n-2),~ 0,~-\J)$. According to this definition,
if $ \hat{g}=\Omega^2 g$ then
we have $ I^{\hat{g}}_A:=(n-2)\widehat{Y}_A-\widehat{\J} X_A$ or
$[I^{\hat{g}}_A]_{\hat{g}}=((n-2),~ 0,~-\widehat{\J}) $. In terms of
the splitting determined by $ g$, $I^{\hat{g}}_A $ is given by
$[I^{\hat{g}}_A]_{g}=((n-2),-(n-2)\Up_a,~-\widehat{\J}-(n/2-1)\Up^b\Up_b)
$.  By \nn{Rhotrans} 
and $ \Up_a=\nabla_a \Up$ this becomes
$[I^{\hat{g}}_A]_{g}=((n-2),-(n-2)\nd_a\Up,~-\J+\Delta \Up) $, and so
\begin{equation}\label{Itrans}
I^{\hat{g}}_A =I^g_A- D_A \Up. 
\end{equation}
This observation is due to Eastwood who also pointed out
\cite{Eastpriv} that on conformally flat structures this yields
Branson's curvature as follows. For each metric define $ Q^g_B $ by
$Q^g_B:=  \Box_{2k-2} I^g_{B}$.  Then, by \nn{Itrans}, $
Q^{\hat{g}}_{B}= Q^g_B- \Box_{2k-2} D_B \Up.  $ Now since the
structure is conformally flat, $ \Box_{2k-2}D_A \Up= -X_A P_{2k}\Up$
(see Theorem~\ref{L24Maybb} or e.g.\ \cite{Gosrni}). Thus we have 
$$
Q^{\hat{g}}_{B}= Q^g_B+ X_B P_{2k} \Up.  
$$
It follows that $X^BQ^{\hat{g}}_{B} =X^BQ^g_B $ is a conformal
invariant of weight $ n_0-2$. On a conformally flat structure there
are no conformal invariants of the structure and so this vanishes.
Since this vanishes, $Z^{Bc}Q^g_B$ is also conformally invariant and so
must vanish. This shows that for any conformally flat metric,
$Q^g_B=X_B Q^g $ for some Riemannian invariant $ Q^g$ and also that $
Q^{\hat{g}}=Q^g+P_{2k}\Up$.  That is, it transforms according to
\nn{Qtrans}.   On conformally
flat structures one can always locally choose a metric that is flat
whence all Riemannian invariants vanish. Using this we deduce that $
Q^g$ is Branson's curvature, that is $ Q^g=Q^g_{2k}$.

Via the theorem we can generalise Eastwood's cunning construction to
the curved case. Note \nn{Itrans} holds on any conformal manifold. Let
us define the operator $F_{C}{}^B: \ce_{B}[k-1-n/2]\to
\ce_{C}[1-k-n/2]$ by $ F_{C}{}^B:=(k-2)^{-1}(n-2k+2)^{-1}D^KE_{CK}{}^{AB}D_A$
where $ E$ is the operator defined in Theorem~\ref{L24Maybb}. Now on a
dimension $n_0= 2k$ manifold we simply define $ Q^{g}_C:=F_{C}{}^B
I^g_B$. From \nn{Itrans} and the theorem we have immediately 
\begin{equation}\label{QAtrans}
Q^{\hat{g}}_C=Q^{g}_C +X_C P_{2k}\Upsilon.
\end{equation}

It remains to verify that for any metric $g$ in the conformal class,
$ Q^{g}_C$ is indeed $ X_C Q^g_{2k}$. In any dimension $ n$ and
given any metric $ g$, if $ f\in \ce[w]$ let us write $ D_A f=D^1_A
f+wD^0_{A}f$, where $ D^1_A f:= (n+2w-2)Z_A{}^a\nd_a f -X_A\Delta f$ and $
wD^0_{A}f$ is the remaining order zero part. That is, $ D^0_{A}f = (n+2w-2)Y_A
f-X_A \J f $. Let $ w=k-n/2$, and assume $ n$ is odd or $ n\leq2k$.  Then $-
X_AP_{2k}f=F_A{}^BD_B f= F_A{}^BD^1_B f+wF_A{}^BD^0_B f $. Let $ \xi^g$
be the section of $ \ce[1]$ corresponding to $ g$. Recall that $ \xi^g
$ is parallel for the Levi-Civita connection of $g$. Since $(\xi^g)^w$
is a section of $ \ce[w]$, we have 
$$
X_A w Q^g_{2k,n}(\xi^g)^w = w
F_A{}^BD^0_B (\xi^g)^w.
$$
 Now $X^A F_A{}^BD^0_B(\xi^g)^w$ can be expressed as a universal
expression which is polynomial in $\bg$, $\bg^{-1}$, $C$, $\V$, $\nd$, and
the $ \nd$ covariant derivatives of $ C$ and $ \V$.  The coefficients
in this universal expression are real rational functions of the
dimension $ n$ which are regular for all odd $n$ and all $ n\geq
2k$. Furthermore, from the left-hand-side of the display, this
expression vanishes in even dimensions $ n<2k$ and for all odd
dimensions. Thus it must vanish in dimension $ n_0=2k$.  Similarly we
can conclude $Z^{Aa}F_A{}^BD^0_B(\xi^g)^w $ vanishes if $ n$ is odd or
$ n\leq 2k$.  Thus in dimension $ n_0$, $ w=0$, $ D^0_{B}(\xi^g)^w=
D^0_{B} 1= I^g_{B}$, and as $ Q^g_{2k}:=Q^g_{n_0,n_0}$, we have
$$
Q^{g}_A=F_A{}^B I^g_B =X_A Q^g_{2k} . 
$$
Thus we have the following.
\begin{proposition}\label{newQ}
$Y^A F_A{}^B I^g_B$ is a formula for Branson's Q-curvature
$Q^g_{2k} $. 
\end{proposition}
Note that if we take this as a definition for $Q$ then the
transformation property \nn{Qtrans} arises from \nn{QAtrans} which in
turn is an immediate consequence of \nn{Itrans}. The formula itself is
direct and requires no dimensional continuation.  The only subtlety in
the construction was in establishing that $Q^g_A $ has the form $X_A
Q^g_{2k}$. We employed a dimensional continuation argument to
establish this above, but it turns out that there is an elementary
direct proof using the ambient construction \cite{BrGogau}.  Finally
note that if we write $I^A_g =h^{AB} I^g_B$, then $ I^A_g F_A{}^B
I^g_B=(n-2)Q^g_{2k} $. 

It is straightforward to convert this tractor formula for $ Q^g_{2k}$ into 
a formula in terms of $ \nd$, $C$, $\V$, the metric, and
its inverse.  We simply expand $Y^A F_A{}^B I^g_B$ using the formula
for $ F_A{}^B$ as a partial contraction polynomial in $D_A$, $
W_{ABCD}$, $ X_A$, $h_{AB}$, and its inverse $ h^{AB}$, as obtained
from Proposition~\ref{tract4gjms}, and apply \nn{Dform}, \nn{box},
\nn{tractcurv}, and \nn{WDef} along the same lines as the calculations
in Section~\ref{convent}. In fact, as a means of checking against
formulae or calculational errors it is prudent to calculate the entire
tractor valued expression $ F_A{}^B I^g_B$ and verify from this that
only the bottom slot is not zero. That is that $ X^A F_A{}^B
I^g_B=0=Z^{Aa}F_A{}^B I^g_B$. Doing this for $ Q^g_{4}$ we obtain the known formula 
$$
Q_4=-2 
   \tensor{\V}{\down{i}\down{j}} \tensor{\V}{\up{i}\up{j}} + 
  2 \Mvariable{\J}^{2} -
  \tensor{\Mvariable{\J}}{\down{;}\down{j}\up{j}}.
  $$
  In terms of the
  Ricci curvature Rc and the scalar curvature Sc, this becomes $Q_4=
  -\frac{1}{2} \tensor{\Mvariable{Rc}}{\down{i}\down{j}}
  \tensor{\Mvariable{Rc}}{\up{i}\up{j}} + \frac{1}{6}
  \Mvariable{Sc}^{2} - \frac{1}{6}
  \tensor{\Mvariable{Sc}}{\down{;}\down{i}\up{i}} $.  For $ Q_6$ the
  formulae are more severely tested by calculating
  $E_{AB}{}^{CE}D^{~}_CI^g_{E}$. For this case $
  E_{AB}{}^{CE}D^{~}_CI^g_{E}=\Box D^{~}_A
  I^g_{B}+\frac{2}{n-4}W_A{}^C{}_B{}^E D^{~}_C I_E^g$ and the calculation
  verifies all components vanish except for the coefficient of $
  X_AX_B$, the negative of which is
$$
Q^g_6=-(8 \tensor{\V}{\down{i}\down{j}\down{;}\down{k}} 
    \tensor{\V}{\up{i}\up{j}\down{;}\up{k}} + 
  16 \tensor{\V}{\down{i}\down{j}} 
    \tensor{\V}{\up{i}\up{j}\down{;}\down{k}\up{k}} - 
  32 \tensor{\V}{\down{i}\down{j}} 
    \tensor{\V}{\up{i}\down{k}} \tensor{\V}{\up{j}\up{k}} - 
  16 \tensor{\V}{\down{i}\down{j}} \tensor{\V}{\up{i}\up{j}} \Mvariable{\J} + 
$$
$$
  8 \Mvariable{\J}^{3}
- 8 
    \tensor{\Mvariable{\J}}{\down{;}\down{k}\up{k}}
    \Mvariable{\J}
+
  \tensor{\Mvariable{\J}}{\down{;}\down{j}\up{j}\down{k}\up{k}} + 
  16 \tensor{\V}{\down{i}\down{j}} 
    \tensor{\V}{\down{k}\down{l}} 
     \tensor{\Mvariable{\C}}{\up{i}\up{k}\up{j}\up{l}})
.
$$
Note that these examples agree with setting $n=4 $ in \nn{Q4sc} and
$n=6$ in \nn{QSix}.

Using $ I^g_A$ it is easy to construct examples of other functionals of the
metric that have transformation laws of the same form as \nn{Qtrans}. We state this as a proposition.
\begin{proposition} \label{Qthings}
In dimension $ n_0=2k$, for each natural 
conformally invariant operator $ G_A{}^B:\ce_{B}[-1]\to
\ce_{A}[1-n_0]$ there is 
a Riemannian invariant $D^A G_A{}^BI^g_{B} $ with a 
conformal transformation of the form  
$$
D^A G_A{}^BI^{\hat{g}}_{B} = D^A G_A{}^BI^g_{B}+ \delta T^g_{2k}d \Upsilon,
$$
where $T^g_{2k}$ is a Riemannian invariant differential operator such
that the composition $\delta T^g_{2k}d $ is a conformally invariant
operator between functions and densities of weight $ -n$. 

If $ G_A{}^B$ is formally self-adjoint, then $\delta T^g_{2k}d$ is
formally self-adjoint.
\end{proposition}
\begin{proof}
  It is clear from \nn{Itrans} that $D^A G_A{}^BI^{\hat{g}}_{B} =
  D^A G_A{}^BI^g_{B} - D^A G_A{}^BD_B \Up$. Note that $D^A G_A{}^BD_B
  $ is a composition of conformally invariant operators.  Since $ \Up$
  is a function (i.e. is a density of weight 0), $ D_B \Up$ factors
  through $ d\Up$. From Proposition~\ref{intparts} it follows that the
  formal adjoint of $D^A G_A{}^BD_B $ also factors through the
  exterior derivative $d$. Thus the conformally invariant operator
  $D^A G_A{}^BD_B $ has the form $-\delta T^g_{2k}d $.
  
  Since $ \delta T^g_{2k}d\Upsilon=-D^A G_A{}^BD_B \Upsilon $, the last
  part of the proposition is immediate from
  Proposition~\ref{intparts}.
\end{proof}
An example in dimension 6 is to take $ G_A{}^B$ to be the order zero
operator $|C|^2 \delta_A{}^B$, where $ | C|^2=C^{abcd}C_{abcd}$. Then $D^A
G_A{}^BI^g_{B} = -4 \Delta |C|^2$, and $ D^A G_A{}^BI^{\hat{g}}_{B}=
-4\Delta |C|^2 + 16 \nabla^a|C|^2\nabla_a \Upsilon$. 
We can easily make many other examples via
the tractor objects already seen above. Other examples in dimension 6
are to take $G_A{}^B $ to be $ W_{ACDE}W^{BCDE}$ or $ D^E
W_{A}{}^B{}_{E}{}^FD_F$. In dimension 8 we could take $G_A{}^B $ to be 
$\delta^B_A D^PW_{PCDE}W^{ECDQ}D_Q $ and so on. 
Note all these examples have $ G$
formally self-adjoint.

It is a trivial exercise to verify that $D^A G_A{}^BI^g_{B} $ is
always a divergence, and so none of the invariants from the proposition
yield non-trivial global invariants.  Thus we could adjust the
definition of $ Q^g_{2k}$ by adding such functions without affecting
it as a representative of $ n_0^{\rm th}$ de Rham cohomology and also
without affecting the form of the transformation law \nn{Qtrans}. Such
changes would of course alter what we meant by $ S^g_{2k}$, but in any
case $ \delta S^g_{2k}d$ would remain an invariant operator on
functions. Such potential modifications are important from several
points of view. The transformation law \nn{Qtrans} is satisfied in
dimension 2 by the scalar curvature, or more precisely by $-{\rm
  Sc}/2$. In this context it is usually called the Gauss curvature
prescription equation. As mentioned earlier, $ Q_{2k}$ lends itself to
higher dimensional analogues of this curvature prescription problem. 
For the same reason, in any case where $D^A
G_A{}^BI^g_{B}$ is non-trivial,  $ Q^g_{2k}+D^A G_A{}^BI^g_{B} $ yields
a distinct, and apparently equally natural, curvature prescription
problem.  Of course then $ Q^g_{2k}+D^A G_A{}^BI^g_{B} $ does not
arise by Branson's construction from the GJMS operator $ P_{2k}$.  But
it is easily verified that it does arise via Branson's argument
applied to the conformally invariant operator
$$
P'_{2k}:= P_{2k}-D^A G_A{}^BD_B
:\ce[k-n/2]\to \ce[-k-n/2],
$$ 
and according to either either
construction the conformal transformation formula in dimension $n_0=2k $ is  
$$
 Q^{\hat{g}}_{2k}+D^A G_A{}^BI^{\hat{g}}_{B}=   Q^g_{2k}+D^A G_A{}^BI^g_{B} +
P'_{2k}\Up .
$$
It is possible, for example, that there are settings where such natural
modifications to the GJMS operators will yield operators
which are positive but the relevant GJMS operator fails
to be positive.

\section{The ambient metric construction} \label{ambsect}

The ambient metric construction of Fefferman-Graham associates to a
conformal manifold $ M$ of signature $ (p,q)$ a pseudo-Riemannian
so-called {\em ambient manifold} $ \tilde{M}$ of signature
$(p+1,q+1)$. The ambient manifold $ \tilde{M}$ is $ {\Cal Q}\times I$,
where $ I=(-1,1)$.  Henceforth we identify $ \Cal Q$ with its
natural inclusion $\iota: {\Cal Q} \to \tilde{M}$ given by $ {\Cal Q} \ni
q \mapsto (q,0)\in \tilde{M}$.  Observe that $\Cal Q $ carries a
tautological symmetric 2-tensor $ g_0$ given by $g_0=\pi^*g$ at the
point $(p,g)\in {\Cal Q}$. This satisfies $\delta_s^* g_0=s^2 g_0$,
where $ \delta_s$ is the natural $ {\Bbb R}_+$-action on $ \Cal Q$
given by $\delta_s (p,g)=(p,s^2 g)$. We will also write $ \delta_s$
for natural extension of this action to $\tM$ and denote by $ \X$ the
infinitesimal generator of this, i.e., for a smooth function $ f$ on
$\tM $, $ \X f(q)=\frac{d}{ds} f(\delta_s q)|_{s=1}$. The metric on
the ambient manifold $ \tM$ will be denoted $ \h$ and is required to
be a homogeneous extension of $ g_0$ in the sense that
\begin{equation}\label{ambmet12}
\iota^*\h=g_0 \quad \delta^*_s\h=s^2 \h ~\mbox{ for }~s>0 . 
\end{equation}
The idea of the Fefferman-Graham construction is to attempt to find a
formal power series solution along $ \Cal Q$ for the Cauchy problem of
an ambient metric $ \h$ satisfying \nn{ambmet12} and the condition
that it be Ricci-flat, i.e.\ Ric$(\h)=0$. It turns out that only a
weaker curvature condition can be satisfied in the even dimensional
case. The main results we need are contained in Theorem~2.1 of
\cite{FG2}: If $n$ is odd then, up to a $ \Bbb R_+$-equivariant
diffeomorphism fixing $ \Cal Q$, there is a unique power series
solution for $ \h$ satisfying \nn{ambmet12} and Ric$(\h)=0$.  If $ n$
is even then, up to a $ \Bbb R_+$-equivariant diffeomorphism fixing $
\Cal Q$ and the addition of terms vanishing to order $ n/2$, there is
a unique power series solution for $ \h$ satisfying \nn{ambmet12} and
such that, along $ \Cal Q$, Ric$(\h)$ vanishes to order $ n/2-2$ and
that the tangential components of Ric$(\h)$ vanish to order $n/2-1$.
We should point out that we only use the existence
part of the Fefferman-Graham construction. The uniqueness of the GJMS
operators, the covariant derivatives of the ambient curvature and so
forth are a consequence of the existence of tractor
formulae for these objects. 

By choosing a metric $ g$ from the conformal class on $ M$ we
determine a fibre variable on $ {\Cal Q}$ by writing a general point
of $ \Cal Q$ in the form $(p,t^2g(p))$, where $ p\in M$ and $ t>
0$.  Local coordinates $ x^i$ on $ M$ then correspond to coordinates $
(t,x^i)$ on $ \Cal Q$.  These extend \cite{FG2,GJMS} to coordinates
$(t,x^i,\rho)$ on $ \tM$, where $ \rho$ is a defining function for $
\Cal Q$ and such that the curves $\rho\mapsto (t,x^i,\rho)$ are geodesics
for $ \h$.  In these coordinates the ambient metric takes the form 
\begin{equation}\label{metricform}
\h=t^2g_{ij}(x,\rho)dx^i dx^j+2\rho dt dt +2tdt d\rho.
\end{equation}  
This form is forced to all orders in odd dimensions. In even
dimensions it is forced up to the addition of terms vanishing to order
$n/2$.  In order, in even dimensions, to recover the order $n$ GJMS
operators via the procedure of \cite{GJMS} we need also to assume that
the metric has this form up to the addition of terms vanishing to
order $ n/2+1$. Although we only need this form to that order, to
simplify our discussion we will assume that the form \nn{metricform}
holds to all orders in even dimensions too. This simply involves some
choice of extension for the Taylor series of the components $ g_{ij}$,
and then with this assumption the identities discussed in the
remainder of this subsection hold to all orders in all dimensions.
We write  $ \aNd$ for
the ambient Levi-Civita connection determined by $ \h$. 

In terms of the coordinates one has $ \X=t\frac{\partial}{\partial t}$, and if we let $
Q:=\h(\X,\X)$, then $ Q=2\rho t^2$ and is a defining function for $\Cal
Q $. In terms of this we have that, when $ n$ is even, the ambient
construction determines $ \h$ up to O$(Q^{n/2})$. Let us use upper
case abstract indices $A,B,\cdots $ for tensors on $ \tM$. For
example, if $ v^B$ is a vector field on $\tM $, then the ambient
Riemann tensor will be denoted $\aR_{AB}{}^C{}_{D}$ and defined by $
[\aNd_A,\aNd_B]v^C=\aR_{AB}{}^C{}_{D}v^D$. Indices will be raised and
lowered using the ambient metric $ \h_{AB}$ and its inverse $ \h^{AB}$
in the usual way.  We will soon see that this index convention is
consistent with our use of these indices for tractor bundles.

The homogeneity property of $\h$ in \nn{ambmet12} means that $ \X$ is
a conformal Killing vector, and in particular $ {\Cal L}_{\sX}\h=2\h$,
where $ \Cal L$ is the Lie derivative. It follows that $\aNd_{(A} \X_{B)}=\h_{AB} $. 
On the other hand, from the explicit coordinate form of the metric, we have that 
$ \aNd_B Q =2\X_B$, and so $\aNd_A \X_B$ is symmetric.
Thus
$$
\aNd_A \X_B =\h_{AB}
$$
which, in turn, implies  
\begin{equation}
\label{XRrel}
        \X^A\aR_{ABCD}=0.
\end{equation}

In terms of our notation the theorem of \cite{FG2} (mentioned above)
means that in even dimensions the ambient Ricci curvature $\aR_{BF}$
can be written in the form
$$
\aR_{BF}= Q^{n/2-2}\X_{(B}\aK_{F)}+Q^{n/2-1}\aL'_{BF}
$$
for appropriately homogeneous ambient tensors $\aK_F$ and $
\aL'_{BF}$. In fact the choice to extend the metric $ \h$ so that it
has the form \nn{metricform} restricts $ \aK_A$ significantly. From
\nn{XRrel} we have that $\X^A\aR_{AC}$ vanishes to all orders. With
the contracted Bianchi identity $2\aNd^A \aR_{AC}=\aNd_C \aS$ (where
$\aS$ denotes the ambient Ricci scalar curvature) this implies that
$\aK_A=\X_A \aK$ for an ambient homogeneous function $ \aK$.  Although
it is not strictly necessary, it will simplify our subsequent
calculations to restrict the ambient metric a little more.  An
elementary calculation verifies that we can adjust the components
$g_{ij}$ in \nn{metricform} so that $ \aK={\rm O}(Q)$. Thus finally we
have that that in even dimensions the metric has the form
\nn{metricform} and
\begin{equation}\label{Ricciform}
\aR_{BF}= Q^{n/2-1}\aL_{BF}
\end{equation}
for an appropriately homogeneous ambient tensor $\aL_{BF}$. (The
authors are appreciative of discussions with A.\ \u{C}ap and
C.R. Graham in relation to this point.)

\subsection{Recovering tractor calculus} \label{recovtract}

Recall that a section of $\ce[w]$ corresponds to a real-valued
function $ f$ on $\Cal Q$ with the homogeneity property $f(p,s^2g)=s^w
f(p,g)$, where $p\in M$ and $g$ is a metric from the conformal class $[g]$.
Let $ \cce_{\Cal Q}(w)$ denote the space of smooth
functions on $\Cal Q $ which are homogeneous of degree $w$ in this way.
We write $ \cce(w)$ for the smooth functions on $ \tM $ which are
similarly homogeneous, i.e. $ \tf\in \cce(w)$ means $\X^A\aNd_A f =w \tf
$. The construction of the GJMS operators in \cite{GJMS} exploits
this relationship between $ \ce[w]$ and $ \cce(w)$. We will use here
the analogous idea at the level of tensors on $ \tM$. This is
developed more fully in \cite{CapGoFG}, and here we just summarise the
basic ideas needed presently.

Writing $ \delta'_s$ for the derivative of the action $ \delta_s$, let
us define an equivalence relation on the ambient tangent bundle by $
U_{q_1}\sim V_{q_2} $ if and only if there is $ s\in {\Bbb R}_+$ such
that $V_{q_2}=s^{-1}\delta'_s U_{q_1}$. Corresponding to this we have
the equivalence relation on $ \tM$ by $ q_1\sim q_2$ if and only if $
q_2=\delta_s q_1$.  It is straightforward to verify that the space $
T\tM/\sim$ is a rank $n+2$ vector bundle over $ \tM/\sim$.  Sections
of this bundle correspond to smooth sections $ V:\tM \to T\tM$ with
the homogeneity property $ V(\delta_s p)= s^{-1}\delta'_sV(p)$, or
they could be alternatively characterised by their commutator with the
Euler field $ \X$, $ [\X,V]=-V$.  We will let $ \cce^A(0)$ ($
\cce^A_{\cq}(0)$) denote the space of sections of $ T\tM$ ($
T\tM|_{\cq}$) which are homogeneous in this way, and we will write $
\cce^{AB}(w)$ ($ \cce^{AB}_{\cq}(w)$) to mean $ \cce^A\otimes \cce^B
\otimes \cce(w) $ ($ \cce^A_{\cq}\otimes \cce^B_{\cq} \otimes
\cce_{\cq}(w) $ respectively) and so forth. (The reason for the weight
convention will soon be obvious.) We will write $ \cce^\Phi(w)$ to
mean an arbitrary tensor power of $ \cce^A(0)$ (or symmetrization
thereof and so forth) tensored with $ \cce(w)$ and we will say
sections of $ \cce^\Phi(w)$ are tensors homogeneous of {\em weight} $
w$. 
(We use the term ``weight'' here to distinguish from the homogeneity
``degree'' \cite{CapGoFG} as 
exposed by the Lie derivative along the field $\X$.)
Of course this construction is formal at the same order as the
construction of $ \tM$, but upon restriction to $ \Cal Q$, $
T\tM/\sim$ yields a genuine rank $ n+2$ vector bundle over $ M={\Cal
  Q}/\sim$ that will be denoted by $ \Cal T$ or $ {\Cal T}^A$.

It is immediate from the homogeneity property of $ \h$ that if $ U$
and $V$ are sections of $\cce^A(0)$, then the function $ \h_{AB}U^A
V^B$ is in $ \cce(0)$. Restricting to $ \Cal Q$ we see that $
\h_{AB}U^A V^B$ descends to a function on $ M$. From the bilinearity
and signature of $ \h$ it follows that $ \h$ descends to give a
signature $(p+1,q+1)$ metric $h^{\Cal T}$ on the bundle $ \Cal T$. We
can use this to raise and lower indices in the usual way.

Observe that $ \X^A\in \cce^A(1)$. Thus if $ \phi\in \cce(-1)$, then $
\phi \X^A \in \cce^A(0)$. The same is true upon restriction to $ \Cal
Q$, so we have a canonical inclusion $\ce[-1] \hookrightarrow {\Cal
  T}$ with image denoted by $ {\Cal T}^1$.  We write $X_{\Cal T}^A $ for
the natural section of $ {\Cal T}^A[1]:={\Cal T}^A\otimes \ce[1]$
giving this map, and so on $ \Cal Q$, $\X^A $ is the homogeneous
section representing $X_{\Cal T}^A $.  Clearly then $ V^A\mapsto
h^{\Cal T}_{AB}X^A_{\Cal T}V^B$ determines a canonical homomorphism $
{\Cal T}\to \ce[1]$, and we let ${\Cal T}^0$ denote the kernel.  Recall
that $ Q$ was defined to be $\h_{AB}\X^A\X^B$ and that this was a defining
function for $ \Cal Q$. Thus $ X^A_{\Cal T}$ is a null vector for the
metric $ h^{\Cal T}$, and it follows immediately that $ {\Cal T}^1
\subset {\Cal T}^0$. There is a simple geometric interpretation of
$\ct^0 $ and $ \ct^1$.  Observe that ${\Cal T}^0[1]$ corresponds to
sections of $
\cce^A_{\cq}(1)$ that are annihilated by contraction with $\X_A$.  On
$ \cq$ we have that $ \X_A= \frac{1}{2}\aNd_A Q$, so along $\cq $ the
sections of $ \cce^A_{\cq}(1)$ corresponding to $\ct^0[1]$ are precisely
those taking values in $ T\cq \subset T\tM|_{\cq}$ and which are
invariant under the action of $ \delta'_s$. Then, since $ \X$ is the
Euler vector field, it follows that $ \ct^1[1]$ corresponds to
functions in $ \cce_{\cq}^A(1)$ taking values in the vertical subbundle of $
T\cq$.  Of course the map $ \cq \to M$ is a submersion, and so 
$ \ct^0[1]/\ct^1[1]$ is naturally isomorphic to
$\ce^a=TM$. Tensoring by $ \ce[-1]$ we have $ \ct^0/\ct^1\cong
\ce^a[-1]$, and we can summarise the filtration of $ \ct$ by the
composition series
$$
\ct= \ce[1]\lpl \ce^a[-1] \lpl \ce[-1] .
$$

It is now straightforward to observe that the ambient Levi-Civita
connection $ \aNd $ also descends to give a connection on $ \Cal T$.
First, from the defining property that $ \aNd $ preserves the metric it
follows that if $ U^A \in \cce^A(w)$ and $ V^A\in \cce^A(w')$, then $
U^A\aNd_A V^B \in \cce^B(w+w'-1)$. Then since $ \aNd $ is torsion free,
we have that $ \aNd_{\sX} U - \aNd_U \X-[\X,U]=0$ for any tangent vector
field $ U$.  So if $ U\in \cce^A(0)$, then $\aNd_{\sX} U =0$, as, in
that case, $[\X, U] =-U$. So sections of $ \cce^A(0)$ may be
characterised as those which are covariantly parallel along the
vertical Euler vector field.  These two results imply that $ \aNd $
determines a connection $ \nd^{\ct}$ on $\ct$. For $ U\in \ct$ let $
\tU $ be the corresponding section of $ \ce^A_{\cq}(0)$. Similarly a
tangent vector field $ V$ on $M$ has a lift to a field $ \tV\in
\cce^A(1)$, on $ \Cal Q$, which is everywhere tangent to ${\Cal Q}$.
This is unique up to adding $f\X$, where $ f\in \cce(0)$. We extend $
\tU$ and $\tV $ homogeneously to fields on $ \tM$.  Then we can form $
\aNd_{\tV} \tU$. This is clearly independent of the extensions. Since
$ \aNd_{\sX} \tU =0$, it is also independent of the choice of $ \tV$ as
a lift of $ V$. Finally, it is a section of $ \cce^A(0)$ and so
determines a section $ \nd^{\ct}_V U $ of $ \ct$ which only depends on
$ U$ and $ V$. It is easily verified that this defines a covariant
derivative on $ \ct$.

Let us summarise. By the above construction the ambient manifold and
metric construction of Fefferman and Graham naturally determines a
rank $ (n+2)$ vector bundle $ \Cal T$ on $ M$. This vector bundle
comes equipped with a signature $(p+1,q+1)$ metric $ h^{\ct}$, a
connection $ \nd^{\ct}$, and a filtration determined by a canonical
section $X_{\ct} $ of $ \ct[1]$.  Furthermore if $ v^a$ is a smooth
tangent field on $ M$ and $ \phi$ is a smooth section of $ \ce[1]$,
one easily verifies from the above that the image of $ v^a\nd^{\ct}_a
(\phi X^B)$ lies in $\ct^0 $ and that composing with the map to the
quotient $ \ct^0/\ct^1$ recovers $ \phi v^b$. This is a non-degeneracy
property of the connection. This with the fact that $ \nd^{\ct}$
preserves the metric means that $ \ct$ is a tractor bundle with a
tractor connection in the sense of \cite{Cap-Gover}.  Since $ \aNd$ is
Ricci flat it follows that $ \nd^{\ct}$ satisfies the curvature
normalisation condition described in \cite{Cap-Gover2,Cap-Gover}.
(This is shown explicitly in \cite{CapGoFG}.)  From this and the
non-degeneracy we can conclude that $\ct^A$ and $ \nd^{\ct}_a$ are a
normal tractor bundle and connection corresponding to the defining
representation of SO$(p+1,q+1)$.  That is we can take, $ \ct^A=
\ce^A$, $ X_{\ct}^A=X^A$, and $ \nd_a^{\ct}$ to be the usual tractor
connection as in Section~\ref{tractorsect}. We henceforth drop the
notation $ \ct$.

We can also recover the operators introduced in the tractor setting.
Observe that the operator $\D_{AP}:=2\X_{[P}\aNd_{A]}$ annihilates the
function $ Q$ on $\tM$.
Thus $ \D_{AP}$ gives an operator $
\cce^\Phi_{\cq}(w)\to \cce_{[AP]}^\cq \otimes \cce^\Phi_\cq (w)$,
and it is a
trivial matter to show that this descends to $ D_{AP}: \ce^\Phi[w]\to
\ce_{[AP]}\otimes \ce^\Phi[w]$ as defined in
Section~\ref{tractorsect}. (Here, of course, $ \ce^\Phi[w]$ is {\em
  the} weight $ w$ tractor bundle corresponding to $ \cce^\Phi_\cq(w)$.)
Now we can formally follow the construction of $ D_A$. First one
calculates that, for $ \tV\in\cce^\Phi(w)$, and using \nn{XRrel}, we
have $ \h^{AB}\D_{A(Q}\D_{|B|P)_0} V =-\X_{(Q}\D_{P)_0} V$, where
\begin{equation}\label{ambDform}
\D_A V= (n+2w-2)\aNd_A V- \X_A \aDe V ,   \quad \aDe:=\aNd^B\aNd_B .
\end{equation} 
Then
we observe the map $\cce_P(w-1) \to \cce_{(PQ)_0}(w)$ given by $
\tS_P\mapsto \X_{(Q}\tS_{P)_0} $ is injective. It follows immediately
that, along $ \cq$, \nn{ambDform} is determined by the equation
$\h^{AB}\D_{A(Q}\D_{|B|P)_0} V =-\X_{(Q}\D_{P)_0} V$ and so is
precisely the operator
$\D_A:\cce^\Phi_{\cq}(w)\to \cce_A^\cq\otimes
\cce^\Phi_{\cq}(w-1) $,
which descends to $ D_A:\ce^\Phi[w]\to \ce_A\otimes
\ce^\Phi[w-1] $.  In particular this is true when $ w=1-n/2$, and so
$\aDe:\cce^\Phi(1-n/2) \to \cce^\Phi(-1-n/2)$ descends to the generalised
Yamabe operator $ \Box: \ce^\Phi[1-n/2]\to \ce^\Phi[-1-n/2]$. We will take
\nn{ambDform} as the definition of $ \D_A$ on $ \tM$. Although we
will not need it here, let us point out that $ \D_{AP}$ as defined
above acts more generally on sections of tensor bundles on $ \tM$ and
not just sections which are homogeneous. Following through the argument
above in this more general setting yields a generalisation of the
operator $\D_A$ on tensor bundles given by $ \D_A=n\aNd_A +2\X^B\aNd_B\aNd_A
-\X_A \aDe$.  This still has the property that along $ \cq$
it acts tangentially.  

Observe that  $
\h^{AB}\D_{A(Q}\D_{|B|P)_0} V$ is only of the form $- \X_{(Q}\D_{P)_0}
V$ to order $ Q^0$ along $ \cq$ and that although $ \D_A$ acts
tangentially to $ \cq$ to this order, it does not commute with $ Q$.
In fact for any tensor field $ V$, homogeneous of weight $w$ on $ \tM$, 
from \nn{ambDform} we have
\begin{equation}\label{DQ}
\D_A Q V= Q \D_A V + 4Q\aNd_A V.
\end{equation}
So, along the $ Q=0$ surface $ \cq$, $ \D_A$ acts tangentially, but, $
\D_A$ does not act tangentially to other $ Q=$ constant surfaces.
Nevertheless this allows us to conclude that if $ U$ and $ V$ are
tensors of the same rank (and with $U+QV$ homogeneous of some weight),
then
$$
\D_{A_1}\cdots \D_{A_\ell}(U+QV) = (\D_{A_1}\cdots \D_{A_\ell}U) +Q W 
$$
for some tensor $ W$. Thus, along $ \cq$, $\D_{A_1}\cdots
\D_{A_\ell}U $ is independent of how $ U$ is extended off $ \cq$.
The identities
\begin{equation}\label{XDidsAA}
        \X_A \D^A V= w(n+2w-2) V - Q \aDe V  
\end{equation}
and
\begin{equation}\label{XDidsBB}
        \D^A\X_A V =(n+2w+2)(n+w)V -Q \aDe V
\end{equation}
will also be useful.
Here $ V$ is a tensor which is homogeneous of weight $ w$.

We are now in a position to show directly how the tractor field
$W_{ABCD} $ is represented in the ambient setting.  Let us for the
while restrict to $ n\neq 4$. Note that the curvature of the ambient
connection $ \aR_{ABCD}$ is a section of $ \cce_{ABCD}(-2)$ and so
determines a section of the tractor bundle $ \ce_{ABCD}[-2]$. We will  
write $ R_{ABCD}$ to denote this section. 
Let $\tV\in
\cce^\Phi(w)$. From \nn{ambDform} we obtain
\begin{eqnarray*}
[\D_A,\D_B] \tV &=
&
  (n+2w-2)(n+2w-4)[\aNd_A,\aNd_B]\tV
\\
&&
-2(n+2w-2)\X_{[A}[\aDe,\aNd_{B]}] \tV. 
\end{eqnarray*}
Now let $V^A\in \ce^A $. We  write $ \tV=\tV^A\in \cce^A_{\cq}(0)$ for the
corresponding field on $\cq$, and extend this homogeneously to a
field on $ \tM$.
Then, along $\cq$, we have (see remark below)
$$
[\D_A,\D_B] \tV^C =
(n-2)(n-4)\aR_{AB}{}^C{}_E\tV^E+4(n-2)\X_{[A}\aR_{B]F}{}^C{}_E \aNd^F
\tV^E .
$$
Thus, since $ \X^F\aR_{BFCE}=0=\X_F\aNd^F \tV^E$, this implies 
$$
[D_A,D_B] V^C =
(n-2)(n-4)\tW_{AB}{}^C{}_E V^E+4(n-2)X_{[A}\tW_{B]F}{}^C{}_E Z^F{}_f\nd^f
V^E .
$$
Comparing this with \nn{Dcomm} (with $w $ set to 0 in that
expression) we can at once conclude that
$X_{[A}W_{BC]DE}V^E=(n-4)X_{[A}\tW_{BC]DE}V^E$. Since this holds for
any section $ V^A$ of $ \ce^A$, it follows from the definition of $
W_{ABCD}$ that $X_{[A}\Om_{BC]DE}=X_{[A}\tW_{BC]DE}$. Contracting with
$Z^F{}_f$ we have immediately $X_{[A}\tW_{B]F}{}^C{}_E Z^F{}_f=
X_{[A}\Om_{B]F}{}^C{}_E Z^F{}_f$. Substituting this in the above
display and once again comparing to \nn{Dcomm} we now have that
$W_{BCDE}V^E=(n-4)\tW_{BCDE}V^E$ for all $ V^E$, and so
\begin{equation}\label{WvsaR}
W_{BCDE}=(n-4)\tW_{BCDE} .
\end{equation} 

\noindent{\bf Remark:} Note that 
$$
[\aDe,\aNd_{B}] \tV_C = 2 \aR_{EBCF}\aNd^E\tV^F+(\aNd^E \aR_{EBCF})\tV^F
+ \aR_{BF}\aNd^F\tV_C .
$$
From the contracted Bianchi identity $\aNd^E
\aR_{EBCF}=2\aNd_{[C}\aR_{F]B}$, so in odd dimensions the last two
terms of the display vanish to all orders. In even dimensions recall
we have that, along $ \Cal Q$, $\aR_{BF}$ vanishes to order $ n/2-1$ and so in all even dimensions, other than 4, these last terms also vanish along $\cq$.

\section{The GJMS operators} \label{GJMS}

Using the properties of $ \D_A$, we observed in the previous section
that if $ V$ is a tensor homogeneous of weight $ 1-n/2$ then, along $
\cq$, $ \aDe V$ is independent of how $ V$ is extended off $\cq$. So $
\aDe$ gives an operator $ \aDe:\cce^\Phi(1-n/2)\to \cce^\Phi(-1-n/2)$,
and this descends to the generalised conformally invariant Laplacian
(or Yamabe operator) as in \nn{key}. The observation that the
conformally invariant Laplacian (on densities) can be obtained from an
ambient Laplacian in this way goes back to \cite{HughHurd} in the
conformally flat dimension 4 setting and to \cite{FG2} for the general
curved case. For the generalised conformally invariant Laplacian we can
also show this directly using the result
\begin{equation}\label{andQ}
 \aNd_A Q=2\X_A
\end{equation} 
from above. From this it follows
 that if $ U$ is a tensor field homogeneous of weight $w$ (i.e. $ U\in
 \cce^\Phi(w)$), then
\begin{equation}\label{delQ}
[\aDe,Q]U =2(n+2w+2) U . 
\end{equation}
Thus if $ V\in \cce^\Phi(1-n/2)$ and $ U$ is a tensor of the same rank
and type but homogeneous of weight $ -1-n/2$, then 
\begin{equation}\label{delQ2}
 \aDe (V+QU) = \aDe
V +Q\aDe U
.
\end{equation}
So clearly $ \aDe V$ is independent of how $ V$ extends off $ \cq$. In
\cite{GJMS}, Graham, Jenne, Mason, and Sparling establish a remarkable
generalisation of the result for densities which we state here in our
current notation.
\begin{proposition}\label{GJMSprop} For $ n$ even and $ k\in \{1,2,\cdots ,n/2\}$ or $ n$ 
  odd and $k\in {\Bbb Z}_+ $, let $ f\in \cce_{\cq}(k-n/2)$, and let
  $\tf\in\cce(k-n/2)$ be a homogeneous extension of $f$.  The
  restriction of $ \aDe^k \tf$ to $ \cq$ depends only on $ f$ and the
  conformal structure on $M$ but not on the choice of the extension $
  \tf$ or on any choices in the ambient metric.  Thus there is a
  conformally invariant operator
$$
  \aDe^k : \cce_\cq(k-n/2)\to \cce_\cq(-k-n/2) ,
  $$
  and this descends to a natural conformally invariant differential
  operator
$$
P_{2k}: \ce[k-n/2]\to \ce[-k-n/2]
$$
on $M$.
\end{proposition}
\noindent
As mentioned in the introduction, we call the operators $ P_{2k}$ the
GJMS operators.

In this section we will describe a way that one can directly rewrite
these operators in terms of $ \D_A$, $ \X_A$, the curvature $ \aR$, and
just one $ \aDe$. As observed above, each of these corresponds to an
object in the tractor calculus. Before we begin we need one more
result from \cite{GJMS}.  (This follows from Proposition~2.2 and
Section~3 from there).
\begin{proposition} \label{harm} 
For $ n$ even and $ k\in \{1,2,\cdots ,n/2\}$ or $ n$ 
  odd and $k\in {\Bbb Z}_+ $, let $ f\in \cce_{\cq}(k-n/2)$. Then $f$
  has an extension $ \tf\in \cce(k-n/2)$ uniquely determined
  modulo O$ (Q^k)$ by the
    requirement that $\aDe \tf=0 $ modulo O$ (Q^{k-1})$. The extension
  is independent of any choices in the ambient metric.
\end{proposition}

We are ready to consider an example. Let $\tilde{f}\in \cce(2-n/2)$,
and let $ f$ denote the section of $ \ce[2-n/2]$ that it determines.
Consider $ \aDe \bD_A \tilde{f} =\aDe (2\aNd_A\tilde{f}-\X_A\aDe
\tilde{f})$. Since in all dimensions the ambient Ricci curvature
vanishes along $ \cq$, we have $[\aDe,\aNd_A] \tilde{f}=0 $. So with the 
operator equality $ [\aDe, \X_A] = 2\aNd_A $ we immediately
see that
$$
\aDe \bD_A \tilde{f} = -\X_A \aDe^2 \tilde{f}
.
$$
Thus $ \Box D_A f =-X_A P_4 f$, where $ P_4$ is the fourth-order
GJMS operator (which agrees with the Paneitz operator). Note that
according to the earlier proposition above, the right-hand side is
independent of how $f$ extends off $ \cq$. So the left-hand
side is likewise independent of the choice of extension.
In fact this is already clear from \nn{DQ} and \nn{delQ2}. 

This suggests attempting to recover the higher order GJMS operators
from $ \aDe \bD_A \cdots \bD_B \tf$. On conformally flat structures
this is immediately successful. 
\begin{proposition} \label{flatcase} 
On conformally flat structures, if $\tf\in \cce(k-n/2)$,  $ k\in {\Bbb Z}_+$,
then 
  $$
  \aDe \bD_{A_{k-1}}\cdots \bD_{A_1} \tf =(-1)^{k-1}
  \X_{A_1}\cdots \X_{A_{k-1}}\aDe^k\tf.
$$
\end{proposition}
\begin{proof}
We are only interested in local
results and differential operators.  So without loss of generality we
suppose that we are in the setting of the flat model for which the
ambient space is simply $ {\Bbb R}^{n+2}$ equipped with the flat
metric $ \h$ given by a fixed bilinear form of signature $(p+1,q+1) $
and the standard parallel transport. The latter also gives the ambient
connection in this setting. In the standard coordinates, $
\X=\X^I\partial/\partial\X^I$ at the point $\X^I$, and  the
identities of the previous section hold as genuine equalities rather
than just formally.

 We have the operator identity $[\aDe,\X_A]=2\aNd_A$ on
  sections of $ \cce^\Phi(w)$.  Since the structure is conformally flat,
  we also have $ [\aDe,\aNd_A]=0$.  It follows that
$
[\aDe^m,\X_A]=2m\aDe^{m-1}\aNd_A .
$
Thus if $ \tf\in\cce^\Phi(m+1-n/2)$, we have
$$
-\aDe^m \bD_A \tf=- \aDe^m [2m\aNd_A \tf-\X_A \aDe \tf]= \X_A
\aDe^{m+1} \tf.
$$
The proposition now follows by induction on $k$. 
\end{proof}

To relate $\aDe \bD_{A_{k-1}}\cdots \bD_{A_1} \tf$ and $ \aDe^k
\tf$ in the general case we must take account of the curvature of the
ambient manifold.
Since this is Ricci flat we have that if $\tV_B\in
\cce_A(w)$, then $ [\aDe, \aNd_A]\tV_B= -2 \aR_A{}^P{}_B{}^{Q}\aNd_P\tV_Q$.
More generally if  $ \tV_{BC\cdots E}\in \cce_{BC\cdots E}(w)$, then 
\begin{equation}
\label{L17May01c}
\begin{array}{rll}
\lefteqn{[\aDe, \aNd_A]\tV_{BC\cdots E}=}&&
\\
&& -2\aR_A{}^P{}_B{}^{Q}\aNd_P\tV_{QC\cdots
  E}-2\aR_A{}^P{}_C{}^{Q}\aNd_P\tV_{BQ\cdots E} -\cdots
\\
&&
-2 \aR_A{}^P{}_E{}^{Q}\aNd_P\tV_{BC\cdots  Q}
.
\end{array}
\end{equation}
In even dimensions the ambient metric is only Ricci flat and
determined by the conformal structure on $ M$ to finite order, as
described above. For example for even $n$ \nn{L17May01c} only holds
mod O$(Q^{n/2-2})$ (or mod O$(Q^{n/2-1})$ if $\tV$ has rank 0).  For
simplicity in the following discussion we will often ignore this point
and assume the given calculations do not involve sufficient transverse
derivatives of the ambient metric to encounter this problem. We will
return to a careful count of tranverse derivatives later in the
section. We will also henceforth restrict to $ n\neq 4$. This also
simplifies matters.  And there is no loss, as the results for $ n=4$
have been obtained above.

It follows from the last display that if $ \tf\in \cce(w)$ (and $ \ell<n/2$ if $ n$ is even), then
\begin{equation}
\label{Deltanablas}
\begin{array}{lll}
\lefteqn{\aDe \aNd_{A_\ell}\cdots \aNd_{A_1} \tf =}\ \ \ \ \ \ \ \ &
\\
&&
-2\aR_{A_{\ell}}{}^P{}_{A_{\ell-1}}{}^{Q}\aNd_{P}\aNd_{Q}
\aNd_{A_{\ell-2}}\cdots \aNd_{A_1} \tf-\cdots
\\
&&-2\aR_{A_{\ell}}{}^P{}_{A_{1}}{}^{Q}
\aNd_{P}\aNd_{A_{\ell-1}}\cdots \aNd_{A_2}\aNd_{Q}\tf
\\
&&-2\aNd_{A_\ell} \aR_{A_{\ell-1}}{}^P{}_{A_{\ell-2}}{}^{Q}\aNd_{P}\aNd_{Q} 
\aNd_{A_{\ell-3}}\cdots \aNd_{A_1} \tf-\cdots
\\
&&-2\aNd_{A_\ell}\cdots \aNd_{A_3}
\aR_{A_{2}}{}^P{}_{A_{1}}{}^{Q}\aNd_{P}\aNd_{Q} \tf
+
\\
&& \aNd_{A_\ell}\cdots \aNd_{A_1} \aDe \tf ,
\end{array}
\end{equation}
where here all $ \aNd_{A}$'s act on all tensors to their right and the result is mod O$(Q^{n/2-\ell}) $ if $ n$ is even.  We
may apply the Leibniz rule to \nn{Deltanablas}.  The term
$\aNd_{A_\ell}
\aR_{A_{\ell-1}}{}^P{}_{A_{\ell-2}}{}^{Q}\aNd_{P}\aNd_{Q}
\aNd_{A_{\ell-3}}\cdots \aNd_{A_1} \tf$, for example, becomes
$$
\begin{array}{l} 
(\aNd_{A_\ell}
\aR_{A_{\ell-1}}{}^P{}_{A_{\ell-2}}{}^{Q})\aNd_{P}\aNd_{Q}
\aNd_{A_{\ell-3}}\cdots \aNd_{A_1} \tf +
\\
\aR_{A_{\ell-1}}{}^P{}_{A_{\ell-2}}{}^{Q}\aNd_{A_\ell}\aNd_{P}\aNd_{Q}
\aNd_{A_{\ell-3}}\cdots \aNd_{A_1} \tf .
\end{array}
$$
Often we will not require the details of contractions or the value
of coefficients, and so we might write the last result
symbolically as $\aNd \aR \aNd^{\ell-1}\tf= (\aNd \aR) \aNd^{\ell-1}\tf + \aR
\aNd^\ell \tf $. (In this informal notation we will write $ \aNd$ to
indicate a $\aNd_A$ which is not part of a $ \aDe$. For example, it
may have a free index or be contracted to the ambient curvature
$\aR$.)  We may repeatedly apply the Leibniz rule to (\ref{Deltanablas}) in
this way until all of the terms on the right-hand side are of the form
(omitting indices) $(\aNd^p \aR)\aNd^q \tf$. We might write the result
symbolically as
\begin{equation} \label{Deltanablas2}
        \aDe \aNd^\ell \tf=  
        \aNd^\ell \aDe \tf +\sum (\aNd^p \aR)\aNd^q \tf
        .
\end{equation}
Note that each term of the second sort on the right-hand side has $
q\geq 2$
and $ p+q =\ell$.  Although in these symbolic formulae we
omit the details of the contractions and the coefficients, we really
want to regard these expressions as representing precise formulae. The
idea of this notation is simply to manifest explicitly only the aspects
of the formulae that we need for our general discussion.

Now observe that 
$$
\begin{array}{lll}
        \lefteqn{(n+2w-2\ell-2)\aNd_{A_{\ell+1}}\aNd_{A_\ell}\cdots
        \aNd_{A_1}\tf=}
\\
&&
        \bD_{A_{\ell+1}} \aNd_{A_\ell}\cdots \aNd_{A_1}\tf+
        \X_{A_{\ell+1}}\aDe \aNd_{A_\ell}\cdots
\aNd_{A_1} \tf ,
\end{array}
$$
or, in our symbolic notation, $(n+2w-2\ell-2) \aNd^{\ell+1} \tf= \bD
\aNd^\ell \tf+ \X \aDe \aNd^\ell \tf$.  We can substitute (\ref{Deltanablas2})
into the right-hand side of this and so observe that if
$n+2w-2\ell-2\neq 0 $, then we can replace a term $\aNd^{\ell+1}\tf $
by the expression $ \bD \aNd^\ell \tf + \X \sum (\aNd^p \aR)\aNd^q \tf +
\X \aNd^\ell \aDe \tf $.  Suppose $ w=k-n/2$.
Then $n+2w-2\ell-2=2(k-\ell-1)$, and we have
\begin{equation}\label{Dexp}
2(k-\ell-1) \aNd^{\ell+1}\tf = 
\bD \aNd^\ell \tf + \X \sum (\aNd^p \aR)\aNd^q \tf + \X \aNd^\ell \aDe\tf.
\end{equation}
In each term of the sum we again have $ q\geq 2$ and $ p+q=\ell$.
Note that the left-hand side of \nn{Dexp} has at most $\ell+1$
transverse derivatives of $\tf$.  Apart from the term $\X \aNd^\ell
\aDe\tf $, which we will deal with below, the right-hand side has at
most $\ell$ transverse derivatives of $ \tf$, as $ \D$ acts
tangentially to $\cq$.  Our strategy below will be to replace $
\aNd$'s with $ \D$'s beginning from the left.

We may apply similar reasoning to $\aR$.  Since $\aR$ has weight $-2$,
we have
$ (n-2m-4) \aNd^{m}\aR = \bD \aNd^{m-1} \aR + \X \aDe \aNd^{m-1} \aR .  $
Here we have used the same informal notation that we used with $\tf$,
above.  By (\ref{L17May01c}) we may write this as
$ (n-2m-4) \aNd^{m}\aR = \bD \aNd^{m-1} \aR + \X \sum (\aNd^p
\aR)\aNd^q \aR + \X \aNd^{m-1} \aDe \aR .  $
Now note that since $ \aR$ is Ricci flat, we have
\begin{equation}
\begin{array}{c}
\label{L17Maya}
\aDe \aR_{BCDE} = \\
2
\left(
\aR^{A}{}_{CB}{}^{F}\aR_{FADE}+
        \aR^{A}{}_{CD}{}^{F}\aR_{BAFE}+
        \aR^{A}{}_{CE}{}^{F}\aR_{BADF}
\right)
        ,
\end{array}
\end{equation}
from the Bianchi identity. In odd dimensions this holds to all
orders.  In general we have $\aDe \aR_{BCDE}=
2(\aNd_B\aNd_{[D}\aR_{E]C}-\aNd_C\aNd_{[D}\aR_{E]B}) +{\rm O}(\aR^2)$
where ${\rm O}(\aR^2) $ indicates the quadratic term in the display.
Using, once again, that in even dimensions $\aR_{AB}=
Q^{n/2-1}\aL_{AB}$ it follows that
$(\aNd_B\aNd_{[D}\aR_{E]C}-\aNd_C\aNd_{[D}\aR_{E]B})$ vanishes to
order $ n/2-3$, and so \nn{L17Maya} holds to that order.

Thus we get the simplification   
\begin{equation}\label{Rexp}
        (n-2m-4) \aNd^{m}\aR = 
        \bD \aNd^{m-1} \aR + \X \sum (\aNd^p \aR)\aNd^q \aR 
        ,
\end{equation}
where in each term of the sum, $ p+q=m-1$. In even dimensions we need
$m< n/2-2$. This follows immediately from the previous paragraph. That 
is, we need $(n-2m-4)>0$.

Our effort to relate $\aDe \bD_{A_{k-1}}\cdots \bD_{A_1} \tf$ and $ \aDe^k
\tf$ involves another identity, viz
\begin{equation}
\label{L11May01}
        \aDe(\aNd^t\aDe^{u}\aR)E
        =
        (\aDe\aNd^t\aDe^{u}\aR)E
        +(\aNd^t\aDe^{u}\aR)\aDe E
        +2(\aNd^{t+1}\aDe^u\aR)\aNd E
        .
\end{equation}
Here $E$ is any expression (for a linear operator) which, in terms of
our informal symbolic notation, is a polynomial in $\aNd$, $\aDe$,
$\aR$, and $\tf$.  We also need the following fact which follows from
the above:
\begin{lemma}\label{curv}
  Suppose $n$ is odd or $t+u\leq n/2-3$.  Then on
  $ \cq$ there is an expression for 
  $\aNd^t\aDe^u \aR$ as a partial contraction
  polynomial in $\D_A$, $ \aR_{ABCD}$, $ \X_A$, $\h_{AB}$, and its
  inverse $ \h^{AB}$. This expression is rational in $ n$, and each
  term is of degree at least 1 in $\aR_{ABCD} $.  
\end{lemma}
\begin{proof}
  Repeatedly use \nn{L17May01c}, \nn{L17Maya}, and \nn{L11May01} to
  rewrite $\aNd^t\aDe^u \aR$ as a sum of terms of the form
  $(\aNd^{v_1}\aR)\cdots(\aNd^{v_j}\aR)$.  In doing this we convert
  some $\aDe$'s into pairs of $\aNd$'s via \nn{L11May01}, but at most
  one $\aNd$ from each pair acts on any given $\aR$.  Thus in even
  dimensions, $v_i\leq n/2-3$, $i\in\{1,\cdots,j\}$, and we may
  construct the desired partial contraction polynomial by repeatedly
  applying \nn{Rexp} to the terms
  $(\aNd^{v_1}\aR)\cdots(\aNd^{v_j}\aR)$. In even dimensions, using
  \nn{delQ} and  $\aNd_AQ=2\X_A$ with the restriction $t+u\leq n/2-3$,
  we see that that \nn{L17May01c}, \nn{L17Maya}, and \nn{Rexp} all hold to
  sufficient order.
\end{proof}

Now let $ f\in\cce_\cq(k-n/2)$. Suppose $\tf\in\cce(k-n/2)$ is any
homogeneous extension of $ f$ as in Proposition~\ref{harm}.  We will
consider $ \aDe \bD_{A_{k-1}}\cdots \bD_{A_1} \tf$, where $ k$ is a
positive integer.  If $ n$ is even, we assume that $ k\leq
n/2$. Let us systematically rewrite this in terms of $(-1)^{k-1}
\X_{A_1}\cdots \X_{A_{k-1}}\aDe^k\tf$ and curvature coupled terms via
the following steps:\\
{\bf Step 1:} Observe that
$$\aDe \bD_{A_{k-1}}\cdots \bD_{A_1} \tf =\aDe
(2\aNd_{A_{k-1}}-\X_{A_{k-1}}\aDe)\cdots
(2(k-1)\aNd_{A_1}-\X_{A_1}\aDe)\tf.$$
Expand this out via the
distributive law
without changing the order of any of the operators. \\
{\bf Step 2:} Move all $ \X$'s to the left of any $ \aNd$ or $ \aDe$ via
the identities $ [\aNd_A,\X_B]=\h_{AB}$ and
$ [\aDe , \X_A]= 2\aNd_A$ (which hold to all orders).  \\
{\bf Step 3:} Move all $\aDe$'s to the right of any $\aNd$'s (other
than those implicit in $ \aDe$) via (\ref{Deltanablas}), and
\nn{L11May01}. In even dimensions one of course needs to be careful,
since (\ref{Deltanablas}) is valid only if $ \ell<n/2$ and holds mod
O$(Q^{n/2-\ell})$. Elementary counting arguments (along similar lines
to the discussion in the next paragraph) quickly establish that for
terms encountered we have $ \ell<k$ satisfied and with no more than $(
k-\ell-1)$ transverse derivatives of the result. Since we assume $
k\leq n/2$ when $ n$ is even the use of \nn{Deltanablas} is valid. Next
by the proof of Proposition \ref{flatcase}, we may cancel all terms
not explicitly involving the curvature except for the term $(-1)^{k-1}
\X_{A_1}\cdots\X_{A_{k-1}}\aDe^k\tf$.  (The proof of Proposition
\ref{flatcase} involves only the identities used in Steps 1 and 2 with
just the difference that these are applied in a different order.)  We
thus obtain
\begin{equation}\label{form}
        (-1)^{k-1}\X^{k-1}\aDe^k \tf + \sum \h^s \X^x
        (\aNd^{p_1}\aDe^{r_1}\aR) \cdots (\aNd^{p_d}\aDe^{r_d}\aR) \aNd^q
        \aDe^r \tf
        ,
\end{equation}
where $d\geq 1$ in each term of the right-hand part.
\vspace{2.5mm}

At this point let us take stock of what we have.  For each term in the
result of Step~1, the sum of the number of $\aDe$'s in the term and
the number of $\aNd$'s in the term is exactly $k$.  In Steps~2 and 3
some $\aDe$'s may have been exchanged for $ \aNd$'s via the identity $
[\aDe , \X_A]= 2\aNd_A$ or for $ \aR$'s via the commutator
$[\aDe,\aNd_A]$, and similarly we may have lost some $ \aNd$'s by $
[\aNd_A,\X_B]=\h_{AB}$.  On the other hand, we may have converted some
$\aDe$'s into pairs of $\aNd$'s via (\ref{L11May01}); note that at
most one $ \aNd$ from each pair acts on $ \tf$, and similarly at most
one $ \aNd$ from each pair acts on any given $\aR$.  Thus for each
term of the sum in (\ref{form}) we must have $d+q+r\leq k$.  Since
$d\geq 1$, it follows that $k-q-r\geq 1$.  Note that
each $\aR$ in (\ref{Deltanablas}) is followed by at least two
$\aNd$'s.  Thus at each step in the construction of the right-hand
part of (\ref{form}), each of the rightmost two $\aNd$'s of each term
arose from Steps 1 and 2, and not from the use of (\ref{L11May01}).  It
follows that at least two of the $\aNd$'s in $\aNd^q\aDe^r\tf$ did
\textit{not} arise from (\ref{L11May01}).  Thus $q\geq 2$, and for any
$ i\in \{1,\cdots, d\}$, $p_i+r_i+3\leq k$.  Now suppose $n$ is even.
Then, by assumption, $k\leq n/2$, and for $ i\in \{1,\cdots, d\}$ we
have $p_i+r_i\leq n/2-3$.  Since the ambient metric is determined modulo
terms of O$(Q^{n/2})$, it follows immediately that the metric
connection $ \aNd$ is determined modulo terms of O$(Q^{n/2-1})$.
Its curvature $ \aR$ is similarly determined modulo O$(Q^{n/2-2})$.
Now when $\aDe=\aNd^A\aNd_A$ acts on functions, its rightmost $ \aNd$
is really just the exterior derivative.  Thus as an
operator on functions, $\aDe$ is determined modulo terms of
O$(Q^{n/2-1})$.  It now follows from \nn{andQ}
and \nn{delQ} that, as an operator on $ \cce(k-n/2)$, all terms in the sum of
(\ref{form}) are determined uniquely modulo O$(Q)$.  If $ n$ is odd,
the ambient metric is determined to infinite order so certainly the
same is true.

Next, by Proposition~\ref{harm} we can assume that $\tf$ satisfies
$\aDe \tf =0$ modulo O$(Q^{k-1})$, and given $f$, this determines $
\tf$ uniquely modulo O$(Q^{k})$.  This will simplify our arguments.
The end result will be independent of this choice.  Since $k-q-r\geq
1$, we see immediately that all terms in (\ref{form}) with $r\geq 1$
will vanish modulo O$(Q)$.  We will thus delete these terms.  From the
inequality $k-q-r\geq 1$ and, in even dimensions, the inequality
$p_i+r_i\leq n/2-3$, it follows that we can carry out the next step.
\vspace{2.5mm}

\noindent {\bf Step 4:}  First rewrite (\ref{form}) as
\begin{equation}\label{L24Mayaa}
        (-1)^{k-1}\X^{k-1}\aDe^k \tf + \sum \h^s \X^x
        (\aNd^{p_1}\aDe^{r_1}\aR) \cdots (\aNd^{p_d}\aDe^{r_d}\aR) \aNd^q
        \tf
        .
\end{equation}
Then repeatedly use \nn{Dexp} and Lemma~\ref{curv} to eliminate all
$\aNd$'s and $ \aDe$'s from the right-hand part of this
expression.  The use of (\ref{Dexp}) introduces additional
$\aDe$'s.  But terms containing these $\aDe$'s vanish modulo O$(Q)$,
and we cancel them as soon as they appear.  We obtain as result
\begin{equation}
\label{L17Mayb}
        (-1)^{k-1}\X^{k-1}\aDe^k \tf
        =
        \aDe \D^{k-1}\tf +\sum \h^s \Psi\tf,
\end{equation}
where, in terms of our informal symbolic notation, the operator $\Psi$ is a
polynomial in $\X$, $\D$, and $\aR$.  The exponent $s$ here is not
claimed to bear any relationship to the $s$ from earlier.
The only differential operator of non-zero order used in the formula
is $ \D$.  Thus although we used the $ \tf$ satisfying $\aDe \tf =0$
modulo O$(Q^{k-1})$ to obtain (\ref{L17Mayb}), observe now that it
follows immediately from \nn{DQ} that each term depends only on $f$
and is otherwise independent of the extension $\tf$.  Thus for any
extension $\tf$, (\ref{L17Mayb}) holds modulo O$(Q)$.\\
{\bf Remark:} At this point it is worthwhile to justify our use of
(\ref{Dexp}) in Step~4.  First note that in each term in
(\ref{L24Mayaa}) we have $q\leq k-1$, by the counting given above.
Thus in (\ref{Dexp}), $\ell+1$ will always be at most $k-1$, $\ell$
will be at most $k-2$, and $k-\ell-1$ will be nonzero.  We may
therefore solve for $\aNd^{\ell+1}\tf$ in \nn{Dexp}.  On the other
hand the use of (\ref{Dexp}) may generate additional curvature terms
$(\aNd^p\aR)\aNd^q\tf$.  But $p+q=\ell$, where $q\geq 2$.  Thus in
even dimensions, $p\leq \ell-2\leq k-4\leq n/2-4$,
and we may apply Lemma~\ref{curv} to $\aNd^p\aR$.

In the final step we will use the fact that $ (n-4)\aR$ descends to
the tractor field $W$, $ \X$ descends to $ X$, $ \h$ descends to $ h$,
and that $ \aDe :\cce^\Phi(1-n/2)\to \cce^\Phi(-1-n/2)$ descends to
$\Box: \ce^\Phi[1-n/2]\to \ce^\Phi[-1-n/2]$.

\noindent {\bf Step 5:} In the right-hand side of (\ref{L17Mayb})
make the following formal replacements: $ \tf$ with $f$, $ \aDe$ with
$ \Box$, $ \X$ with $ X$, $ \h$ with $ h$, $ \aR$ with $ W/(n-4)$ (in
dimensions $ n\neq 4$) and $ \D$ with $ D$. The result is a tractor
formula for $(-1)^{k-1}X^{k-1} P_{2k} f $. We state this as a proposition.
\begin{proposition}\label{tract4gjms}
There is a tractor calculus expression for the GJMS operators of the form
\begin{equation}
\label{tcalexp}
\begin{array}{lll}
\lefteqn{X_{A_1}\cdots X_{A_{k-1}}P_{2k} f =}
&&
\\
&&
(-1)^{k-1}\Box D_{A_{k-1}} \cdots D_{A_{1}} f + \Psi_{A_{k-1} \cdots
A_{1}}{}^{PQ}D_PD_Q f
        ,
\end{array}
\end{equation}
where $ f\in \ce[k-n/2]$ and $ \Psi$ is a linear differential
operator 
$$
\Psi_{A_{k-1} \cdots
A_{1}}{}^{PQ}:\ce_{PQ}[k-2-n/2] \to \ce_{A_{k-1} \cdots A_{1}}[-1-n/2],
$$
expressed as a partial contraction polynomial in $D_A$, $ W_{ABCD}$, $
X_A$, $h_{AB}$, and its inverse $ h^{AB}$. The expression for $
\Psi$ is rational in $ n$, and each term is of degree at
least 1 in $W_{ABCD} $.
\end{proposition}
\begin{proof}
It is clear from the argument of this section that 
$$
X_{A_1}\cdots X_{A_{k-1}}P_{2k} f =(-1)^{k-1}\Box D_{A_{k-1}} \cdots
D_{A_{1}} f + \Psi_{A_{k-1} \cdots A_{1}} f,
$$
where $ \Psi_{A_{k-1} \cdots A_{1}}$ is a linear differential operator
on $ f$ expressed as a partial contraction polynomial in $D_A$, $
W_{ABCD}$, $ X_A$, $h_{AB}$, and its inverse $ h^{AB}$.  It is also
clear that that this expression for $ \Psi$ is rational in $ n$ and
that each term is of degree at least 1 in $W_{ABCD} $.  Furthermore,
recall that in Step~4 we used (\ref{Dexp}) and Lemma~\ref{curv} to
convert the expression $\aNd^q\tf$ of (\ref{L24Mayaa}) into an
expression in $\D$, $\X$, $\aR$, $ \h$, and $ \h^{-1}$. 
Since $q\geq 2$ in (\ref{Dexp}) and
(\ref{L24Mayaa}), it follows that each term of this tractor expression
ends in two consecutive $\D$'s.  The result now follows.
\end{proof}
We conclude this section with examples. 

\subsection{Examples} \label{examples}

The simplest example of our procedure is the Paneitz operator $ P_4$,
which we treated at the outset of this section. Recall that we obtained $
\Box D_A f =-X_A P_4 f$, and it is clear that the tractor expression on
the left-hand side of this is independent of any choices in the ambient
construction. This is as guaranteed by the argument following Step~3.

The next simplest case is of course the operator $P_6$. By assumption
then, $n\neq 4 $. Let $f$ denote a section of $ \ce[3-n/2]$. Let $
\tilde{f}$ be a section of $ \cce(3-n/2)$ such that its restriction to
$ \Cal Q$ agrees with $ f$ and such that, $ \aDe
\tilde{f}=Q^2 g$ for some smooth $ g\in \cce(-3-n/2)$.  Expanding out
$ \aDe \bD_A \bD_B \tilde{f}$ according to Steps 1 and 2 gives
$$
\begin{array}{lll}
\lefteqn{\aDe \bD_A \bD_B \tilde{f} =}
&&
\\
&&
        \X_A \X_B \aDe^3 \tilde{f}
        +2\X_B[\aNd_A,\aDe]\aDe \tilde{f} +
\\
&&
         2\X_A[\aNd_B,\aDe]\aDe\tilde{f} 
        +4\X_A\aDe [\aNd_B,\aDe]\tilde{f} -8[\aNd_A,\aDe ]\aNd_B\tilde{f}.
\end{array}
$$
Since  $[\aDe,\aNd_A] $ vanishes on functions mod O$(Q^2)$, 
the third step reduces to  
$$
\begin{array}{lll}
        \lefteqn{\aDe \bD_A \bD_B \tilde{f}=
        \X_A \X_B \aDe^3 \tilde{f}-
        8[\aNd_A,\aDe ]\aNd_B \tilde{f}=}
&&
\\
&&
        \X_A \X_B \aDe^3 \tilde{f} -16\aR_A{}^C{}_B{}^E\aNd_C\aNd_E \tilde{f}
        ,
\end{array}
$$
along $\cq$. 
The fourth step is simply the observation that on $ \Cal Q$ (with $
\tf$ as above) we have
$$
8 \aNd_C\aNd_E \tilde{f} = \bD_C \bD_E \tilde{f}  
$$
and so  
$$ 
\X_A \X_B
\aDe^3 \tilde{f} =\aDe \bD_A \bD_B \tilde{f}
+2 \aR_A{}^C{}_B{}^E \bD_C \bD_E \tilde{f}.
$$
As we have observed generally, at this stage none of the terms on
either side depend on how $ \tilde{f}$ extends off $ \Cal Q$. Thus
finally we have
$$
\Box D_A D_B f +\frac{2}{n-4}W_A{}^C{}_B{}^E D_C D_E f =X_A X_B P_6 f,
$$
where $P_6$ is the sixth-order GJMS operator. Thus as promised, we
have recovered the tractor formula found by other means in
Section~\ref{powersLap}.  

Our final example is $P_8$.  By following Steps~1 through 4, above,
and by applying (\ref{XRrel}),
we\newcommand{\strutbb}{\rule{0mm}{4mm}}
obtain
$$
\begin{array}{rll}
        \lefteqn{-\X_A \X_B \X_C \aDe^4 \tf=}&&
\\
        &&
        \aDe \D_A \D_B \D_C \tf
        +
        2\aR_A{}^P{}_B{}^Q \D_P \D_Q \D_C \tf
        \strutbb
        +
        2\aR_A{}^P{}_C{}^Q \D_P \D_B \D_Q \tf
        \strutbb
\\
        &&
        -
        \frac{2}{(n-6)}\X_A
        (\D_E \aR_B{}^P{}_C{}^Q)
        \D^E \D_P \D_Q \tf
        \strutbb
\\
        &&
        +
        4\X_A \aR_B{}^P{}_C{}^Q
        \aR_P{}^E{}_Q{}^F
        \D_E \D_F
        \tf
        \strutbb
        -2\X_A \aDeR
        \D_P \D_Q
        \tf
        \strutbb
\\
        &&
        -
        \frac{2}{n-6}
        \X_A \X_E
        \aDeR
        \D^E \D_P \D_Q \tf
        \strutbb
\\
        &&
        +
        \frac{4}{(n-6)}\X_A \X^E
        (\D_E \aR_B{}^P{}_C{}^Q)
        \aR_P{}^F{}_Q{}^G
        \D_F \D_G
        \tf
        .
        \strutbb
\end{array}
$$
Here $\aDeR$ denotes the tractor field
$$
\begin{array}{l}
        \hspace{3.50mm}
        2\left(
        \aR^{AP}{}_B{}^F \aR_{FAC}{}^Q
        +
        \aR^{AP}{}_C{}^F \aR_{BAF}{}^Q
        +
        \aR^{APQF}
        \aR_{BACF}
        \right)
        .
\end{array}
$$

To demonstrate explicitly that $ P_8$ is formally self-adjoint, a variation
on this formula is preferred. It is a straightforward exercise to
rewrite the above equation as follows.
$$
\begin{array}{rll}
        \lefteqn{\X_A\X_B\X_C\aDe^4\tf=}
\\
        &&
        -\aDe\D_A\D_B\D_C\tf
        \strutdd
        -2\aR_A{}^P{}_B{}^Q\D_P\D_Q\D_C\tf
        \strutdd
        -2\aR_A{}^P{}_C{}^Q\D_P\D_B\D_Q\tf
        \strutdd
\\
        &&
        -
        \frac{4}{n-6}\X_A
     (\aDeR)\D_P \D_Q \tf
        \strutdd
        +
        \frac{2}{n-6}\X_A\D^E \aR_B{}^P{}_C{}^Q\D_E\D_P \D_Q \tf
        \strutdd
\\
        &&
        +
        4\frac{n-2}{n-6}\X_A
        \aR_B{}^P{}_C{}^Q \aR_P{}^S{}_Q{}^T\D_S\D_T \tf
        \strutdd
        .
\end{array}
$$
This together with \nn{L17Maya} yields the tractor formula of
Proposition~\ref{p468}.

%
%

\end{document}